\documentstyle[12pt,a4wide,epsf,./mathsymb]{article}
\begin{document}

\title{
\begin{flushright}
{\normalsize SUSX-TH-96-011}\\[-3mm]
{\normalsize \tt hep-th/9608098}\\[1cm]
\end{flushright}
Universality and Critical Phenomena in String Defect
Statistics}
\author{Karl Strobl and Mark Hindmarsh\\
{\it Centre for Theoretical Physics,}\\
{\it University of Sussex, Brighton BN1 9QH, U.K.}}

\date{August 1996\\[4mm]
{\small\tt PACS: 11.27.+d 61.30.Jf 61.72.Lk 98.80.Cq}}
\maketitle

\begin{abstract}
The idea of biased symmetries to avoid or alleviate cosmological problems
caused by the appearance of some topological defects is familiar
in the context of domain walls~\cite{ZelKobOk},
where the defect statistics lend themselves
naturally to a percolation theory description \cite{Coulson},
and for cosmic strings~\cite{V91,MHKS}, where the proportion
of infinite strings can be varied or disappear entirely depending on
the bias in the symmetry.

In this paper we measure the initial configurational
statistics of a network of string defects after a symmetry-breaking phase
transition with initial bias in the symmetry of the ground state.
Using an improved algorithm, which is useful for a more general class of
self-interacting walks on an infinite lattice, we extend
the work in \cite{MHKS} to better statistics and a different ground state
manifold, namely $\R P^2$, and explore various different discretisations.

Within the statistical errors, the critical exponents of the Hagedorn
transition are found to be quite possibly universal and identical to the
critical exponents of three-dimensional bond or site percolation.
This improves our understanding of the
percolation theory description of defect statistics after a biased
phase transition, as proposed in \cite{MHKS}. We also find strong evidence
that the existence of infinite strings in the Vachaspati Vilenkin algorithm
is generic to all (string-bearing)
vacuum manifolds, all discretisations thereof, and all regular
three-dimensional lattices.
\end{abstract}

\newpage
Many symmetries in nature are not exact, including internal symmetries
in field theories.
A simple example of an approximate symmetry is a spin
system in an external field, e.g.~a nematic liquid crystal with
diamagnetic molecules \cite{deG}.
In particle physics, an example of an approximate symmetry is
the Peccei-Quinn symmetry $U(1)_{\rm PQ}$,
associated with the axionic degree of freedom, which has the
degeneracy of its ground state manifold lifted at the QCD scale.

Condensed--matter systems, as well as the vacuum in the early Universe,
are -- in the process of being cooled -- subject to phase transitions.
If these transitions
are accompanied by symmetry breaking, they may lead
to the formation of defects: domain walls, strings or vortices, monopoles,
or a combination of these, depending on the topology of the set of
equilibrium states after the phase transition.
In cosmology such defects are associated with internal
symmetries of a field theory, while in condensed matter systems
there are defects associated with the rotational symmetry of the
ground--state. Such defects in a nematic liquid crystal are called
disclinations \cite{deG}. Along a closed path around line disclinations,
the orientation of the molecules rotates by an angle
$\pi$, while in a point defect the molecules
are in a ``hedgehog" configuration (or a continuous deformation
thereof), directed away from a central
point. On the central points of the disclinations the molecules
cannot have any alignment directions.

Topological defects form through what in
cosmology is called the Kibble mechanism~\cite{Kibble-mechanism}:
The simple requirement of causality prevents regions of the Universe
which are separated by more than twice the horizon distance
$\int_0^ta(t')dt'$ from being correlated. The actual correlation
length can of course be much smaller.
It is this lack of correlation which  allows the initial
conditions to trap topological defects after phase transitions.
Defects are also a convenient and compelling way to seed large-scale
structure formation in the early Universe.
A crucial ingredient for the usual string-seeded structure formation
scenario is that the string configurations develop a scaling solution
\cite{VilShel,MarkTom}, which in turn seems to depend on the initial
scale-invariance of the network.
The presence or absence of infinite strings (or, in a closed universe,
of strings that wrap around it) also seems to affect the string density in
the scaling solution~\cite{NeilPedro}: moreover, the question whether any
respectable fraction of the string-mass ought to be in infinite strings
is still controversial~\footnote{The usual Vachaspati-Vilenkin algorithm
yields lattice dependent results for this fraction, which can be attributed
to the assumption of an unphysical lattice-dependent lower cut-off in the loop
length-distribution~\cite{MHKS,thesis}. It has been claimed that
a complete absence of infinite strings can be achieved by similar
algorithms on generalised graphs, corresponding to an irregular
discretisation of allowed string-positions, obtained
through modelling the collisions of true vacuum-phase bubbles after a
first order phase transition~\cite{Julian}. However, the reason for
the very small fraction of infinite strings in~\cite{Julian} has not yet
been identified, and various proposals are on the
market~\cite{RobinsonYates,fractal}.},
and far from being decidable by analytical means.

What is desired for strings, is a cosmological disaster for domain walls:
infinite walls would come to dominate the expansion very quickly, which is
incompatible with stronger bounds from the cosmic microwave background
\cite{Okun1975}.
A convenient way to escape this problem has been
to make the symmetry between the disjoint (sets of) vacuum states
an approximate one~\cite{Coulson}.
In particle physics such adjustments are possible and in fact
necessary for instance within attempts to explain the family
hierarchy~\cite{Frampton+Kephart}. Similar biases in other symmetries can
affect the configurational statistics for all kinds of topological defects.
Attempts to circumvent the monopole
problem by allowing monopoles to annihilate with anti--monopoles at
a sufficient rate (facilitated through an acclaimed tendency for the
two to accompany each other) have been made~\cite{LeePro},
however, effects which soft symmetry breaking might have
to facilitate the monopole anti--monopole correlations to a higher
degree have not been considered. In this paper we will not address
monopoles, but attempt to
establish such a description in the case of cosmic strings and
topological line defects in solids,
i.e.~for softly spoiled symmetries with a non--trivial first homotopy
group. The technical details of the lattice algorithm, and the proof
that it conserves (in most cases) the essential invariants of the
continuum theory, is presented in a separate paper \cite{numerics}.

In \cite{MHKS} it
was pointed out that the existence of infinite strings may be understood
as a percolation phenomenon, and that associated critical behaviour
can be observed at a transition into a phase where the string network
only exhibits loops.
In this paper we show that this is indeed a
percolation transition, and can only be obtained -- once the vacuum
manifold and the discretisation thereof and of space-time are chosen --
by biasing the symmetry of the ground state. Although still no proof
is available, we find strong
evidence that the existence of infinite strings in the
Vachaspati-Vilenkin algorithm at perfect vacuum symmetries is generic to all
vacuum manifolds, all discretisations thereof, and to all regular
three-dimensional lattices.

We also find that the critical exponents for configurational parameters
near the percolation threshold are universal for different vacuum
manifolds, and identical (to within our statistical errors) to the
corresponding critical exponents in
standard bond or site percolation theory. For the case of an
$\R P^{\infty}$ symmetry and a minimally discretised $U(1)$ symmetry,
plausibility
arguments for this correspondence are brought forward.

Section \ref{section:scaling} introduces the well--known scaling concept
and methods to make it break down through a bias in the vacuum symmetry.
There, and in the subsequent section, we
point out some aspects of the scaling concept which -- to our knowledge --
have not been mentioned in the existing literature on the subject.
In particular, a clear distinction between the different manifestations of
scaling in the loop and the infinite-string ensemble is made, and a
correlation to scaling in percolation theory is illustrated.
In section \ref{section:2} we show ways to control and estimate statistical
errors and present results of measurements for perfect vacuum
symmetries. Section \ref{section:3} explains the theoretical basis on which
one expects the percolation transition to occur.
Section \ref{section:Hagedorn} presents results for this Hagedorn-like
transition
at which the infinite strings start to appear (as one decreases the bias in
the symmetry). We extract critical exponents associated e.g.~with the
divergence of the average loop length, a suitably defined susceptibility,
and a correlation length (for correlations in the string configurations,
not the vacuum field). Compared to ref.~\cite{MHKS}, the
accuracy of the results is greatly improved, results for the $\R P^2$
symmetry are new, and statistical errors are estimated.
Section \ref{section:polymers} discusses issues of the universality of
the critical exponents, and a percolation theory understanding of the
Hagedorn transition is developed.
A renormalisation group understanding
of the scaling concept is developed, and problems with the RG method in
calculating the critical exponents are addressed, as
are cosmological implications.

\section{Fundamentals of String Statistics}
\label{section:scaling}

\subsection{The numerical methods}

We will numerically evaluate the statistics of
cosmic $U(1)$ strings (or, equivalently, vortices
of superfluid $^4$He) and of $\R P^2$ strings (like e.g.~line
disclinations in nematic liquid crystals), in lattice-based simulations
of the Kibble mechanism.
The numerical method will be presented in ref.~\cite{numerics}, and contains
on perhaps essential improvements to the usual Vachaspati-Vilenkin (VV)
algorithm \cite{VV}. The most important improvement in our calculations
is that our lattice size is formally infinite, i.e.~we can avoid specifying
any boundary conditions, and can trace much longer strings with a given
amount of computer memory than was possible before.
The VV algorithm in general discretises space such that the lattice
spacing corresponds to the smallest physical length beyond which field
values can be considered to be uncorrelated, i.e.~the lattice spacing $a$
is of the order of, but perhaps slightly larger than the correlation
length $\xi$ of the field at the symmetry-breaking phase transition,
but certainly no larger than the cosmological horizon. Details of the
field dynamics are then inessential to the statistical properties of a
large ensemble of such lattices, and vacuum field values are assigned
randomly and independently at each space-point to create a Monte-Carlo
ensemble of field configurations on the lattice.
It should be pointed out that the regular lattice we use has been
criticised, as it does not allow variations in the size of correlated
domains. In particular, some of the results of Borrill's
simulations~\cite{Julian} are rather different from ours, for reasons which
are still poorly understood. However, they may well suffer from important
finite size effects.

Line defects are then considered to have formed if a closed walk along
lattice links maps, through the field map, onto a non-contractible
loop on the vacuum manifold. The assumption of a `geodesic rule'
\cite{RudSri93,HinDavBra94} for the
interpolation of the field between the lattice points is not only
intuitively acceptable, but in this formalism it is also essential in
order to guarantee string flux conservation, an essential symmetry of
the problem~\cite{numerics,thesis}. Refs.~\cite{numerics,thesis}
also prove that
only the dual lattice to the tetrakaidekahedral lattice can preserve
a uniqueness in the identification of the paths of single strings and
rotational symmetry of the Monte Carlo ensemble at the same time.
This lattice has been used in this context in
refs.~\cite{MHKS,thesis,Ed}, and to study simulations of
monopoles and textures~\cite{LeePro,LeeProtexture}.

\subsection{Long--Range Correlations in Topological Defects}

Strings are usually modelled by random
walks, either Brownian or self-avoiding.  A self-avoiding random walk
(SAW)\footnote{As usual in the literature, we
use the abbreviation SAW to mean self--avoiding {\it random} walk.
There are obviously infinitely many ways of building
self--avoiding walks, each leading to possibly quite different statistics.
As one example, straight walks are self--avoiding but
obviously exhibit quite different statistics to SAWs in dimensions higher
than one.}
models an excluded volume effect, and is known to apply well to
polymers \cite{deG}.  However, it is not clear that either kind of
walk represents the configurational statistics of cosmic strings or
superfluid vortices, for there are long-range
interactions which could change the Hausdorff dimension. That there are
super--horizon correlations in the configurations of topological defects
is not exactly new \cite{LeePro}. In the case of cosmic strings it can be
demonstrated by arriving at a simple contradiction when assuming no long--range
correlations: Take a closed circular walk through three--space spanning many
correlation volumes. How many strings does one expect to
encircle with such a walk? Since there is a well-defined string density
per correlation area, the number of encircled strings should increase
proportionally with the area $A$ enclosed by the walk. If they are
uncorrelated,
the net flux through this area will be partially cancelled by strings of
opposite orientation, and one expects an average net flux of around
$\sqrt{A}$ going through the loop formed by our walk. On the other hand,
the net flux is given also by seeing how often the field winds around
$U(1)$ while one follows the walk. This number, however, is expected to
increase as the square root of the length of the walk, since the field values
are uncorrelated on some length scale which is small compared to the
size of the loop, and many of the windings will cancel out.
One therefore has to conclude that, if the Kibble mechanism is responsible for
the formation of strings (i.e.~if the {\it field values} are uncorrelated
beyond some scale initially), the string network will have long--range
correlation, favouring for instance flux cancellation for oriented strings
on large surfaces. One should therefore expect deviations from Brownian
behaviour in the string statistics. This is
also favoured, because a cosmic string clearly {\it is}
self--avoiding~\footnote{Because a string is defined by the topology of the
field map, it will always follow the same way again, once it has turned
back onto itself. A cosmic string is therefore always forming a loop or has to
be infinite and self--avoiding.}, which does not mean, however, that the
statistics are those of a self--avoiding {\it random} walk (SAW),
because the nature of the self--avoidance is dictated by the field map.
A SAW shows correlations only
on very short scales (typically scales of the lattice constant). One of the
reasons why strings
also cannot be randomly self--avoiding, is that the field map carries the
memory of the position of all the other strings, which cannot be crossed.
We will show that neither a Brownian walk nor a random self--avoiding
walk model cosmic strings accurately, and that the walk statistics depend on
the vacuum manifold creating the strings.

\subsection{The Scaling Hypothesis}

A $U(1)$ string in the Vachaspati--Vilenkin algorithm on a tetrahedral
lattice is self--avoiding \cite{thesis}, irrespective of the discretisation
(if any) of $U(1)$ used in the algorithm.
One might therefore expect the network of cosmic strings to have the
statistical properties of a self-avoiding random walk.
A SAW builds up an
excluded volume as it follows its path, which is, in a statistical
sense, spherically symmetric and clustered around the
origin~\cite{Madras}. The SAW
therefore has a stronger tendency to move away from the
origin than the Brownian walk, which is allowed to intersect itself
arbitrarily often. This property is expressed in the fractal
dimension $D$ of the walk, which is the exponent relating
the average string length $l$ between two points on the same
string to their relative distance $R$ by
\begin{equation}
l\propto \langle R\rangle^D\,\,\,\,\,,\,\,\,\,\,
\label{eq:dimension}
\end{equation}
where the brackets $\langle\,\cdot\,\rangle$ denote some averaging procedure
over a large ensemble of walks~\footnote{We show in another paper
that several different averaging procedures, in particular the ones
of the kind $$\langle R \rangle = \lim_{N\rightarrow\infty}\sqrt[n]
{N^{-1}\sum_{i=1}^N |R^n|}=\sqrt[n]{\langle |R^n|\rangle}$$ produce the same
results for the fractal dimension, such that in particular
$l\propto\langle |R|^D \rangle\propto\langle |R|\rangle^D$~\cite{scaling}.}.
It is well known that the dimension for a
Brownian walk is $D=2$, and for a
self-avoiding random walk in three dimensions it is
$D=1/\nu=1/(0.5877\pm0.0006)$
(see ref.~\cite{Sokal} and
references therein for a summary of different methods used
to obtain that result).
However, the original string formation simulations \cite{VV,V91} are
consistent with $D=2$. The reason for this was seen in the fact that
they simulated a dense string network:
A single string, as we trace out its path,
experiences a repulsion from all of the segments of other strings, which
do not have any statistical bias towards the origin.
Therefore the repulsion
from the forbidden volume will also have no directional
bias. Thus the fractal dimension of the string could also be argued to be
(close to) two, although the string is self--avoiding. In polymer
physics, this effect has been known for some time to occur in a dense
solution of polymers~\cite{deG}.
In a statistical sense, the network
of cosmic strings was argued to be equivalent to a dense network of
polymers \cite{Scherrer}.
A polymer in a dilute solution will
exhibit the configurational statistics of a self--avoiding random walk, while
in a dense solution of polymers, each one has the structure of a Brownian
random walk. Thus, taking this lesson from polymer physics, one
would expect the scaling of the string size $R$
with length $l$ in the initial configuration of cosmic strings to correspond
to a SAW on scales smaller than the mean separation between different strings,
and to a Brownian walk on scales larger than this. In the cosmic string case,
however,
the mean separation is of the order of the correlation length itself, which
is the same as the lattice spacing. So we expect the bias towards a SAW to
fall off with distance roughly as fast as the lattice discretisation errors,
which makes this short--distance effect immeasurable. We shall anticipate
the results of the following chapters: the fractal dimension of a string
at formation is in general not the same as for a Brownian walk. It is only
for $U(1)$ strings that measurements are
consistent with the exact value of two, in the extremely long-distance
limit ($\approx 10^5$ to $10^6$ lattice units)
Of the other strings which have been measured, none have
fractal dimensions higher than the $U(1)$ strings, but all have distinctly
larger $D$ than the SAW.

As is customary, we can introduce the scaling hypothesis in order to estimate
a few other properties of the string network,
which states that, in terms of its
statistical properties, the string network looks the same on all scales
much larger than the correlation length of the vacuum
field\footnote{Scale invariance is
phenomenologically the same as the existence of a large--scale (IR)
renormalization group fixed point. However, renormalization group arguments
for topological objects are hard to find. To our knowledge, there
exists no analytic work which would lend firm support to the scaling
hypothesis. In fact, most analytic work gets
intractable if the scaling hypothesis is not put in a priori. One would
expect a proof of the scaling hypothesis to contain renormalization
group (RG) arguments. We will develop
percolation theoretical RG arguments in favour of this
hypothesis in the Appendix.}.
In fact, Brownian random walks are scale invariant. With
the scaling hypothesis, the expected
distribution of closed loops can be derived~\cite{Vil}.
{}From dimensional arguments,
the number of closed loops with size from $R$ to
$R+dR$ per unit volume can be written as
\begin{equation}
dn=f\left(\frac{R}{\xi}\right)\,\frac{dR}{R^4}\,\,\,\,\,.\,\,\,\,\,
\label{eq:scaling_function}
\end{equation}
If the system is scale invariant,
the distribution should be independent of the
correlation length $\xi$, and one expects
\begin{equation}
dn\propto R^{-4}dR\,\,\,\,\,.\,\,\,\,\,
\label{eq:dn}
\end{equation}
The length distribution of loops for strings with a fractal dimension of
$D$ is therefore
\begin{equation}
dn\propto l^{-b}dl\,\,\,\,\,,\,\,\,\,\,
\label{eq:lengthdistr}
\end{equation}
with
\begin{equation}
b=1+3/D\,\,\,\,\,,\,\,\,\,\,
\label{eq:scaling}
\end{equation}
or more generally $1+d/D$, with $d$ being the dimension of the space wich the
walk is embedded in.
It was originally expected \cite{VV} that it follows from scale invariance
that there should be no infinite strings. This turned out not to be the case,
since, as we will discuss in section~\ref{section:U1perfect},
ensembles of infinite strings and ensembles of loops manifest scale invariance
in entirely different ways, namely in the validity of the
Eqs.~(\ref{eq:dimension}) and (\ref{eq:scaling}), respectively.
Infinite strings
can still look statistically the same on all scales much larger than
the lattice spacing: a Brownian walk
is scale invariant and has a non-zero probability
not to return to the origin in $d>2$ dimensions.
The origin of the scale invariance of the string network seems
to be connected with the
absence of long-range correlations in the order parameter \cite{VV}.
However, scale invariance does not necessarily imply that the network
is Brownian as originally stated. Strictly speaking, scale invariance holds
when all the scaling properties of a network, such as Eqs.~(\ref{eq:dimension})
and (\ref{eq:lengthdistr}) are power laws: Only power laws do not change
upon linear rescaling of the variables. In this sense, scaling is satisfied
whenever the Eqs.~(\ref{eq:dimension})
and (\ref{eq:lengthdistr}) hold. However, to make scaling also work
in spite of finite size effects prohibiting us from identifying the very
long loops, Eq.~(\ref{eq:scaling})
is taken as the manifestation of scale invariance. This is plausible:
Eq.~(\ref{eq:scaling}) implies that loops exhibit (on scales larger much than
the lattice spacing but much smaller than the loop size)
the same fractal dimension as infinite strings, so that on scales
where one counts some number of loops wrongly as infinite strings, the
distinction between the two becomes unnecessary\footnote{In this sense,
Eq.~(\ref{eq:scaling}) is a more stringent definition of
scale invariance, because it allows to be ignorant about
effects on scales which a particular observation may not reach. If
we defined scale invariance by some omniscient observer which can distinguish
loops even if they exceed the observed scale in size, then there is no
reason for the exponents of the loop distribution to be in any relation to
the exponents of the distribution in infinite strings.}.
One does of course not need $D=2$
in order to have a scale-free distribution of loop sizes $R$.
It is important to note that, because of Eqs.~\ref{eq:dimension} and
\ref{eq:lengthdistr}, although they are the criterion for scale-invariance,
there are some observables which are not scale-invariant, if they happen to
be dependent on the UV cutoff. An example of this is the fraction of string
mass in loops, as discussed in~\cite{MHKS} and item 3 in the next section.
Whatever numbers one gets for these quantities are probably unphysical,
since there is no known algorithm (least of all VV type algorithms) that would
tell us what the physical UV cutoff on the loop size distribution
Eq.~\ref{eq:lengthdistr} should look like. Scale-invariance can only hold in
the limit $l\gg \xi$.

\begin{figure}[htb]
\centering
\mbox{~}{\hbox{
\epsfxsize=240pt
\epsffile{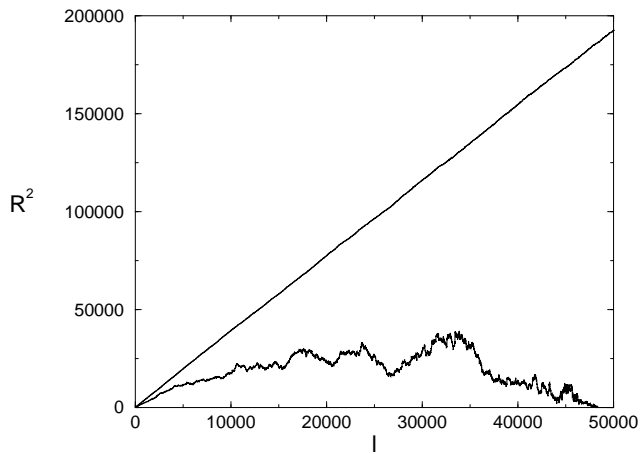}}}\\
\caption{The relation between the average distance of a string element
from the origin and the string length walked until arriving at this element.
The upper line represents averages over the ``infinite" strings only
(i.e.~strings which survive up to the length
$\Lambda=50000$ where this particular
measurement was stopped). The lower line represents averages over the string
loops, and has high statistical errors on the long--loop end,
because of the low number of loops, and even higher systematic errors,
because the ratio of loops wrongly counted as infinite, to the correctly
counted loops increases. The averages were taken over 10,000
strings, 6334 of which happened to be ``infinite". Only 54 loops survived
up to length 10,000, and only 10 to length 30,000.
One sees that the relation for $R^2$ vs.~$l$ for the infinite string
part is almost perfectly linear,
suggesting Brownian statistics. The vacuum manifold $U(1)$ was discretised by
three equidistant angles.}
\label{fig:linlinU1003}
\end{figure}

\begin{figure}[htb]
\centering
\mbox{~}{\hbox{
\epsfxsize=240pt
\epsffile{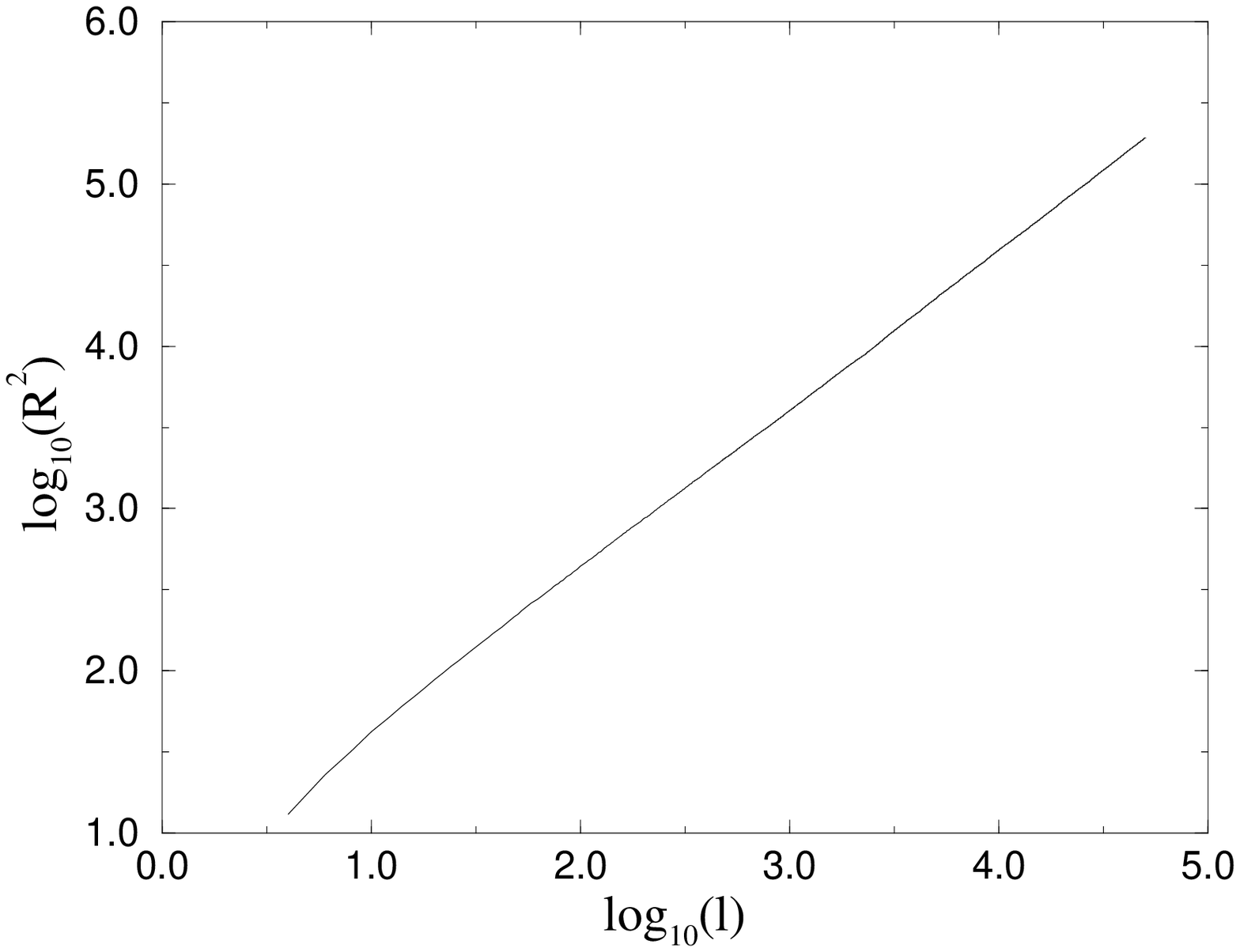}}}\\
\caption{log--log plot of the ``infinite string" contribution in
Fig.~\ref{fig:linlinU1003}. We see that the short--length limit has
lattice discretisation errors, and that a true power law is approached
only for $l\stackrel{>}{\sim}30$. The measurements are
listed in Table \ref{table:DU1}.}
\label{fig:loglogU1003}
\end{figure}

If $D=2$, one would expect a linear relationship between walk length and
average $R^2$, which would then, if the probability distribution
for ending up at a point $\vec{x}$
after $l$ steps is Gaussian, be interpretable as the
average $\sigma^2$ of the distribution.
All this is familiar from the Brownian walk, and measurements
seem to indicate that -- in the case of a $U(1)$ symmetry, we are close to
such statistics. Fig.~\ref{fig:linlinU1003} presents the linear--linear graph
of $R^2$ vs.~the walk length $l$, which can be seen to be an almost perfectly
linear relation. The measurements in Figs.~\ref{fig:linlinU1003} and
\ref{fig:loglogU1003} are made using a discretisation of $U(1)$ by three
equidistant angles\footnote{In a sloppy way, we could say we discretise $U(1)$
by $\Z_3$. This only is correct as far as the allowed vacuum angles are
concerned,
but may be misleading, since in an actual $\Z_3$ symmetry the lines which we
identify as geodesics on $U(1)$ would be associated with a finite vacuum
energy (i.e.~they would be crossing domain walls). It is more correct to
say that we
triangulate the vacuum manifold as well as space: in this case with
3 vertices and 3 edges, joining adjacent points. This automatically
encodes the geodesic rule.  This distinction seems trivial, but it allows an
easier generalisation to e.g.~discretisations of $\R P^N$.}.
Such measurements were made in~\cite{MHKS},
and we complete results from there. In particular, we present a
much better error analysis here. Results are represented split between
the infinite string part and the loop contribution. This is a procedure we
will follow throughout this paper, in analogy to conventions
in percolation theory, and we will show that
it is in fact necessary to do so.

\section{Results for Perfect Symmetries}
\label{section:2}

\subsection{Elimination and Estimation of Errors}

Before we turn to the results, a few words of caution are in order. Among
those, we include an explanation of
how we arrive at error estimates for the statistical
error.
\begin{enumerate}
\item Because of the nature of the Monte Carlo averaging,
there are two big sources of error
for very long loops (i.e.~the longest ones permitted in the
simulations). We follow every string
until it hits a certain upper limit of the string length $l\le\Lambda$,
or until it returns to the origin, whatever happens first.
If it does not return to the origin until
we have reached the length $\Lambda$, it is counted towards the ``infinite
string" ensemble. This does not introduce too many problems for the averaging
over infinite strings (as long as there are many), because of the nature
of Eq.~(\ref{eq:dn}), which ensures that only very few of the strings
surviving up to length $\Lambda$ are actually wrongly counted as infinite.
For the loop distribution, which has only very few strings in this regime,
the statistical errors are huge (in the end, usually just before we
reach $l=\Lambda$, we even ``average" over one string only) but the
systematic errors in this regime are equally bad,
because there is a finite number of
strings which {\it should} be in the loop distribution, but are not
identified as loops.
This drives the measured $R^2$ to zero at the length
where the longest of correctly identified loops closes (compare the
lower line in Fig.~\ref{fig:linlinU1003}.
Extraction of configurational exponents on the loop distribution will
therefore be defined through fitting appropriate curves which approach
the actual measurements asymptotically in the intermediate--length
regime only.
\item A word of caution is also necessary for the short--length limits. As seen
for example in Fig.~\ref{fig:loglogU1003}, scaling is not satisfied in
the short walk--length regime. This is
due to two sources of error. Firstly, there
are obviously lattice discretisation artifacts. Secondly, at small
distances the excluded volume effect (from the self--avoiding nature of
the string) is still turned on, with a directional bias away from the
origin. Eventually, repulsion from all other pieces of string should
even out with repulsion from pieces of the same string, with no directional
bias at all, but this happens only at some distance from the
origin.
We will allow for this source of error by cutting off the low length
regime at lengths between 10 and 1000 lattice units~\footnote{A cutoff
of 20 is usually sufficient for extracting the exponents $c$ and $b$ in
Eq.~(\ref{eq:lengthdistr2}), while 500 is very good, and still perfectly
practical, for extracting
the fractal dimension $D$ on the infinite string part. Which cutoff to
choose is decided on a case by case basis by observing where a cutoff
independent measurement can be obtained.} whenever we measure
configurational exponents from the Monte Carlo ensembles.
\item Frequently, we will not
quote the fraction of the total string mass
found in loops. Such numbers are meaningless, unless there is a lower
cutoff for loop lengths much larger that one in units of correlation lengths
(in which case one gets very little mass in loops anyway).
This parameter is lattice--dependent~\cite{Scherrer}, even for
simple Brownian random walks or self--avoiding random walks. This is partly
due to different coordination numbers of different lattices (e.g.~every
vertex on the tetrakaidekahedral lattice is connected to four lattice links,
while this number is six in the simple cubic case. There are more possibilities
for the string to ``stray" in the simple cubic case). This is connected
with another lattice dependence of this number: The length
of the lattice links in the tetrakaidekahedral case is $a\,\sqrt{2}/4$, with
$a$ the edge length of the underlying bcc lattice, while the correlation
length is between $a\,\sqrt{3}/2$ and $a$~\footnote{Since every link
borders three tetrakaidekahedra, we need to take all the distances between
those three as representatives of a correlation length. Two pairs have
the distance $a\,\sqrt{3}/2$, while one pair is separated by $a$. When we
refer to ``walk lengths in lattice units" we mean in units of $a\,\sqrt{2}/4$,
which is the edge length of the tetrakaidekahedral lattice.}.
Since the smallest allowed loops in both lattices
consist of four links, the tetrakaidekahedral lattice allows much smaller
loops (in units of correlation lengths) than the simple cubic lattice.
According
to Eq.~(\ref{eq:dn}), we expect a large contribution to the total string mass
to be in very small loops, so that on a tetrakaidekahedral lattice the total
string mass in loops will be considerably higher than on the cubic lattice.
The problem of the lattice dependence of the mass fraction in loops also
reflects a lack of knowledge about the physics involved in the production
of small loops. Physically, one
would expect a smooth cutoff for short loops, so that
the very small loop contribution in Eq.~(\ref{eq:dn}) gets gradually
suppressed.
We do not know the form of this cutoff, and we expect it to depend not only
on dynamical details of the Kibble mechanism, but also on thermal production
mechanisms for string loops (which should be relevant right at the phase
transition temperature, but quickly become subdominant as the Universe
cools further). In any case, the
physical relevance of knowing the exact contribution
to the total string mass in small loops produced at a phase transition
is highly questionable, as they disappear quickly in any case.
This does not
affect the physical relevance of the other data we can extract from
Vachaspati--Vilenkin type measurements, because the long loops and infinite
strings are not transient.
\item Finally, we need to explain how we arrived at the values for the
statistical error:
To estimate, for instance, the statistical error percentage of the fractal
dimension $D$, measured for a total number of ${\cal N}_{\rm measured}$
strings up to a length $\Lambda$, we take for example 10 sets of ${\cal N}=
{\cal N}_{\rm measured}/10$ strings and measure the variance of the result,
then take 10 sets of ${\cal N}={\cal N}_{\rm measured}/20$
and 10 sets of ${\cal N}={\cal N}_{\rm measured}/40$ strings and so on.
We then measure the variance of the results for all those sets, and, under
the assumption that the error behaves like a power of the size of the string
ensemble, we extrapolate to an ensemble of ${\cal N}_{\rm measured}$ strings.
If all our measurables were Gaussian random variables for any sample within
the ensemble, this power law would just be
$\sigma^{-1}\propto\sqrt{{\cal N}}$, which motivates this approach. Since
configurational exponents are normally not distributed in a Gaussian
distribution within samples of the ensemble,
we decided to allow a generalised power law for the variance.
We present this method by the example of a $U(1)$ manifold discretised
by $N=3$ equidistant vacuum angles.
Fig.~\ref{fig:loglogU1003}, the log--log plot of $\langle R^2 \rangle$
vs.~$l$, has a linear fit suggesting
\[
l=0.232\,R^{2.0212},\,\,\, (N=3) \,\,\,\,\,.\,\,\,\,\,
\]
This was measured for
${\cal N}_{\rm measured}=10,000$ strings being allowed to reach the length
50,000 in lattice units (taking only the ``infinite" strings, and using a
lower cutoff of 500 lattice units).
Similar measurements on several ensembles with less strings yield
the values in Table~\ref{table:varianceU1003}.
\begin{table}[htb]
\centering
\begin{tabular}{|c|c|c|c|}
\hline
number of & number of & average & standard\\
strings   & ensembles & $D$ & deviation\\
\hline
1000      & \mbox{~~~}10 & ~2.027~ & ~0.038~ \\
500       & \mbox{~~~}10 & ~2.022~ & ~0.045~ \\
250       & \mbox{~~~}10 & ~2.035~ & ~0.051~ \\
100       & \mbox{~~~}10 & ~2.021~ & ~0.094~ \\
\hline
\end{tabular}
\caption{The statistical
variances in measurements of the fractal dimension for ensembles
of less and less strings. The statistical error for a large ensemble is
the extrapolation of these values to the appropriate number of strings.}
\label{table:varianceU1003}
\end{table}
On a log--log plot, the statistical variances may be fit by the expression
\[
\sigma\approx
0.0146\,({\cal N}/{\cal N}_{\rm measured})^{-0.383}\,\,\,\,\,,\,\,\,\,\,
\]
so that the $\sigma$
expected in our measurement can be taken to be $\approx 0.015$, which is
simply the intercept of the linear fit in the
log--log plot of the variance against ${\cal N}/{\cal N}_{\rm measured}$,
as displayed in Fig.~\ref{fig:sigmaU1003}.
\end{enumerate}
\begin{figure}[htb]
\centering
\mbox{~}{\hbox{
\epsfxsize=240pt
\epsffile{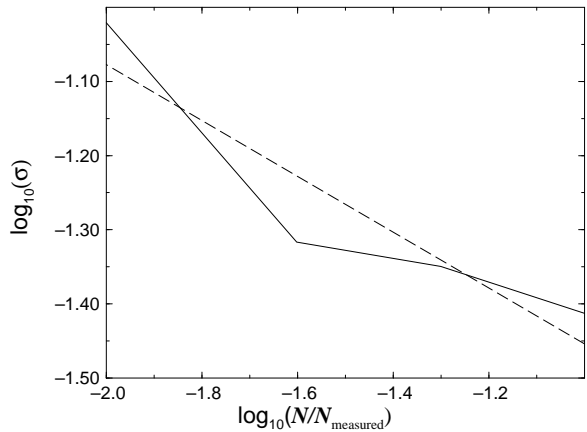}}}\\
\caption{The variances in Table~\ref{table:varianceU1003}. Extrapolating
of the linear fit to ${\cal N}_{\rm measured}$ yields an estimate for
the statistical error in the original measurement, in this case
$\sigma_{\rm measured}=10^{-1.831}\approx 0.015$.}
\label{fig:sigmaU1003}
\end{figure}
\begin{figure}[htb]
\centering
\mbox{~}{\hbox{
\epsfxsize=240pt
\epsffile{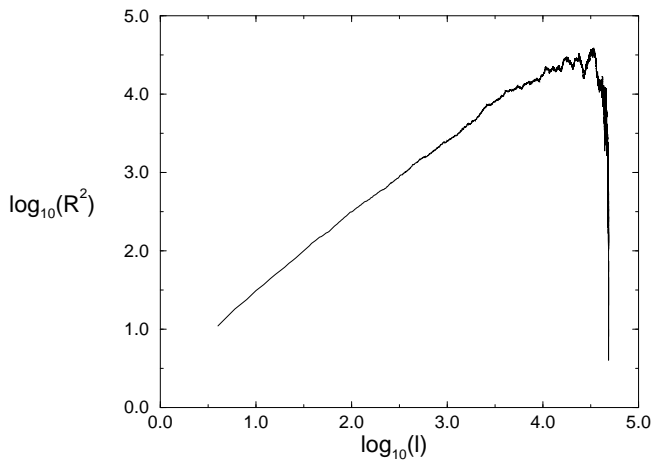}}}\\
\caption{Log--log plot of the loop contribution in
Fig.~\ref{fig:linlinU1003}. Only the short--loop regime can be deemed useful,
because there are very few long loops, making bad statistics. In addition to
that, the counting is biased by the exclusion of all loops longer than
$\Lambda$ or shorter than any given length where the plot is read.}
\label{fig:loglogU1003loop}
\end{figure}

\subsection{Results for a Perfect $U(1)$ Symmetry}
\label{section:U1perfect}

We can now proceed to the presentation of the results. For a perfect $U(1)$
symmetry, We have
used a series of different discretisations of $U(1)$, each
consisting of $N=2^n-1$ equidistant angles. The range of such discretisations
is from $N=3$, the lowest possible number of points on $U(1)$ to give
non--contractible contours, to $N=255$, a rather good approximation to
continuous symmetry, as we shall see from the asymptotic behaviour of
the measurables for large $N$.

\subsubsection{Loops Have No Fractal Dimension}

The linear fit to Fig.~\ref{fig:loglogU1003loop}, the log--log plot of
$\langle R^2 \rangle$ vs.~$l$, averaged over loops only,
obviously requires some
sensible upper cutoff much lower than $\Lambda$, and to some extent
any measurement of the fractal dimension of the loop--ensemble is cutoff
dependent. Nevertheless, a fractal dimension of $D=2$ is inconsistent with
any part of Fig.~\ref{fig:loglogU1003loop}, as is any fractal dimension
measured for the infinite string contribution as listed in
Table~\ref{table:DU1}. A typical subset of this loop
ensemble, reduced to loops with
$l\in [10,1000]$, gives $D=2.14$, whereas a typical infinite
string dimension for, say, strings with $\Lambda=50000$ is
$D=2.02$. The reason for both, this discrepancy and
the cutoff dependence of the loop dimension, is simple to understand:
A single loop cannot be a fractal, but a single infinite string can be!
As it turns out, the measurements allow a slightly
stronger statement: the {\it average} loop is not a fractal, whereas
the {\it average} infinite string is.
The reason for this is simply the finite size of loops: a
single infinite string approaches a scaling behaviour
asymptotically for large scales with no upper length scale arising, whereas
a loop has a cutoff through its finite size.
A finite object has a well
defined fractal dimension only on scales much smaller than the size of the
object and much larger than the lattice spacing. This means, that only
very large loops will exhibit ``scaling", and then only on a finite range
of scales, which makes the scaling concept rather risky to depend on as far
as the fractal dimension of the loops is concerned.
We will therefore average over the infinite--string component of the
string ensemble only, whenever it comes to extracting a fractal dimension.
Fig.~\ref{fig:loglogU1003loop} also suffers from the problem that the
averages are only taken
over those loops which actually survive up the the given length.
An ``average loop" would simply not be there any more after we have walked
the length, say,  $l=10$.

This does not invalidate Eqs.~(\ref{eq:dn}), (\ref{eq:lengthdistr}),
(\ref{eq:scaling}), and (\ref{eq:lengthdistr3}), as these specifically imply
and require the finiteness of loops. We can sum this up in the following way:
The scaling concept enters the loop
distribution through the distribution of loop sizes, rather than the average
properties of a single loop, whereas the scaling properties of the infinite
strings can be interpreted as properties of the average infinite string.

There is another qualifying statement to be made: If we remind ourselves
how Eq.~(\ref{eq:lengthdistr}) was derived from Eq.~(\ref{eq:dn}), we have
used the fractal dimension (in the cases to follow: the fractal dimension
of the infinite--string ensemble), to derive properties of the loop ensemble.
This is justified only because of a lower loop--size cutoff (in addition
to the upper one, which ensures that the averaging is not over too small
an ensemble)
employed in measuring $b$. By having an appropriate lower cutoff,
we make sure that we use this fractal
dimension only for loops long enough to exhibit an
intermediate length scale on which fractal behaviour can be approached,
allowing us to use Eq.~(\ref{eq:dimension}) in deriving
(\ref{eq:lengthdistr}) from (\ref{eq:dn}).
The measurements do then indeed seem to indicate that, with rather minor
deviations, the intermediate length regime
of a properly chosen long--loops ensemble looks like an infinite--string
ensemble with an intermediate--range upper cutoff, and the scaling relation
Eq.~(\ref{eq:lengthdistr}) holds.

\subsubsection{The Infinite--String Ensemble}
\label{section:infinite}

Using linear fits to log--log plots of
$R^2$ vs.~$l$, with the error elimination and error estimation methods
presented above, we arrive at measurements for the fractal dimension of
strings as presented in Table~\ref{table:DU1}.
\begin{table}[htb]
\centering
\begin{tabular}{|c|c|r|r|}
\hline
Number $N$ of & fractal & number & string length\\
discretisation & dimension & of & cutoff\\
points on ${\cal M}$& $D$ & strings & $\Lambda$\\
\hline
3 & $2.021 \pm 0.015$ & 10,000 & 50,000 \\
7 & $2.022 \pm 0.015$ & 10,000 & 50,000 \\
15 & $2.022 \pm 0.015$ & 10,000 & 50,000 \\
31 & $2.022 \pm 0.015$ & 10,000 & 50,000 \\
63 & $2.025 \pm 0.015$ & 10,000 & 50,000 \\
127 & $2.013 \pm 0.015$ & 10,000 & 50,000 \\
255 & $2.007 \pm 0.025$ & 3,000 & 1,000,000 \\
255 & $2.055 \pm 0.006$ & 100,000 & 2,000 \\
\hline
\end{tabular}
\caption{The fractal dimensions of an average string with different,
increasingly finer, discretisations of $U(1)$. The averages are over
infinite strings only. The lower cutoff is $l=500$ in all ensembles.}
\label{table:DU1}
\end{table}
We see that the statistical errors are still
larger than the discretisation errors coming from a particular discretisation
of $U(1)$ (although the string density for instance does depend
on which discretisation one uses). Table~\ref{table:DU1} also suggests
systematic errors: the higher the upper cutoff $\Lambda$, the lower
the measured fractal dimension. This effect has already been observed, in a
diffferent context, for $U(1)$ strings with $N=3$~\cite{Bradley}.
Except for the short strings with $\Lambda=2000$, the measurements are,
however, consistent with each other.
To decide
whether the last line in Table~\ref{table:DU1} is actually the manifestation
of systematic errors or not, let us investigate the possible sources
for such a discrepancy. Either
\begin{itemize}
\item an excluded volume effect is active for intermediate distances,
making the strings slightly self--seeking. This simply forces us to
accept a scale-dependent fractal dimension, which was once of the
conclusions drawn for $N=3$ $U(1)$-strings in ref.~\cite{Bradley}.
\item Another source for the discrepancies in the values of $D$ could be
that we are counting too many strings which are to form loops eventually
(but are not identified as loops, because of the cut-off $\Lambda$). These
would naturally bias the fractal dimension towards a higher number for
smaller $\Lambda$, or
\item our averaging procedure introduces systematic errors.
\end{itemize}
To explore which of these interpretations is the right one, we need
to know how many of the strings which reach the length $\Lambda$ are to be
expected to close onto themselves again to form loops, i.e.~we need to
know the prefactor and the exponent involved in Eq.~(\ref{eq:lengthdistr}).
Once we know the number of wrongly counted strings, we need to subtract their
expected configurational parameters from the ensemble of infinite strings.
As a first approximation, we assume that these strings have the average
loop properties extracted from the loop--ensemble. This will actually
over--compensate for the effect that those strings lower the effective
fractal dimension, because they should
on average of course have larger $R^2$ than the average strings which
are actually counted as loops\footnote{This is easy to see in the
large $l$ limit, where they are obviously not at $R^2=0$, whereas the
strings which are counted as loops are arriving there for
$l\rightarrow\Lambda$.}. Thus, we will over--estimate errors coming from
this source.  The procedure is now obvious: if the number density of
{\it traced} loops is
\begin{equation}
dn=q\,l^{-b+1}\,dl\,\,\,\,\,,\,\,\,\,\,
\label{eq:lengthdistr3}
\end{equation}
then the number of strings expected to exceed the length $\Lambda$,
i.e.~the number of uncounted loops, is
\begin{equation}
n_{\rm loop}^{\rm unc}=\int_{\Lambda}^{\infty}dn(l)=q\,\frac{\Lambda^{-b+2}}
{b-2}\,\,\,\,\,.\,\,\,\,\,
\label{eq:n_loop_unc}
\end{equation}
The corrected $R^2$ is then given by
\begin{equation}
R_c^2\,\left(n_m-n_{\rm loop}^{\rm unc}\right)=
R_m^2\,n_m-R_{\rm loop}^2\,n_{\rm loop}^{\rm unc}\,\,\,\,\,,\,\,\,\,\,
\label{eq:R2correction}
\end{equation}
where the index $c$ stands for ``corrected" and $m$ for ``measured", for the
infinite--string ensemble, and $R^2_{\rm loop}$ is the measured value for
the loop ensemble.
In Table~\ref{table:dnU1} we list the configurational parameters $q$ and
$b$, together with $n_{\rm loop}^{\rm unc}$.
\begin{table}[htb]
\centering
\begin{tabular}{|c|c|c|c|c|c|}
\hline
Number $N$ of  &     &     & number of     & maximum           &\\
discretisation & $q$ & $b$ & ``infinite"   & length of         &
$n_{\rm loop}^{\rm unc}$ \\
points         &     &     & strings $n_m$ & strings $\Lambda$ &\\
\hline
3       & $5576 \pm 1003$       & $2.530 \pm 0.022$     &
6334    &       50,000  & $34 \pm 19$\\
7       & $3398 \pm 495$        & $2.456 \pm 0.018$     &
6394    &       50,000  & $54 \pm 24$\\
15      & $4506 \pm 628$        & $2.507 \pm 0.018$     &
6382    &       50,000  & $37 \pm 16$\\
31      & $3930 \pm 646$        & $2.492 \pm 0.020$     &
6412    &       50,000  & $39 \pm 20$\\
63      & $4270 \pm 715$        & $2.500 \pm 0.020$     &
6346    &       50,000  & $38 \pm 20$\\
127     & $3350 \pm 637$        & $2.461 \pm 0.023$     &
6373    &       50,000  & $50 \pm 30$\\
255     & $796  \pm 201$        & $2.425 \pm 0.035$     &
1903    &  1,000,000    & $6 \pm 6$\\
255     & $36800 \pm 3200$     & $2.468 \pm 0.014$     &
65773   &  2,000        & $2240 \pm 550$\\
\hline
\end{tabular}
\caption{The configurational parameters $q$ and $b$ of the loop distribution
Eq.~(\ref{eq:lengthdistr3}) for $U(1)$ strings,
with the expected number of loops unaccounted for in the measurements.
It should be noted that there are systematic computational
errors which increase as the
number of strings decreases, because the loops need to be grouped in
increasingly large length intervals. The errors quoted here are statistical
errors only. $q$ is not rescaled by the total string number, to make the
extraction of $n_{\rm loop}^{\rm unc}$ more transparent.}
\label{table:dnU1}
\end{table}
Including the (overcompensating) correction
Eq.~(\ref{eq:R2correction}) should give us some idea of the systematic
errors, but it corrects almost all the results of Table~\ref{table:DU1}
for the fractal dimension
down by only $D_c\approx D-0.002$, so that the statistical errors overshadow
the systematic ones by far, except for the ensemble of 100,000 strings with
upper cutoff $\Lambda=2000$, where there are many miscounted loops, but
small statistical errors. In this case, the prediction gets corrected down
to $D_c=2.031 \pm 0.007$. This makes all the measurements of the fractal
dimension at different intermediate and long scales just consistent
with each other, so that we would need somewhat better statistics than we
have accessible at this moment to see whether there is some physical effect
or just a conspiracy of statistical fluctuations suggesting a tendency
for strings to be slightly self--seeking on intermediate scales (this is
indeed implied by a running of the effective fractal dimension, also implying
weak violations of scale invariance, as observed for $N=3$ by Bradley
et.~al.~\cite{Bradley}.
However, we have not yet explored the third possible source of
errors, the averaging procedure:
Table~\ref{table:DU1} was obtained by taking $\langle R^2 \rangle$ over the
ensemble of strings, i.e.~we measured the exponent $\kappa$ in
\[
\frac{1}{N}\sum_i^N R_i^2(l)\propto l^{-\kappa} \,\,\,\,\,,\,\,\,\,\,
\]
and defined $D=2/\kappa$. The question is whether this is the best possible
way of defining a fractal dimension, i.e.~whether this is a good averaging
procedure.
However, if the results
depend on the specifics of the averaging procedure, then the scaling hypothesis
is in trouble, because, if for example the ratio
$\sqrt[n]{\langle R(l)^n \rangle}/\langle R(l) \rangle$
varies with $l$, then the string network obviously does not look the same
on all scales. The ratio of all the moments of the probability distribution
for $R(l)$ has to be such that all the $\sqrt[n]{\langle R(l)^n \rangle}$
stay in a fixed proportion to each other for $l/\xi\gg1$~\footnote{In the
polymer literature, such ratios are called ``(universal) amplitude ratios".
It turns out~\cite{scaling},
that the fractal dimension for a truly scale-invariant walk is
also independent of the definition of $R(l)$ itself, which could be the mean
end--to--end distance (which is what we use), or the radius of gyration
(which is the average separation of {\it all} point pairs on a walk
segment of length $l$),
or the root--mean--square distance of a monomer from the
end--points.}.
So, if the averaging procedure is the reason for the discrepancies
in Table~\ref{table:DU1}, scaling is noticeably violated up to lengths
of several tens of thousands of correlation lengths! The same argument leads to
another important remark: {\it if scaling is violated, Eq.~(\ref{eq:scaling})
is not only violated, but also ambiguous}, because an unambiguous definition
of $D$ requires an unambiguous convention of how the average of $R^2$ is
to be extracted. Such a convention is not necessary if the string network
scales.

This makes it very easy to check that the scaling hypothesis is
satisfied. Firstly, when we compare the mean values of the measurements
for $D$ and $b$, they satisfy Eq.~(\ref{eq:scaling}) extremely well.
The average $b$ in Table~\ref{table:dnU1} is 2.4821, whereas $1+3/D$ with
the average $D$, 2.0235, is 2.4826. All we need to show now is that the
fluctuations in Table~\ref{table:DU1} are not systematic. There are two
ways of doing this: either we improve the statistics of the measurement,
hoping that the values converge toward each other (presumably somewhere
near the range $[2.0235,2.0241]$, which corresponds to the means measured
for $b$ and $D$), or we show that the ratio of $\langle R^2 \rangle$ to
$\langle R \rangle^2$ stays fixed. Here we prefer the latter of the two,
because it will confirm that there are no problems arising from the specific
averaging procedure we used, whereas simply increasing the statistics does not
give us this reassurance\footnote{Strictly speaking, showing that the
ration of $\langle R^2 \rangle$ to $\langle R \rangle^2$ stays fixed
does not prove scaling unless one shows that {\it all} the ratios
$\sqrt[n]{\langle R(l)^n \rangle}/\langle R(l) \rangle$
stay fixed for infinite strings.}.

We did this by reproducing Table~\ref{table:varianceU1003} with exactly the
same ensembles (i.e.~ensembles having the same random number seed), but
using $l \propto \langle R \rangle^D$ instead of
$l \propto \langle R^2 \rangle^{D/2}$, as
is used in all the other measurements. The comparison is shown in
Table~\ref{table:Dcomparison}.
\begin{table}[htb]
\centering
\begin{tabular}{|c|c|c|c|}
\hline
number of & number of & average $D$ with & average $D$ with\\
strings   & ensembles & statistical error, & statistical error,\\
          &           & based on $l\propto\langle R^2(l) \rangle^{D/2}$ &
                        based on $l\propto\langle R(l) \rangle^D$\\
\hline
1000      & \mbox{~~~}10 & ~$2.027\pm 0.012$~ & ~$2.027\pm 0.013$~\\
500       & \mbox{~~~}10 & ~$2.022\pm 0.014$~ & ~$2.019\pm 0.014$~\\
250       & \mbox{~~~}10 & ~$2.035\pm 0.016$~ & ~$2.029\pm 0.018$~\\
100       & \mbox{~~~}10 & ~$2.021\pm 0.030$~ & ~$2.027\pm 0.027$~\\
\hline
\end{tabular}
\caption{Comparison of two different ways of averaging $R$ to obtain the
fractal dimension. It can be seen that both lowest moments of the probability
distribution yield the same fractal dimension to higher accuracy
than expected from the statistical error margins.}
\label{table:Dcomparison}
\end{table}
It can be seen that the measurements of the fractal dimension agree with
each other better than to be expected from statistical errors alone. This
indicates that there is not only no measurable discrepancy between the
scaling of different moments of the distribution for $R(l)$, but also that
there are correlations between those
moments for any finite string ensemble, so that,
unfortunately,
one cannot really exploit more than one moment of the distribution to extract
two or more statistically {\it independent} measurements for $D$ from a single
ensemble. This is an important observation, as it justifies not only to keep
on using the averaging procedure we used from the start, but it tells us that
there is no gain of statistical
accuracy in keeping track of more than one such average.
We conclude that the averaging procedure does not introduce additional
systematic errors in Table~\ref{table:DU1}.\footnote{The source of these
systematic errors remains therefore unidentified (cf.~following sections).}
We will therefore continue to measure $D$ by fitting $D/2$ in
$l\propto\langle R^2 \rangle^{D/2}$ only.

Summing up our analysis of Table~\ref{table:DU1}, we conclude that
our averaging procedure does not introduce systematic errors, but correcting
for the wrong counting of the loops longer than $\Lambda$ as infinite
makes the measurements (just) consistent with each other. Keeping in mind,
however, that - for the above explained reasons - this correction is likely
to be too generous, we have to agree with the conclusions of \cite{Bradley}
that the existence of a slightly scale-dependent fractal dimension has
to be accepted as given. This is further supported by the observations
in section \ref{sec:4.1}. Table~\ref{table:DU1} also suggests
that the very-long string limit of $D$ may be exactly two.

\subsection{Results for a Perfect $\R P^2$ Symmetry}

All the qualitative arguments stay the same for an $\R P^2$ symmetry,
as it is exhibited e.g.~by nematic liquid crystals~\cite{deG}.
For an $\R P^2$ symmetry the vacuum manifold is a sphere with opposite points
identified ($\R P^2$ is therefore identical with $S^2/Z_2$ or
$SO(3)/O(2)$). In nematic liquid crystals
the occurrence of this symmetry
is easily understood: the molecules are mirror--symmetric rods or discs,
and the ground state of the theory is reached when all rods have the same
orientation. If the phase change
can propagate faster than the fluctuations in the rotational degrees of
freedom of the molecules,
this is only achievable locally~\cite{Che+90}.

\begin{figure}[htb]
\centering
\mbox{~}{\hbox{\epsfxsize=150pt
\epsffile{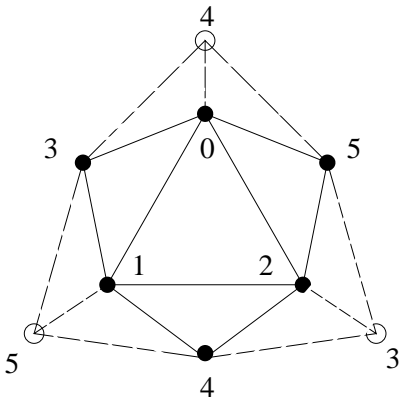}}}\\
\caption{The minimal discretisation of $\R P^2$ and its geodesics. It is
obtained by discretizing the points on the sphere by the vertices of an
embedded icosahedron. We
have to imagine that we look at the
icosahedron facing one of its triangles head--on.
The ``sphere" is completed by identifying opposite points. Where necessary
to identify all geodesics, points of the lower ``half--sphere" have been drawn,
connected by dashed lines.
Every point
can be connected with any other point by exactly one of the links, so that
the geodesic rule is unique. The non--contractible paths are the
ones that go along
an odd number of broken lines, because broken lines lead onto
the other half--sphere.}
\label{fig:RP2}
\end{figure}

We have used only a minimal discretisation and
a continuous $\R P^2$ group to compare measurements of configurational
parameters. The minimal discretisation consists of the vertices of an
icosahedron embedded in the sphere, as depicted in Fig.~\ref{fig:RP2}.
The uniqueness of the geodesic rule and the definitions of non--contractible
paths on this discretisation of $\R P^2$ are both immediately obvious from
Fig.~\ref{fig:RP2}. Non--contractible paths are those which follow an odd
number of those links which cross the equator.
Flux conservation is easily established, too: every tetrahedron edge has
either one of the broken lines associated with it (i.e.~it carries the
field values into the other half--sphere), or a solid line. Changing any
one of the links with respect to this behaviour changes the flux in two
triangles. Thus, the total
flux can only be changed in steps of two (or zero), and the number of triangles
having strings going through them is always even.
By going through the different combinations, it is easy to convince oneself
that the thus constructed strings are also
self--avoiding, i.e.~that no tetrahedron has four faces penetrated by strings.

In the continuous case, the geodesic rule can be realised as follows: Let
the random field assignment on a spatial vertex be a random vector of the
upper unit half--sphere. If the vacuum manifold ${\cal M}$
were to be this half--sphere only, the length of the geodesic between two
points on ${\cal M}$ would just be the angle between the two corresponding
vectors. If ${\cal M}=$$\R P^2$, the geodesic is therefore either this angle
or its complement, whatever is smaller (the probability that a pair of points
is connected by a geodesic of length exactly $\pi/2$ is zero).
Whenever we need to take the complement of
the angle between the two vectors on the upper half sphere, the geodesic will
therefore cross the equator. This happens if the two field angles have a dot
product smaller than zero. Since there are three vertices to each face
of the tetrahedra on our lattice, we need to take all three pairwise dot
products. If the curve drawn by the geodesics has crossed the equator an even
number of times, then it is contractible, otherwise a string has to pass
through the corresponding triangle\footnote{Of course, a
similar criterion has to be
possible for any $\Z_2$ string, and was used for the $\Z_2$ strings
appearing in the breaking of $SO(3)$ in \cite{KibZ2}, using a bounding
sphere instead of a bounding circle. If a closed path on the $SO(3)$
manifold crosses the bounding sphere an odd number of times, it is
non--contractible.}.
Therefore, if the field values on the
vertices of a particular triangle are (in the ``vector on the upper
half--sphere" representation) the vectors $\vec{v}_i\,\,,\,\,i=1,2,3$, then the
string flux through the appropriate triangle is
$n=\frac{1}{2}\left[1-{\rm sign}[(\vec{v}_1\cdot\vec{v}_2)\,
(\vec{v}_2\cdot\vec{v}_3)\,(\vec{v}_1\cdot\vec{v}_3)]\right]$. It cannot have
negative sign, because $\R P^2$ strings are
non--orientable\footnote{This is a direct consequence of the non--orientable
nature of the source field: It is apparent that the sum of two
non--contractible paths on $\R P^N$ is always contractible, as
a non--contractible path is one that ends in the antipode of the starting
point. The concatenation of two non--contractible paths therefore ends in
the starting point itself. This means that any $\R P^N$ string is any other
string's anti--string in the sense that any two strings (parallel to each
other) can form objects which are no longer topologically stabilised.}.
Flux conservation is easily proved~\cite{numerics}, but a continuous representation
of the $\R P^2$ symmetry suffers from the same uniqueness problems as a $U(1)$
string on a cubic lattice, because a single tetrahedron can carry two
strings\footnote{Imagine for instance the vectors $(theta_1,\phi_1)=
(0,0), (theta_2,\phi_2)=(\pi/2-\epsilon,0), (theta_3,\phi_3)=(\pi/2-
\epsilon,2\pi/3),  (theta_4,\phi_4)=(\pi/2-\epsilon,4\pi/3)$, which for
a range of small $\epsilon$ has every face penetrated by a string.}. To avoid
random matching of open string segment (which might
introduce an unnatural bias towards Brownian statistics on large
scales~\cite{MHKS}), we chose to connect the free ends
in such a way that, in case of ambiguities, every string goes through a pair of
faces which share an edge of length $a$, i.e.~the edge length of the bcc
lattice.
The measured ensembles of $\R P^2$ strings are listed in
Table~\ref{table:rpcontperf}.
\begin{table}[htb]
\centering
\begin{tabular}{|c|c|c|c|c|}
\hline
& Number of strings & cutoff $\Lambda$ & fractal dimension $D$ & $b$\\
\hline
continuous & 3000 & 100,000 & $1.979 \pm 0.023$ & $2.59\pm 0.05 $\\
& 100,000 & 2000 & $1.971 \pm 0.001$ & $2.643 \pm 0.014$ \\
\hline
discrete & 10,000 & 10,000 & $1.975 \pm 0.014$ & $2.62 \pm 0.03$ \\
\hline
\end{tabular}
\caption{Measurements for the fractal dimension of continuous $\R P^2$
strings.}
\label{table:rpcontperf}
\end{table}
The continuous $\R P^2$ strings do not seem any more Brownian than the
ones which are forced to be self--avoiding. This may indicate (as in the
case of $U(1)$) that the discretisation of the vacuum manifold does not
significantly affect the measurements for perfect symmetry,
maybe because we have not allowed random reassignments of string
pairs to each other, but
we have not checked whether a random solution to the problem of uniqueness
would indeed bias the statistics towards Brownian configurations.
In any case, no discretisations of $\R P^2$, other than the minimal one,
have been investigated at this stage. In fact, no
discretisation which would be finer than the minimal one, but still force
self--avoidance, is known to us. Interestingly enough, a discretisation
produced by embedding a tetrakaidekahedron into the two--sphere is uniform.
Uniform distribution of the lattice points on the sphere is a necessary
criterion for unbiased data (cf.~the discussion in the following sections).
However, it is easy to convince oneself that
many of the vector pairs in that scheme are at right angles to
each other, introducing ambiguities in the definition of the string flux
through a triangle. Another discretisation, achieved by
embedding a dodecahedron in the sphere,
produces a discretisation which does not exhibit these ambiguities,
but does allow two strings to penetrate a tetrahedron. The appropriate
proof is developed in the Appendix.

We should mention that $SO(3)$ strings have been measured to
have a similar tendency to have lower fractal dimension, and therefore
higher values for $b$. Kibble~\cite{KibZ2} arrives at values of
$D=1.950\pm 0.037$, and $b=2.546\pm 0.065$. It is therefore possible that such
deviations are generic for either $\Z_2$ strings or for higher dimensional
vacuum manifolds. We will further discuss this issue later.

\section{String Percolation and Biased Symmetry Breaking}
\label{section:3}

\subsection{Low String Density}

Drawing lessons from polymer statistics, the fact that our algorithm
generates nearly Brownian strings could be a result of the dense
packing of strings. From what we have measured so far, there is a strong
caveat to that statement: The continuous $\R P^2$ strings are actually denser
($1/\pi$ strings per face~\cite{V91}) than the continuous $U(1)$ strings
($1/4$), but exhibit more ``self--avoiding" statistics.
This trend also holds for the minimally discretised ensembles ($5/18$
for $\R P^2$, and $2/9$ for $U(1)$).
So how does the string density affect string statistics?

We have already shown that for minimally discretised $U(1)$-strings, a
Hagedorn-like transition~\cite{Hag,BowWij,MitTur} occurs below a critical
string
density~\cite{MHKS}. According to Vachaspati~\cite{V91}, we can achieve
variations in the string density by inducing correlations
in the order parameter by lifting the degeneracy of the manifold of
equilibrium states. This reduces the probability of a string
penetrating the face of a lattice (Thus we can generate an ensemble
with the {\it average\ } string density fixed at will.
Physically one can think of this as applying an external field, which
spoils the symmetry of possible ground states),
but {\it increases\ }the dimension $D$, which
argues against the identification of strings with polymers.
There is a critical density below which there are no ``infinite''
strings.
In the low density phase there is a scale $c$
which appears in the loop length distribution,
\begin{equation}
dn = a l^{-b} e^{-cl} dl\,\,\,\,\,,\,\,\,\,\,
\label{eq:lengthdistr2}
\end{equation}
as a cut-off. As the critical density is approached from below,
$c\to 0$, and the mean square fluctuation in the loop length
\begin{equation}
S = \langle l^2\rangle - \langle l \rangle^2\,\,\,\,\,,\,\,\,\,\,
\label{eq:S}
\end{equation}
diverges (see exponents $\gamma$ and $\psi$ in Table~\ref{table:U1}).

This divergence signals a phase transition, in some ways analogous to the
Hagedorn transition in relativistic string theory at finite temperature.
This has been implicated in
many branches of physics. Previous
studies \cite{Cop+} deal with string dynamics and can
treat the ensemble in thermal
equilibrium~\footnote{The ensemble in \cite{Cop+}
is therefore very appropriate for
situations where the critical temperature is approached slowly.}.
Vachaspati's algorithm enables
us to measure directly the string statistics such as the critical
density, the dimension, and the critical exponents,
and to test the validity of the hypothesis of scale invariance for the
{\it initial conditions}, which cannot be expected to be thermalised.

\subsection{Low String density and the Hagedorn transition}
\label{section:RPI}

{}From the ``microscopic" point of view,
Vachaspati~\cite{V91} argued for such a Hagedorn-type transition to occur at
low string densities with the following reasoning:
consider a string formation simulation on a cubic lattice. The probability
of a string passing through a certain face 1 of the cell is $p_s$. Since the
plaquette opposite face 1 is causally disconnected, the probability for it
to have a string passing through it is also $p_s$, regardless of the
actual situation at face 1.
Therefore, the probability for a string to bend after entering a cell is
$1-p_s$. Now, if we reduce $p_s$, the bending probability increases and
the chances of the string closing up to form a loop also increases.
As Vachaspati argues,
``This tells us that by reducing the probability of string formation, or
equivalently, by decreasing the string density, we can decrease the
infinite string density and increase the loop density".

Vachaspati then goes
on to construct a model with $\Z_2$ strings (i.e.~non--orientable
strings), in which he assigns
either $+1$ with the probability $1-p$ or $-1$ with probability $p$ to
each link of the lattice (on a periodic lattice). A string is said to pass
through a plaquette if the product of the field values on the associated
links is $-1$. This is, although reminiscent of it, not quite identical with
the way we constructed our $\R P^2$ strings in the previous section, because
whether a $+1$ or a $-1$ is ``assigned" to a link in the continuous $\R P^2$
case depends on the relative angles between the three vectors involved, so
that the assignment to the links are not entirely
uncorrelated. If they were, then the probability of an $\R P^2$ string
passing through a triangular (or in fact any) plaquette should be
$\frac{1}{2}$,
whereas it is (for a triangular plaquette and continuous $\R P^2$)
$\frac{1}{\pi}$~\cite{V91}. The way strings are constructed in
ref.~\cite{V91} is, however, appropriate to model $\R P^{\infty}$ strings.
We can see
this by the following argument: it is well known that the different
components of a random unit vector in $\R^N$, in the limit
$N\rightarrow\infty$,
become mutually uncorrelated Gaussian random variables with standard
deviation $\sigma=\sqrt{N}$. Any two random unit vectors will therefore
have positive or negative sign with equal probability. The relative angles
to a third unit vector, and in particular their signs, are then completely
uncorrelated to this angle, so that taking the sign of the product of
uncorrelated Gaussian random variables would indicate whether a
sequence of geodesics between random points on $S^{\infty}/\Z_2$ will cross the
horizon or not.

With this model Vachaspati observed that, as the symmetry bias is
increased, a lot of string mass is transferred from infinite strings to
loops, so that the loop density actually increases. This is not what one
would e.g.~expect from statistical arguments for a box of
(non--interacting) strings
in equilibrium~\cite{MitTur}, so that one should not
assume {\it a priori} that the string statistics right at the phase
transition will follow statistical mechanics arguments.
Another prediction of statistical mechanics arguments is that, at low
densities, the loop distribution is described by Eq.~(\ref{eq:lengthdistr2}),
with $b=\frac{5}{2}$. Vachaspati, however, measures values consistent
with 2 (within large statistical errors).
As one increases the string density again, approaching the scaling
regime, $c$ approaches zero from above, signalling a phase transition ($c$
can be interpreted as the inverse of some characteristic length scale arising
from the breakdown of scale-invariance).

\begin{figure}[htb]
\centering
\mbox{~}{\hbox{
\epsfxsize=240pt
\epsffile{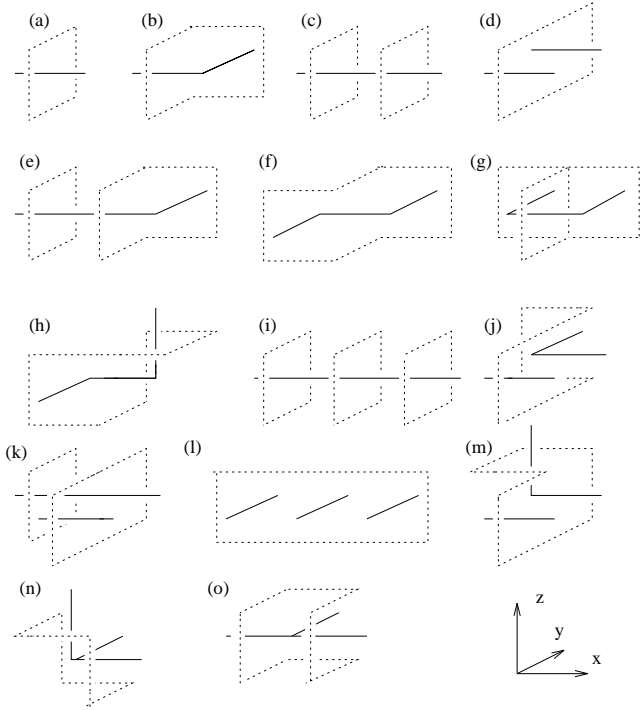}}}\\
\caption{The string loops formed in this model, with one
isolated link with field value $-1$ (a), and with two ((b) to (d)) or three
((e) to (o)) such links adjacent to each other. The fact that infinite
strings disappear looks very reminiscent of a bond percolation problem for
the ``$(-1)$--links", except that two consecutive strings, if they are aligned,
do not surround themselves with pieces of the same string (the prototype is
(c),
other examples are (e), (i), and (k)), whereas neighbouring parallel links do
(the prototype is now (d), with other examples being (j), (k), (l), and (m)).}
\label{fig:Vach}
\end{figure}

Vachaspati's argument, relating the probability of a string forming at
a particular lattice plaquette to the Hagedorn transition,
actually does not go far enough: it supports a notion
that the string is getting wigglier as we decrease the density, which could,
strictly speaking, result in just a rescaling of some of the parameters, but
none of the exponents: the scaling function in Eq.~(\ref{eq:scaling_function})
could converge towards a different constant, the factor $q$ in
Eq.~(\ref{eq:lengthdistr3}) could change, all without changing $D$ or $b$,
which determine the global properties of the network after the local properties
have been absorbed into appropriate prefactors. In particular, the
{\it complete} disappearance of infinite strings is not explained convincingly.
What Vachaspati observes in
Monte Carlo measurements, however, {\it can} be explained on the ``microscopic"
level. Vachaspati varies the string density by decreasing $p$, the probability
for a link to have the value $-1$ assigned to it. Let us take it to the
extreme and assume that all the links that have $-1$ assigned to them are
so rare that they are usually isolated from each other, submerged into a
sea of links with field value $+1$. Then it is obvious that a string loop
of minimal length winds around each of these links, so that there will be
nothing but a few isolated short string loops. We can take it further and
ask ourselves what happens when two such links are adjacent to each other.
If they are consecutive links with the same orientation, they will have their
own loops of length four, if they are in different spatial orientations,
a loop of length 6 will form, as depicted in Fig.~\ref{fig:Vach}.
This figure also illustrates that, because the length of the strings
seems to be intimately linked to the size of the $(-1)$--link clusters,
{\it in the Vachaspati model,
the Hagedorn transition is almost a bond percolation problem}, except
that parallel bonds touching each other (i.e.~bonds along the same line)
do not connect their strings with each other, and parallel bonds which are
just one lattice spacing apart, do. There are more configurations of these
``$-1$--links" that
break this correspondence between the Vachaspati model and bond percolation
(e.g.~a flat cross of four $(-1)$--links produces
two separate loops). Thus, although there is no one--to--one correspondence,
one still intuitively expects the Hagedorn threshold to be close to
the bond percolation threshold.
Indeed, Vachaspati measures
a percolation threshold of $p_c\approx 0.29$, while the threshold for bond
percolation on a simple cubic lattice in three dimensions is
$p_c=0.3116$~\cite{Stauffer}.
There is more to be learned from the correspondence of the Vachaspati model
with bond percolation.
To get a respectable number of large, but isolated lattice animals,
we have to approach the percolation threshold from below. At the threshold,
the percolating cluster has a well defined fractal dimension. Thus
we conclude that scaling must be restored as the
percolation threshold is approached from below, and a fractal dimension
will begin to become well defined.
Lastly, we shall just mention that one can easily derive the general form
of Eq.~(\ref{eq:lengthdistr2}) by similar percolation arguments.

Not only can we now claim
to understand the microscopic aspects of the lattice description
of this Hagedorn-like transition, but we also expect this transition to
have many properties of a percolation transition. We can relate
many variables and critical exponents of the Hagedorn transition
to critical behaviour in standard percolation transitions.
With his model, Vachaspati got qualitative indications of a lot of the
results which are to follow here.
With the infinite--lattice and the hash--table algorithms used in~\cite{MHKS}
and presented in~\cite{numerics,thesis}
we have some advantage when extracting numerical data or attempting
reasonably large ensembles for good statistics.

\subsection{String Percolation in the Vachaspati--Vilenkin Algorithm}

Now we need to go back to the more realistic model: the Vachaspati--Vilenkin
method on a tetrahedral lattice.
The string density can only be varied (once the lattice
is chosen) by lifting the degeneracy in the vacuum states, i.e.~by making
some vacuum states less likely than others.
Once the details of the discretisation
of space and of the vacuum manifold are chosen,
{\it the initial string density, and in particular the density in
infinite strings, can only be changed
by spoiling the vacuum symmetry}.
Some of the recent work on dynamical scaling\footnote{Dynamical
scaling is quite different from scale-invariance. Dynamical
scaling is exhibited if the system looks statistically the same at all
times, on length scales which may vary with time according to some power
law or some other function of time.
This does not imply that the system is scale--invariant.
In fact, in a scale--{\it invariant} system (if it
stays scale--invariant), dynamical
scaling is a misplaced concept,
because there is no length scale which could evolve in time. Parameters
which are {\it not} scale--invariant, and whose dynamical scaling it therefore
makes sense to observe, like e.g.~ the average string--string separation,
are those parameters which are affected by lattice-effects in the VV
algorithm.} in string
networks~\cite{NeilPedro} implies that the ratio of the
densities in string loops and in infinite strings may be freely
variable, based on the realisation that this ratio depends on the lattice
description invoked. Whereas we agree with the general argument\footnote{The
density in loops depends on the lattice. Our tetrahedral lattice allows smaller
loops (in units of correlation lengths) than a cubic lattice, and one expects
more loops to appear because the low cutoff in Eq.~(\ref{eq:lengthdistr})
gets shifted to lower values. There is also a difference in this ratio
depending
on the discretisation of the vacuum manifold, and on the vacuum manifold
itself.}, we will show in section \ref{section:6.1} that there is
probably a lower limit to the amount of infinite
string that has to appear, and that infinite strings would
therefore be a generic feature of the VV algorithm on a regular lattice.
This issue is still controversial, but in
some simple cases, like the Vachaspati model, we can develop a percolation
theory understanding for the emergence of an infinite string network. Had
Vachaspati used a tetrahedral lattice, there would still be infinite strings,
as the bond percolation problem threshold for the bcc lattice is $p_c=0.1803$,
and the symmetric case has $p=0.5>p_c$\footnote{The reason why the bond
percolation
threshold is reduced compared to the simple cubic lattice is, from the
percolation theory viewpoint, that there are more bonds per lattice site.
Within the string network picture, the reason is that we have a finer mesh
and therefore a higher string density.}. In fact, on any three dimensional
lattice $p_c<0.5$, so that the appearance of infinite strings is
lattice--independent. Serious lattice ambiguities would only arise if
strings (under the same physical conditions)
percolate on one lattice, but not on another, i.e.~if $p_c$ lies in between
percolation thresholds of different lattices. No such model is known.

Because of its better correspondence to a physical situation, let us consider
another brief example, taken from the measurements in the next section:
Take the tetrahedral lattice with a minimal
discretisation of $U(1)$. Let us denote the three possible field values by
0, 1, and 2. We introduce a bias in the symmetry, such that the value 2
is assigned with the probability $p_s$, and the other two values have the
probabilities $(1-p_s)/2$. Without loss of generality, let us constrain
ourselves to biases with $p_s\le \frac{1}{3}$. We can produce infinite strings
only if all three field values percolate. In particular, this implies
that $p_s>p_c$, where the critical value $p_c$ is the site percolation
threshold of a bcc lattice, $p_s\approx 0.246$~\cite{Stauffer}. In
the unbiased case $p_s=\frac{1}{3}$. Again, this is higher than the site
percolation on {\it any} sensible lattice, such that
the appearance of infinite strings is a generic feature.
{}From measurements in the next
section, we deduce $p_c=0.2476 \pm 0.0014$. The agreement is almost
suspiciously
good, but certainly justifies the percolation theory arguments for
an intuitive understanding of the Hagedorn transition. Had we taken a simple
cubic lattice, we would still be above the percolation threshold $p_c=0.3116$,
and get infinite strings~\cite{VV}. In both, Vachaspati's $\Z_2$ model and
the minimally discretised $U(1)$ model we get infinite strings irrespective
of the lattice we are using\footnote{To be precise, a diamond lattice would
not allow site percolation for $p=\frac{1}{3}$.
However, because it has hexagonal ``plaquettes"
(the quotation marks are to indicate that the plaquettes not planar), it
is unsuitable not only for a simulation of the Kibble mechanism, but also
for the percolation theory argument developed here. This is because two
consecutive plaquettes are not everywhere connected by link walks of
length one, so that the lattice points with disfavoured vacuum values
do not need to neighbour each other directly to allow strings to percolate,
and site percolation with {\it next}-nearest neighbours should be our
reference in this case.}.
Percolation phenomena have long been know to be independent
of the microscopic details of the lattice. This may lend some support to
the assumption that, in the Vachaspati's $\Z_2$ model (and maybe more
generally for $\Z_2$ strings) and for $U(1)$ strings the emergence of
a network of infinite strings is a generic feature. Although the
correspondence of the
Hagedorn transition to a percolation phenomenon seems rather strong, we suffer
from the same deficiency here as most of percolation theory does: there is
no analytic proof.

In many respects, the best we can hope for is to establish a better
understanding through better and more numerical measurements.
The next section is therefore dedicated to the results of various measurables
of the percolation transition.
We will bring more arguments for the correspondence of the string ensembles
with a percolation theory picture later when
we discuss the results of those measurements.

\section{Numerical Results for Biased String Formation}
\label{section:Hagedorn}

\subsection{$U(1)$ strings}
\label{sec:4.1}

\begin{figure}[htb]
\centering
\mbox{~}{\hbox{
\epsfxsize=140pt
\epsffile{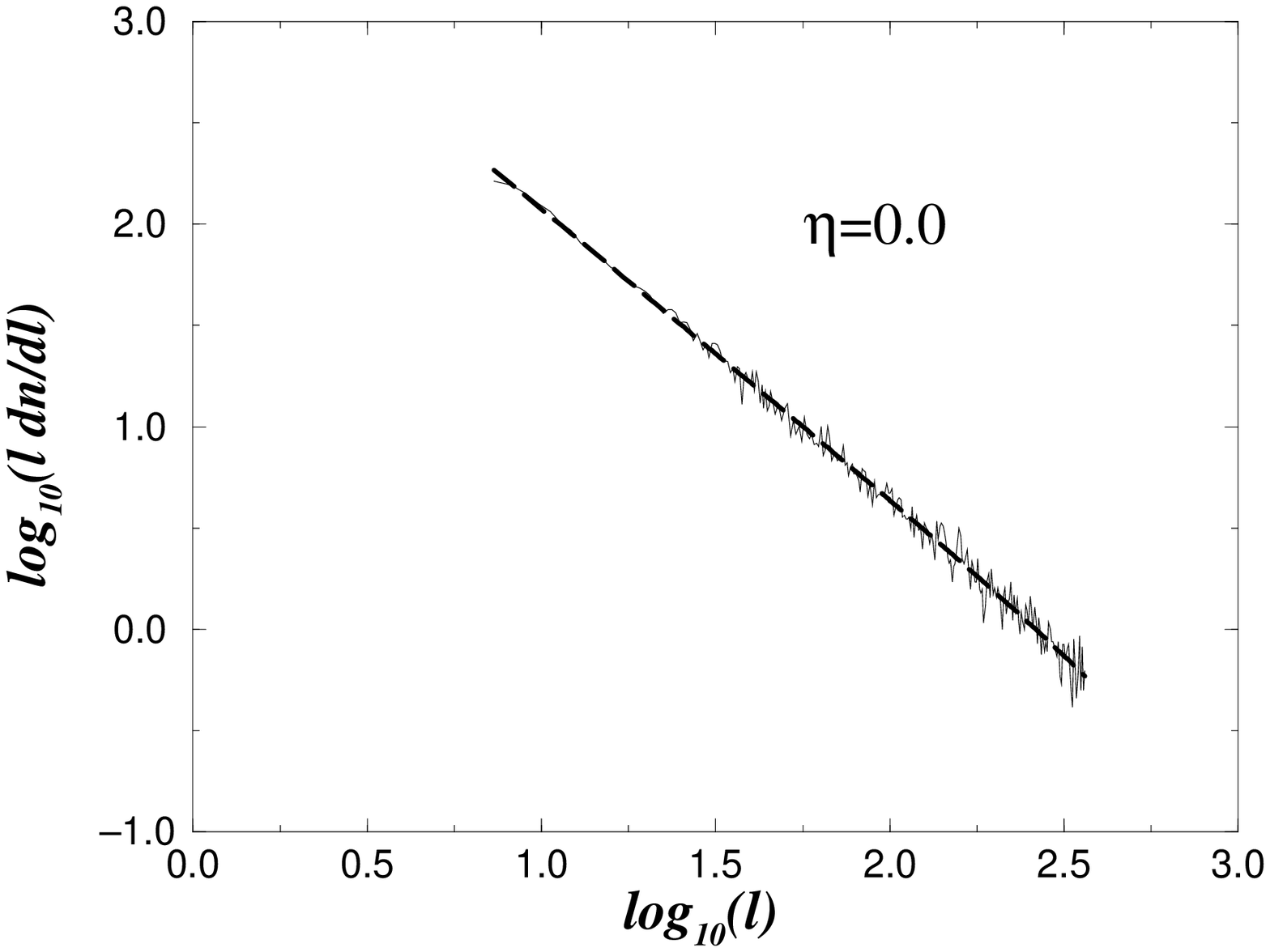}
\epsfxsize=140pt
\epsffile{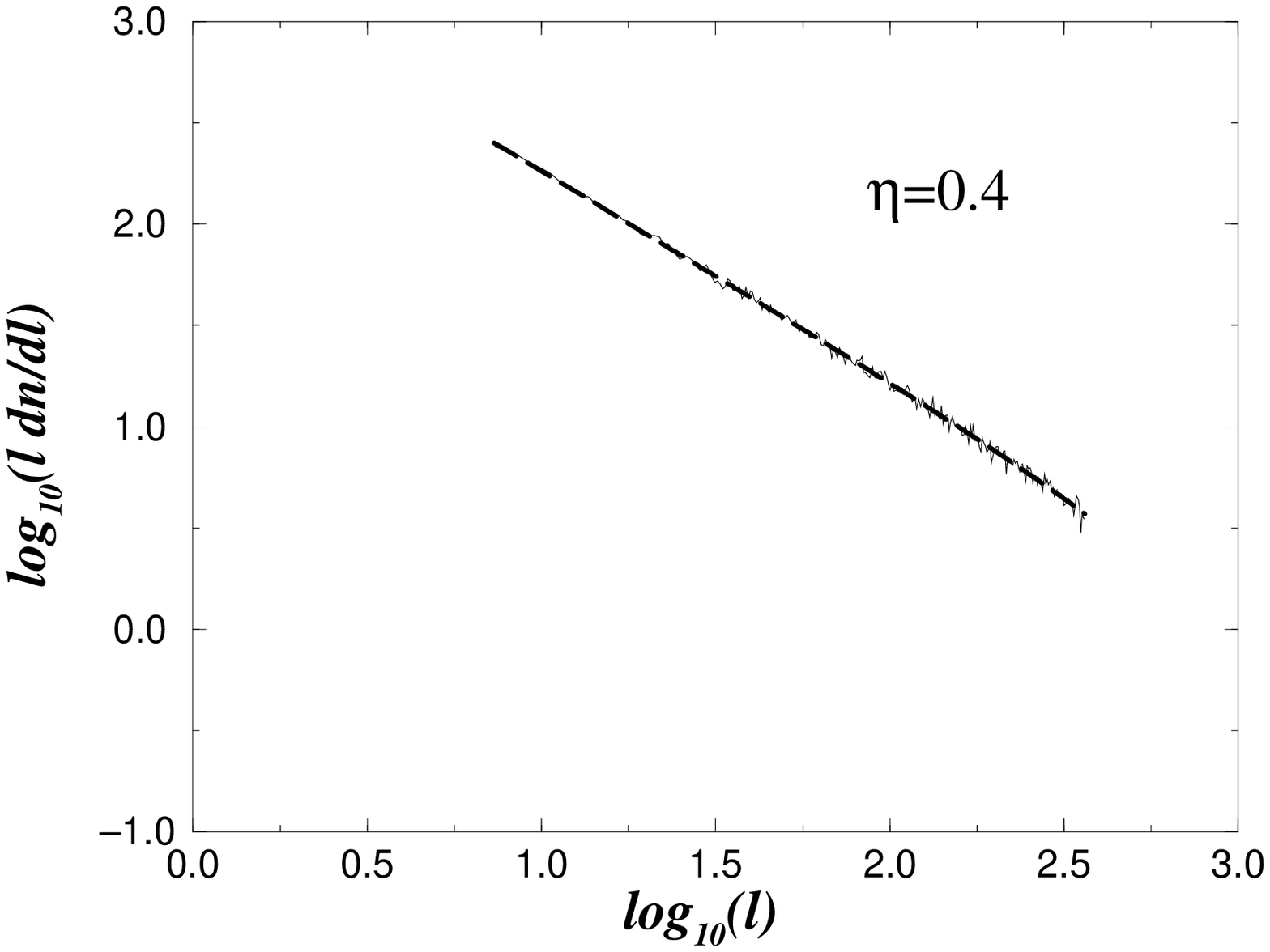}}}\\
\mbox{~}{\hbox{
\epsfxsize=140pt
\epsffile{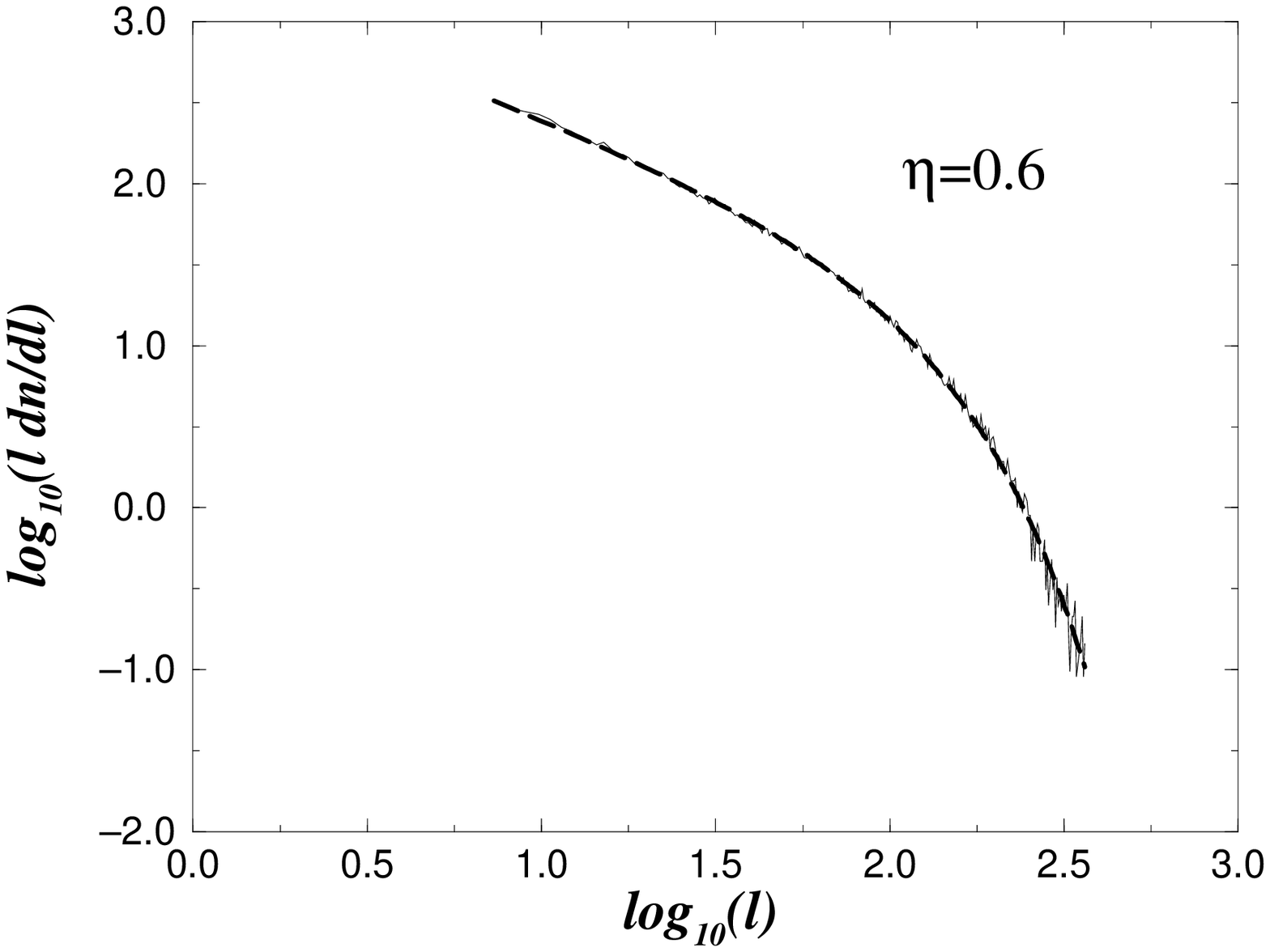}
\epsfxsize=140pt
\epsffile{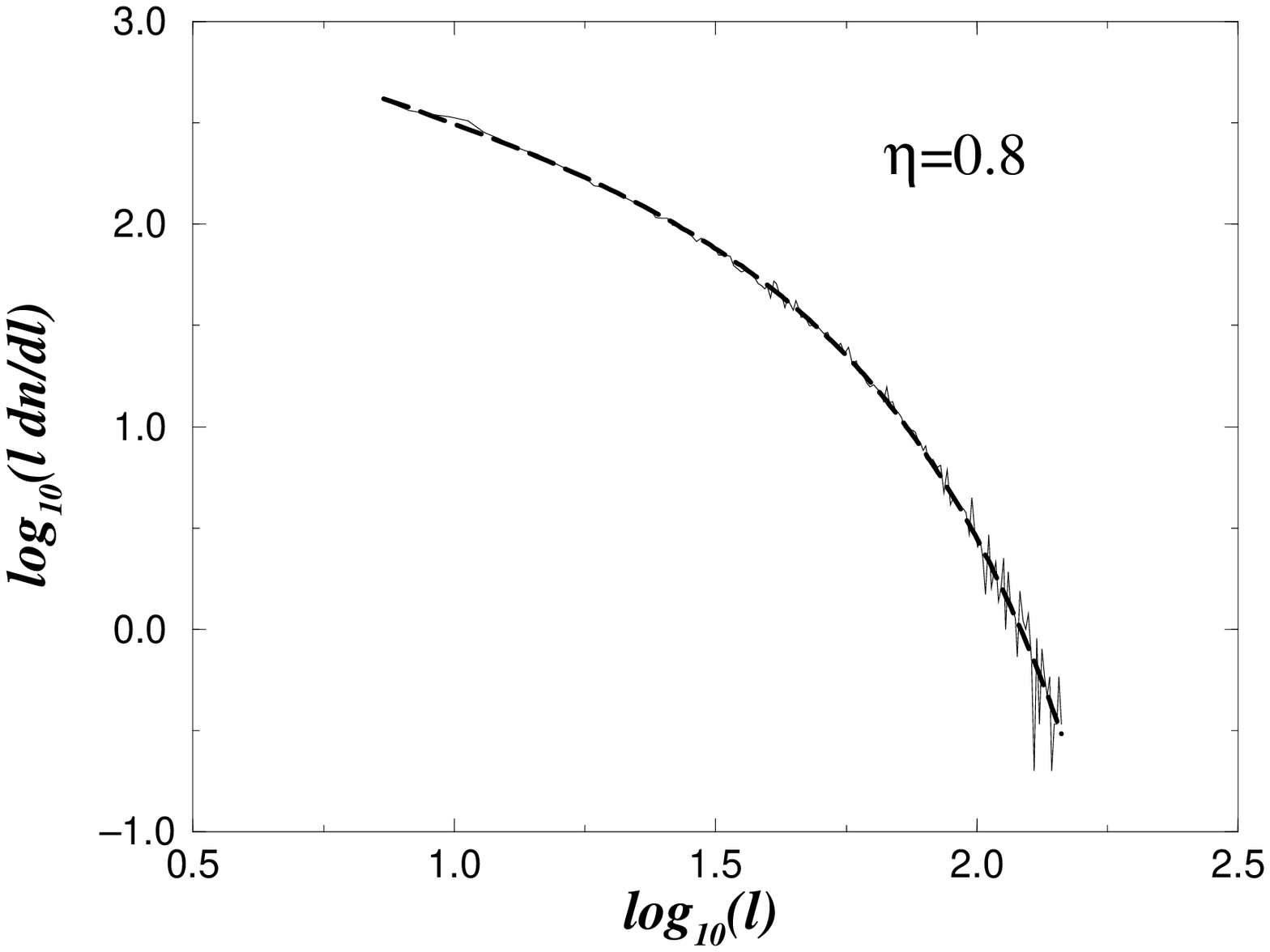}}}\\
\caption{Typical fits to the Eq.~(\ref{eq:lengthdistr2}), with different
values of the bias. Where infinite strings are present, no cutoff can be
identified. Where $\Lambda<1/c$ (i.e.~for $\eta=0.4$), the cutoff cannot
be seen, as the strings in this ensemble are too short. For higher values
of the bias, the cutoff can be recognised clearly. The dashed lines are the
fits using Eq.~(\ref{eq:lengthdistr2}). No deviations from a behaviour of the
type in Eq.~(\ref{eq:lengthdistr2}) can be recognised. The ensemble consists
of 100,000 strings with cutoff $\Lambda=2000$.}
\label{fig:lengdistr2fit}
\end{figure}

\begin{figure}[htb]
\centering
\mbox{~}{\hbox{
\epsfxsize=240pt
\epsffile{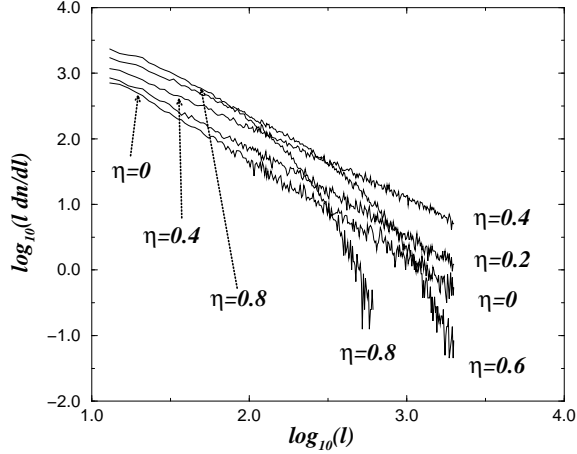}}}\\
\caption{Comparison of the loop distributions for different bias. It can
be seen that the loop density is increased for small bias. For large bias,
the running of the cutoff length can be observed.}
\label{fig:lengthdistr2comp}
\end{figure}

\begin{figure}[htb]
\centering
\mbox{~}{\hbox{
\epsfxsize=240pt
\epsffile{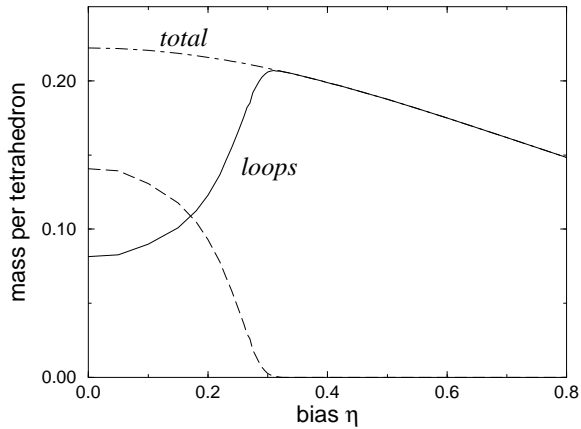}}}\\
\caption{The mass density in string segments belonging to loops (in units
of one per tetrahedron) (solid line), in infinite strings (dashed line),
and the total mass density, given by Eqs.~(\ref{eq:loopdensity}) and
(\ref{eq:C}). The loop density increases as we approach the percolation
threshold from above, and energy from infinite string is transferred into the
loop ensemble.}
\label{fig:loopdensityU1}
\end{figure}

The following convention has been used to introduce a bias for $U(1)$ strings:
We discretised the $U(1)$ manifold by $N=2^n-1$ points,
and assigned the following probabilities to the each of these points
$m\in\{0,1,2,...,N-1\}$
\begin{equation}
p(m)=C^{-1}\exp(-\eta\,\cos(2\pi m/N))\,\,\,\,\,,\,\,\,\,\,
\label{eq:probability}
\end{equation}
where $\eta$ is the bias parameter and $C$ simply normalizes the
probabilities
\begin{equation}
C=\sum_{m=0}^{N-1} \exp(-\eta\,\cos(2\pi m/N))\,\,\,\,\,.\,\,\,\,\,
\label{eq:normalization}
\end{equation}
Unless stated otherwise, we will quote results from the minimal discretization
of $U(1)$ in this section, i.e.~$N=3$. The reason why this is the
best--studied ensemble is the fact that it corresponds most closely to a site
percolation problem, and therefore relates best to the discussion of the
results in the next section.
Firstly, we confirm that Eq.~(\ref{eq:lengthdistr2}) gives an extremely
good fit for the loop distribution beyond the percolation threshold.
Typical such fits are shown in Fig.~\ref{fig:lengdistr2fit}.
In Fig.~\ref{fig:lengthdistr2comp} we compare the loop distributions for
different biases. It can be seen that for low bias the string
density in loops increases with increasing bias, in agreement with Vachaspati's
measurements.
The density in loops as a function of $\eta$ is shown in
Fig.~\ref{fig:loopdensityU1}. The mass density in loops (in units of
segments per tetrahedron) is obtained in the following
way:
Let $p_t$ be the probability for a triangle to carry a string segment.
Since the number of triangular plaquettes on a tetrahedral lattice is
twice the number of tetrahedra, the average number of strings per
tetrahedron is then $2p_t$.
Whenever we start tracing a string, we start at a randomly chosen
string segment (out of all possible segments on the imagined infinite lattice).
Thus we will pick elements belonging
to an infinite string according to their density ratios, such that
\begin{equation}
\rho_{\rm loop}=\rho_{\rm total}\,\frac{N_{\rm loop}}{N_{\rm
total}}\,\,\,\,\,,\,\,\,\,\,
\label{eq:loopdensity}
\end{equation}
where $N_{\rm total}$ is the number of strings we have traced in the ensemble,
and $N_{\rm loop}$ is the number of those which turned out to be loops.
The unknown parameter is $p_t$. For the minimally discretised $U(1)$
manifold, the sum $C$ of Eq.~(\ref{eq:normalization}) is just
\begin{equation}
C=2\,e^{\frac{\eta}{2}}+e^{-\eta}\,\,\,\,\,,\,\,\,\,\,
\label{eq:C3}
\end{equation}
and the probability for a triangle to exhibit all three vacuum values on
its vertices reduces to the very simple form
\begin{equation}
p_t=\frac{3!}{C^3}\,\,\,\,\,.\,\,\,\,\,
\label{eq:C}
\end{equation}
For other discretisations of $U(1)$ we use exact numerical summations to
extract $p_t$, as the functional form becomes a rather complicated sum
of a number of terms that increases as $N^3$ and has to be evaluated for
all values of $\eta$ used in the data set. It turns out that
the defect density increases slightly
when a finer discretisation of the vacuum manifold is used\footnote{This
tendency is also observed in Monte Carlo simulations of texture
formation~\cite{LeeProtexture} and monopole formation~\cite{LeePro}, and
in the $\R P^2$--string measurements in the following section.}.
The string density per tetrahedron
in terms of the bias, for the minimal discretisation
of $U(1)$, is therefore given by
\begin{equation}
\rho_{\rm total}=\frac{2\cdot3!}
{(2e^{\frac{\eta}{2}}+e^{-\eta})^3}
\,\,\,\,\,.\,\,\,\,\,
\label{eq:stringdensity}
\end{equation}
This can be used as a reparameterisation of the bias, so that all variables $X$
scaling like $X\propto|\eta-\eta^\star|^{\chi}$ near the critical point
will also scale as $X\propto|\rho_{\rm total}-\rho_{\rm total}^\star|^{\chi}$,
as it is a smooth
and analytic function of the bias\footnote{Note that e.g.~the mass density
in loops cannot be taken as such a reparameterisation, as it is not smooth
at the critical point.}. Fig.~\ref{fig:loopdensityU1} shows the separate
mass densities in infinite strings and loops.
Note that the energy in loops at the percolation transition
still exceeds the energy in infinite string at zero bias! In analogy with
the polymer literature, we could say that this
transition (when approached from the non--percolating phase) is very
efficient in pumping energy into the entropy terms, i.e.~in utilising
new degrees of freedom as the bias is lowered\footnote{This is why we
call it a Hagedorn-like transition: the Hagedorn transition~\cite{Hag}
is associated with an exponential increase in the degrees of freedom,
such that (in the thermal situation) the Hagedorn temperature is not
reachable, as all the energy -- pumped into the system to further increase
the temperature -- goes into entropic terms of the Helmholtz free energy.
However, since our model does
not deal with a thermalised ensemble (or with any dynamics at all) we can
still reach domains beyond this Hagedorn-like transition.}.

Figure~\ref{fig:loopdensityU1} allows us to measure
the location of the percolation threshold, by fitting a power law of the
form\footnote{Where appropriate, the exponents are named according to their
use in percolation theory. In percolation theory $\beta$ is the critical
exponent associated with the strength of the infinite network. Since the
total mass density is a smooth function of $\eta$, $\beta$ is
also associated with the mass density in loops.}
\begin{equation}
\rho_{\infty}\propto(\eta^\star-\eta)^{\beta}\,\,\,\,\,.\,\,\,\,\,
\label{eq:gammaU1}
\end{equation}
We do this by trying different fixed values of $\eta^\star$ and
taking the one that gives the smallest sum of residuals on a log--log
least squares fit\footnote{This means that the statistical errors are
obtained in a less accurate procedure than in the previous section:
Instead of taking many different ensembles,
we take the fluctuations of $\eta^\star$
to be such that the sum of the error squares in the linear fit to the
plot of $\log(\rho_{\rm loop})$ vs.~$\log(\eta^\star-\eta)$ is allowed
to fluctuate by a factor of two around its minimum. The respective slopes
will usually differ by an amount of the order of the statistical error.
Since we have to vary both a lower and an upper cutoff, as well as the
estimate for $\eta^\star$, when searching for the best fit, this method
reduces the large computational effort which would be involved if we had
extracted statistical errors by measuring many
different ensembles for every symmetry group. Quite often, we get
very large estimated errors in $\eta^\star$ and the
critical exponents because of the many free variables involved. A proper
analysis of corrections to scaling, as done in ref.~\cite{Bradley} for the
minimally discretised $U(1)$ strings, is necessary, but will be done
elsewhere.}. We measure
\begin{equation}
\beta=0.54 \pm 0.10\,\,\,\,\,,\,\,\,\,\,
\label{eq:gamma_exponent}
\end{equation}
where the fit has been done in the region $\eta\in[0.22,0.265]$, and
the errors are associated with the uncertainty in
$\eta^\star$, which is measured to give the best fit at
$\eta^\star=0.279\pm0.005$. If the uncertainty in $\eta^\star$
is large, the errors for $\beta$ get quickly out of hand.
The criterion of whether one gets a good fit or
not is not very efficient in finding $\eta^\star$,
but it is the best we can do\footnote{In
percolation theory there is a useful procedure (cf.~ref.~\cite{Stauffer},
p.~72), which involves observing how the probability to find a
lattice--spanning cluster (as a function of $p$) scales with the size of
the lattice. For example, one could look at how the point where this
probability is $\frac{1}{2}$ scales with the system size and then extrapolate
where this point will end up as the lattice size goes to infinity. This
gives a very good estimate for the percolation threshold only if one keeps
track of all clusters generated on a given lattice. We only trace
one string at a time, not worrying about the rest of the lattice, so that
this method of identifying the percolation threshold does not work.}.

\begin{figure}[htb]
\centering
\mbox{~}{\hbox{
\epsfxsize=240pt
\epsffile{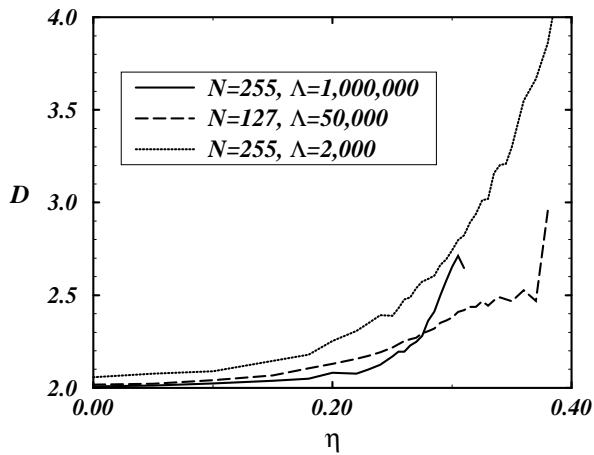}}}\\
\caption{The fractal dimension of the infinite string ensemble as a function
of the bias, plotted for different values of the upper length cutoff $\Lambda$.
The measurements are for nearly continuous $U(1)$.}
\label{fig:fractal_running}
\end{figure}

\begin{figure}[htb]
\centering
\mbox{~}{\hbox{
\epsfxsize=240pt
\epsffile{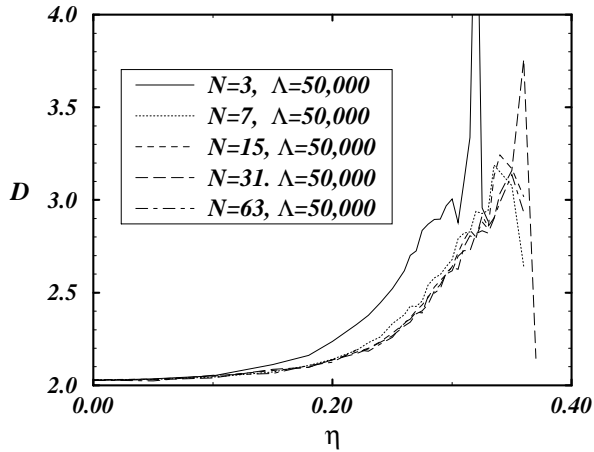}}}\\
\caption{The fractal dimension of the infinite string ensemble as a function
of the bias parameter, plotted for different discretisations of $U(1)$.
The critical biases can be found in Table \ref{table:U1} and in Figure
\ref{fig:rho_crit_N}.}
\label{fig:fractal_running2}
\end{figure}

The fractal dimension is plotted in Fig.~\ref{fig:fractal_running}
for nearly continuous $U(1)$, but with different upper cutoffs $\Lambda$,
and in Fig.~\ref{fig:fractal_running2} for $\Lambda=50,000$, but with
different discretisations of $U(1)$.
Five features are noteworthy:
\begin{itemize}
\item
The statistical errors increase with increasing bias, because the number
of infinite strings in the ensemble becomes smaller.
\item
The fractal dimension stays nearly constant for very small bias.
We observe that the $\Lambda$--dependence of $D$ as measured in
the previous section is {\it not} a statistical fluctuation.
\item
The fractal dimension becomes not only
hard to measure near the percolation threshold,
but also becomes ill--defined beyond it,
because we are counting many strings wrongly as infinite
($\langle l_{\rm loop}\rangle$ diverges at the critical point),
whereas they will eventually turn back onto themselves and form loops.
\item
The behaviour is largely independent of the particular discretisation used,
except for the obvious shift in $\eta^\star$, pronounced only for $N=3$.
\item
The measurements are consistent with a possible assumption that, as
$\Lambda\rightarrow\infty$, $D=2$ right up to the critical point.
\end{itemize}

\begin{figure}[htb]
\centering
\mbox{~}{\hbox{
\epsfxsize=240pt
\epsffile{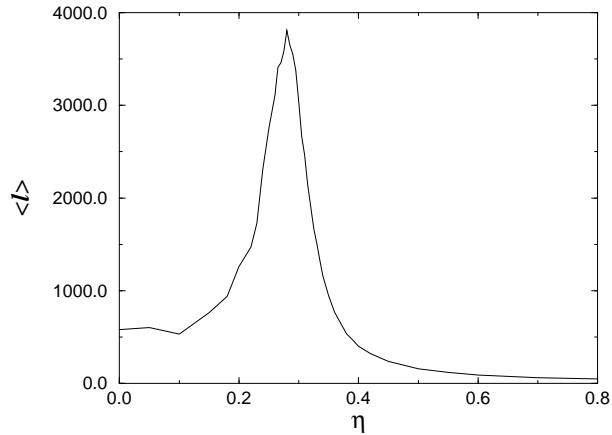}}}\\
\caption{The average length of a loop in the minimal discretisation of $U(1)$,
as a function of the bias $\eta$. Here $\Lambda=50,000$.}
\label{fig:loopsize}
\end{figure}

\begin{figure}[htb]
\centering
\mbox{~}{\hbox{
\epsfxsize=240pt
\epsffile{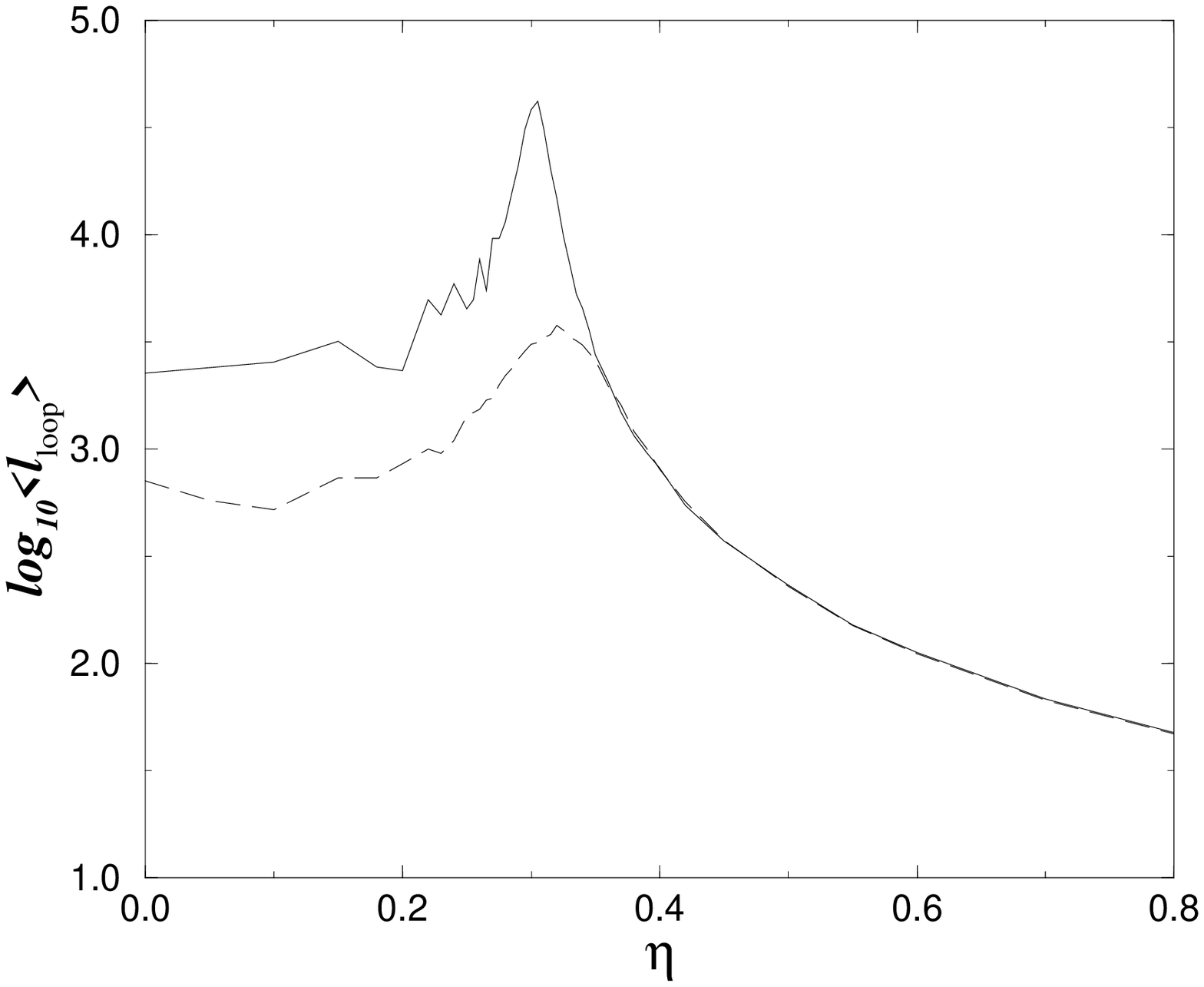}}}\\
\caption{The logarithm of the average length of a loop (with $U(1)$ nearly
continuous: $N=255$, and $N=127$)
as a function of the bias $\eta$, for different cutoffs.
For the solid line $\Lambda=1,000,000$, whereas for the dashed line it is
$\Lambda=50,000$. They should (to very good approximation) have the same
critical bias. However,
the lower--cutoff
peak has an inflexion point, where it deviates from the power--law
behaviour, well before the critical point (at around $\eta=0.36$), where we
need to stop the fits to measure the critical exponents. This effect gets
worse with smaller cutoffs, and gives rise to inaccuracies in estimating the
exact position of $\eta^\star$.}
\label{fig:loopsize2}
\end{figure}

Measurements for the average loop size\footnote{Note that we mean the
average size of a loop that a randomly chosen string segment belongs to.}
are shown in Figs.~\ref{fig:loopsize} and \ref{fig:loopsize2}. In
Fig.~\ref{fig:loopsize2} the effects of a finite cutoff are also explained.
The behaviour is just as one would expect from percolation theory:
In the non--percolating phase, the main contribution comes from gradually
larger clusters as we approach the percolation threshold.

Let us assume the average loop size near the percolation threshold
scales as
\[
\langle l_{\rm loop} \rangle \propto
|\eta-\eta^\star|^{-\gamma}\,\,\,\,\,.\,\,\,\,\,
\]
The best power--law fits to the loop size give
\[
\eta^\star_{(N=3)}=0.279\pm 0.004\,\,\,\,\,,\,\,\,\,\,
\gamma=1.59\pm 0.10\,\,\,\,\,,\,\,\,\,\,
\]
measured in the range $\eta\in [0.33,0.6]$.
Again, large errors are associated with the uncertainty of where exactly the
percolation threshold lies\footnote{This again
amounts to a problem with finite
size effects: unless $\langle l_{\rm loop} \rangle \ll \Lambda$, the
ensemble average will miss out on large contributions from loops wrongly
counted as infinite strings.}.

For bias values below the percolation threshold there is no
critical exponent. In this domain
the average loop length is a divergent function of the upper cutoff, and
an average length becomes ill--defined. This is obvious from
Eqs.~\ref{eq:lengthdistr} and \ref{eq:scaling} and the fact that
$\langle l_{\rm loop} \rangle\propto \int_{\lambda}^{\Lambda}l^{-b+2}\,dl$
while $D>3/2$.
This problem is alleviated in the non--percolating regime
(where the loop distribution is exponentially suppressed by an additional
factor of $e^{-cl}$),
as long as $\xi=1/c\ll \Lambda$, i.e.~for values of $\eta$ not too close
to the critical bias. The same argument holds for any higher moment
$\langle l_{\rm loop}^n\rangle$ of the loop distribution.

\begin{figure}[htb]
\centering
\mbox{~}{\hbox{
\epsfxsize=240pt
\epsffile{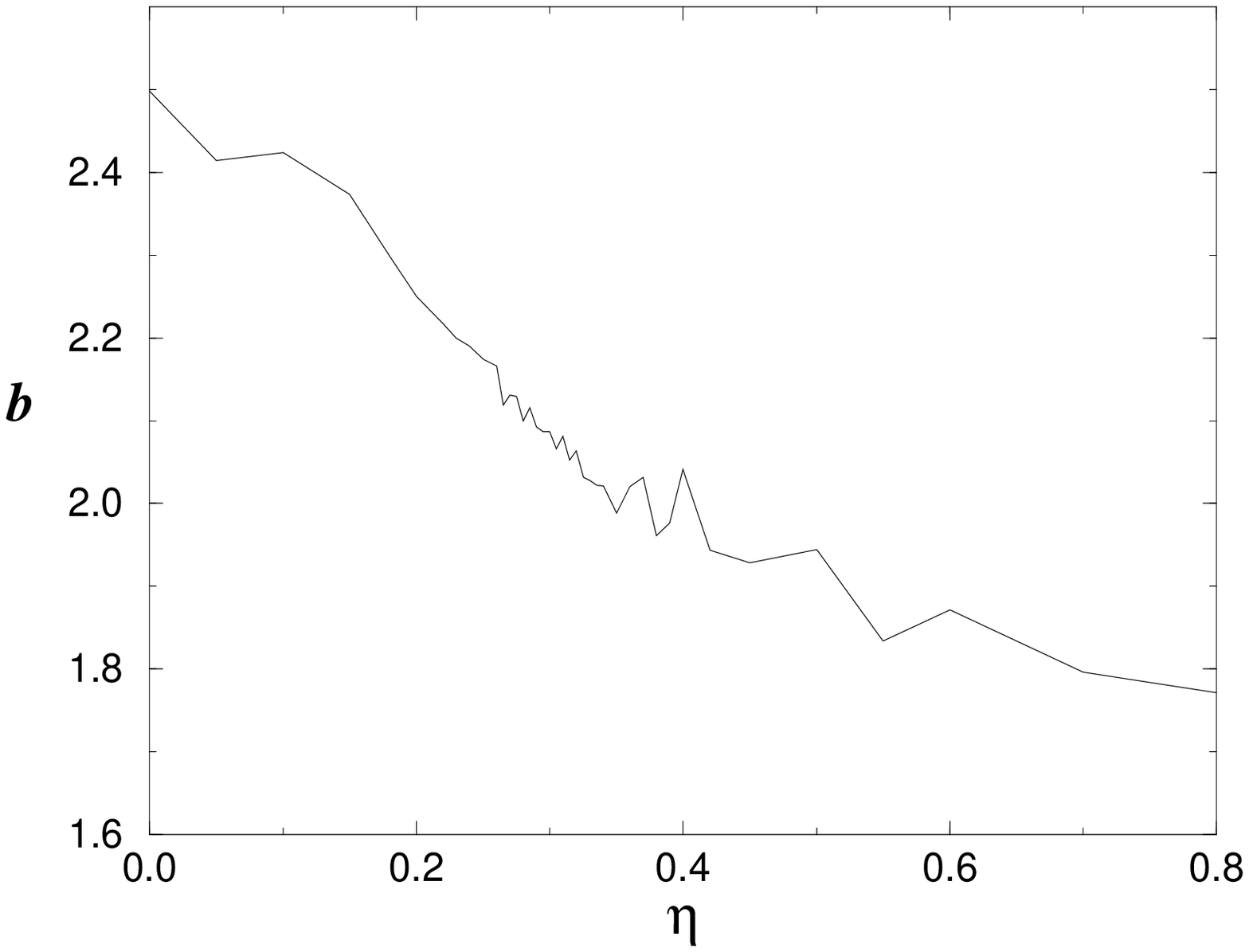}}}\\
\caption{The loop distribution exponent $b$ in Eq.~(\ref{eq:lengthdistr2}), for
an ensemble of 10,000 strings with cutoff $\Lambda=50,000$.}
\label{fig:b_U1003}
\end{figure}

\begin{figure}[htb]
\centering
\mbox{~}{\hbox{
\epsfxsize=240pt
\epsffile{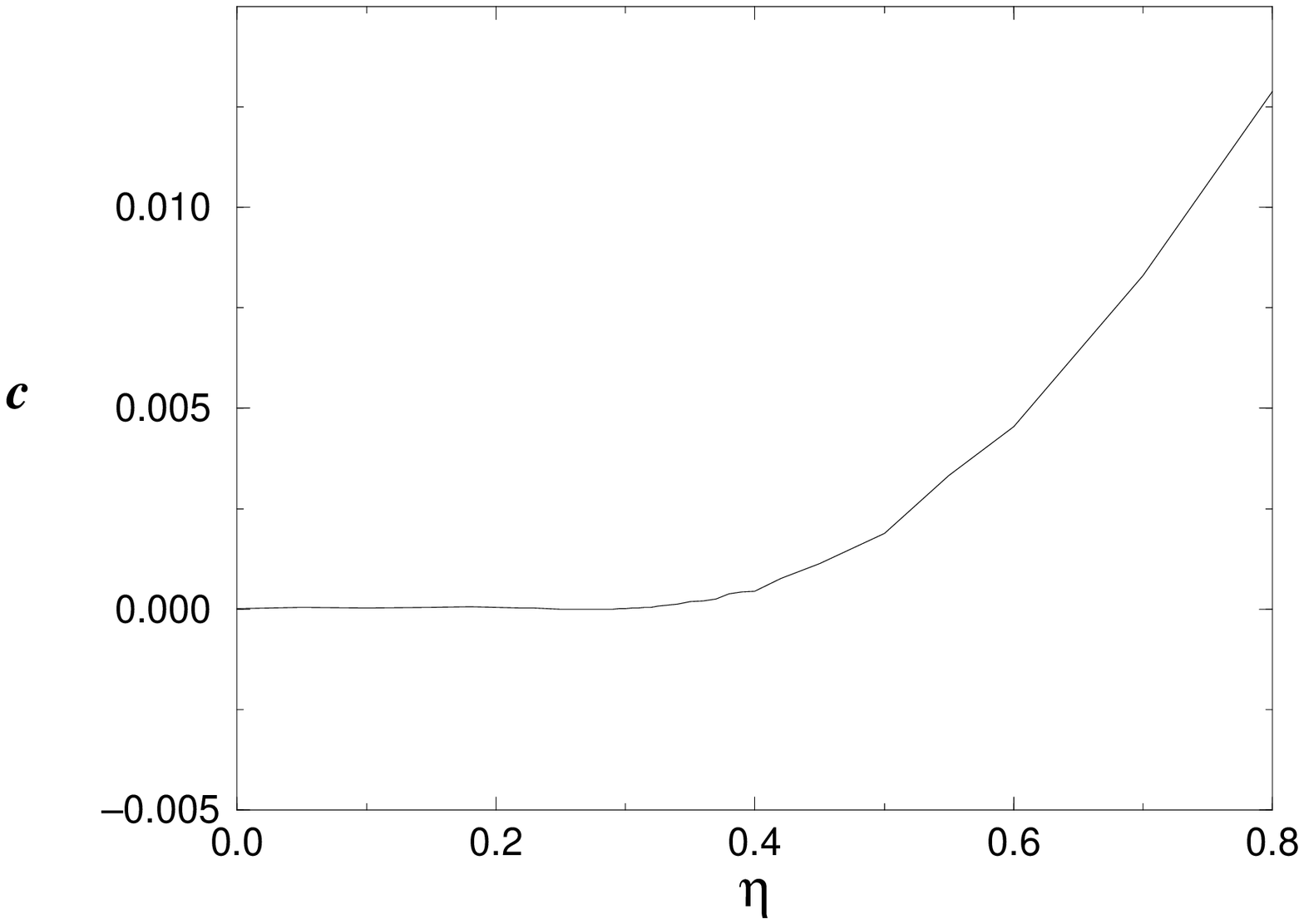}}}\\
\caption{The parameter $c$ in Eq.~(\ref{eq:lengthdistr2}) for the
same ensemble as in Fig.~\ref{fig:b_U1003}.}
\label{fig:c_U1003}
\end{figure}

\begin{figure}[htb]
\centering
\mbox{~}{\hbox{
\epsfxsize=240pt
\epsffile{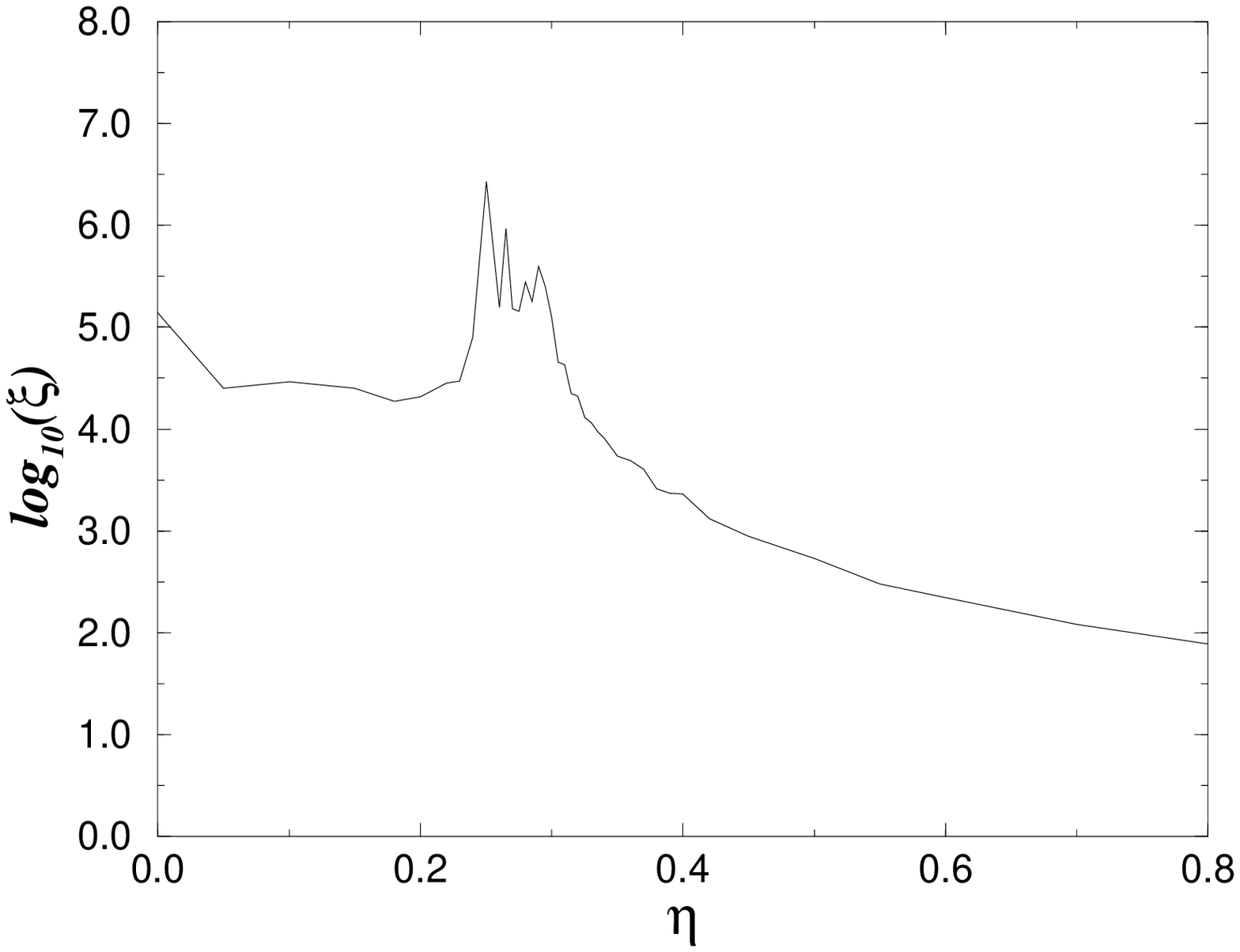}}}\\
\caption{The parameter $\xi=1/c$ in Eq.~(\ref{eq:lengthdistr2}) for the
same ensemble as in Fig.~\ref{fig:b_U1003}. Very large values in the region
$\eta<\eta^\star$ just correspond to statistical fluctuations of $c$
around zero.}
\label{fig:xi_U1003}
\end{figure}

Another way of investigating the ensemble is by means of a partition
function,
which is the (un--normalized) sum over probabilities
$p(l)$
\begin{equation}
Z=\sum_l l^{-b+1}e^{-cl}\,\,\,\,\,.\,\,\,\,\,
\label{eq:Z}
\end{equation}
This is not a thermal partition function, but should
rather be viewed as a generating function for the moments of the loop size
distribution.
If we wanted to use thermodynamics language, then the factor $l^{-b+1}$ would
be
proportional to the density of states for given $l$ (or one could say it
is a suitably defined integration measure, which amounts to the same),
and $\xi=1/c$ fulfils the role of a temperature, as is shown in
the Appendix. This only serves to
show that a thermodynamic nomenclature is inappropriate,
as the threshold for a dynamical Hagedorn transition really
lies at a finite temperature, and
instead of having the temperature diverge at the critical point, one would
have to factor out the divergent terms and pull them into the density of
states. This distinction becomes meaningless in our non--thermal ensemble.
Our ``partition function" can equally well be viewed as
a sum over the density of states only, with a critical temperature (or density)
dependence. Although these names are slightly inappropriate, they give
the right behaviour e.g.~for the average energy $\langle l_{\rm loop}\rangle$
in the non--percolating phase
\[
``E"=\langle l_{\rm loop}\rangle=-Z^{-1}\frac{d}{dc}{\rm
ln}\,Z\,\,\,\,\,.\,\,\,\,\,
\]
Since the parameters $b$ and $c$ are crucial in this interpretation,
measurements for them are shown
in Figs.~\ref{fig:b_U1003} and \ref{fig:c_U1003}.
Although the parameters $b$ and $c$ are comparably easy to measure,
since one does not have to guess $\eta^\star$, there are still large
statistical fluctuations for $b$. For $\eta>\eta^\star$, $b$ is allowed to
be smaller than two, since the scaling relation Eq.~(\ref{eq:scaling}) does
no longer hold. It becomes more difficult to measure $b$ at large bias, but it
is also less important there, since the exponential cutoff dominates the
loop size distribution.
The diverging length scale associated with the phase transition is
$\xi=1/c$, which is plotted in Fig.~\ref{fig:xi_U1003}.
There is another critical exponent associated with the divergence of $\xi$:
\[
\xi\propto(\eta-\eta^\star)^{-\frac{1}{\sigma}}\,\,\,\,\,.\,\,\,\,\,
\]
Implicitly, we measure for the percolation threshold $\eta^\star=
0.280\pm 0.003$.
The critical exponent is
\begin{equation}
\sigma=0.46\pm 0.02\,\,\,\,\,.\,\,\,\,\,
\label{eq:Sigma}
\end{equation}
Before concluding this section, we measure the critical
exponent in
\[
\langle l_{\rm loop}^2\rangle\propto
(\eta-\eta^\star)^{-\psi}\,\,\,\,\,,\,\,\,\,\,
\]
which gives (as well as $\eta^\star=0.277\pm0.004$)
\[
\psi=3.79\pm 0.14\,\,\,\,\,.\,\,\,\,\,
\]
This is the exponent associated with what we called ``susceptibility"
in ref.~\cite{MHKS}, $S=\langle l_{\rm loop}^2\rangle-
\langle l_{\rm loop}\rangle^2$, since the diverging
behaviour of $\langle l_{\rm loop}^2\rangle$ dominates in the expression
for $S$, because $\psi>2\gamma$.
In terms of the partition function, it is
\[
\langle l_{\rm loop}^2\rangle=Z^{-1}\frac{d^2}{dc^2}\,Z\,\,\,\,\,,\,\,\,\,\,
\]
such that $\langle l_{\rm loop}^2\rangle=-Z^{-1}\frac{d}{dc}(
Z\langle l_{\rm loop}\rangle)$. When compared to a Hagedorn transition,
however, we want the derivative of $\langle l_{\rm loop}\rangle$ with
respect to a temperature (i.e.~any smooth reparameterisation of the bias),
such that a proper definition of the susceptibility as the energy cost
per loop associated with decreasing $\eta$ would simply diverge
with an exponent $(-\gamma -1)$\footnote{From $\langle l_{\rm loop}\rangle
\propto(\eta-\eta^\star)^{-\gamma}$ and the ``energy cost" per loop
associated with decreasing the bias becomes $\frac{d}{d\eta}\langle
l_{\rm loop}\rangle$.}. Another reasonable definition of a susceptibility
would be to define it (at least for $\eta>\eta^\star$) as $\frac{d}{d\eta}
\rho_{\rm loop}$, which then diverges with the exponent $(-\beta-1)$.

This concludes the discussion of the critical behaviour of the ensemble
with a minimal discretisation of $U(1)$. Measurements for the other
ensembles with gradually finer discretisations of the vacuum manifold are
listed in Table~\ref{table:U1}~\footnote{Note that the measurements supercede
the ones in~\cite{MHKS}, where the string
ensembles were smaller and fewer compared to the ones available now.}.
A reminder of which exponent is associated
with which variable can be found in Table~\ref{table:reminder}.
\begin{table}[htb]
\centering
\begin{tabular}{|c|}
\hline
$\xi=1/c\propto(\eta-\eta^\star)^{-\frac{1}{\sigma}}$\\
$\langle l_{\rm loop}\rangle\propto(\eta-\eta^\star)^{-\gamma}$\\
$\langle l_{\rm loop}^2\rangle\propto(\eta-\eta^\star)^{-\psi}$\\
$\langle \rho_{\infty}\rangle\propto(\eta^\star-\eta)^{\beta}$\\
\hline
\end{tabular}
\caption{A summary of the definition of the critical exponents.}
\label{table:reminder}
\end{table}
\begin{table}[htb]
\centering
\begin{tabular}{|r|r|c|c|c|c|c|}
\hline
N & $\Lambda$ & $\eta^\star$ & $\sigma$ & $\gamma$ & $\psi$ & $\beta$ \\
\hline
3 & 50,000 & $0.279\pm 0.005$ &
$0.46\pm 0.02$ & $1.59\pm 0.10$ & $3.79\pm 0.14$ & $0.54\pm 0.10$ \\
7 & 50,000 & $0.299\pm 0.005$ &
$0.44\pm 0.02$ & $1.77\pm 0.14$ & $3.85\pm 0.16$ & $0.45\pm 0.07$ \\
15 & 50,000 & $0.295\pm 0.007$ &
$0.39\pm 0.04$ & $1.86\pm 0.13$ & $4.5\pm 0.3$ & $0.39\pm 0.06$ \\
31 & 50,000 & $0.305\pm 0.005$ &
$0.42\pm 0.02$ & $1.75\pm 0.03$ & $4.1\pm 0.3$ & $0.45\pm 0.06$ \\
63 & 50,000 & $0.309\pm 0.005$ &
$0.417\pm 0.017$ & $1.69\pm 0.07$ & $4.0\pm 0.2$ & $0.50\pm0.07$ \\
127 & 50,000 & $0.301\pm 0.004$ &
$0.413\pm 0.013$ & $1.80\pm 0.06$ & $4.07\pm 0.14$ & $0.41\pm 0.05$ \\
255 & 2,000 & & $0.339\pm 0.005$ & N/A & N/A & N/A \\
255 & $10^6$ & $0.300\pm 0.002$ &
N/A & $1.83\pm 0.07$ & $4.10\pm 0.15$ & $0.456\pm 0.025$ \\
\hline
\end{tabular}
\caption{The critical exponents and percolation thresholds for different
discretisations of $U(1)$. The values quoted for $\eta^\star$ are taken
from the measurement of $\gamma$. In the case
of the very low $\Lambda$, the divergences for loop sizes flatten out long
before the critical point, and no exponents can be extracted. For very large
$\Lambda$, however, the number of traced strings  is not large enough to
extract $c$. The errors quoted are statistical errors only.}
\label{table:U1}
\end{table}
Ideally, we should have two ensembles for every discretisation, one with
a small $\Lambda$, allowing us to trace a larger number of strings
to extract the exponent $c$, and one with very large $\Lambda$, to measure
the divergences in the average loop size, to which the loops of length
$l\stackrel{<}{\sim}\Lambda$ contribute the most, as we approach
the critical point. However, we do not notice too much of a reduction of
statistical errors by doing this (as is done for $N=255$), so
these ensembles have not been generated. A cutoff of $\Lambda=2,000$ is
clearly too low to give sensible answers\footnote{The very long string
ensemble agrees with the medium--cutoff ensembles, whereas the ensemble with
cutoff $\Lambda=2,000$ allows hardly any measurements, and the one exponent
that is measurable is obviously affected by severe systematic errors.
This also means that on a finite lattice of about $45^3$ lattice points --
which amounts to a similar cutoff -- measurements like the ones presented
here are virtually impossible. In this way we can understand that -- in lack
of an infinite--lattice formalism of the kind we are using for these
ensembles -- no--one has picked up on Vachaspati's model to investigate
the Hagedorn transition more closely.}.

Another consequence of finite cutoff effects is that
the measurements of $\gamma$ give the best fits to a power law and are
therefore
the best indicators of where the percolation threshold lies. The reason
for this is simple to understand: large loops give the main contribution
towards the length average, and finite cutoff effects will eventually
flatten the peak which should diverge at the critical bias.
However, the
contribution of large loops to $\langle l_{\rm loop}^2\rangle$ is even
more important, and finite cutoff effects will spoil the power law approach
to the critical point at a much earlier stage, such that $\psi$ has much
larger errors associated with it, and it is recommendable to use estimates
for $\eta^\star$ which were obtained from fits of the average loop size
to a power law behaviour.
The same is true for $c$ (and its associated exponent $\sigma$), because
of similar reasons, which, in an intuitive fashion one might simply sum up
as: the smaller the modulus of the
critical exponent, the closer one can go towards the
critical point without feeling finite size effects.

\begin{figure}[htb]
\centering
\mbox{~}{\hbox{
\epsfxsize=240pt
\epsffile{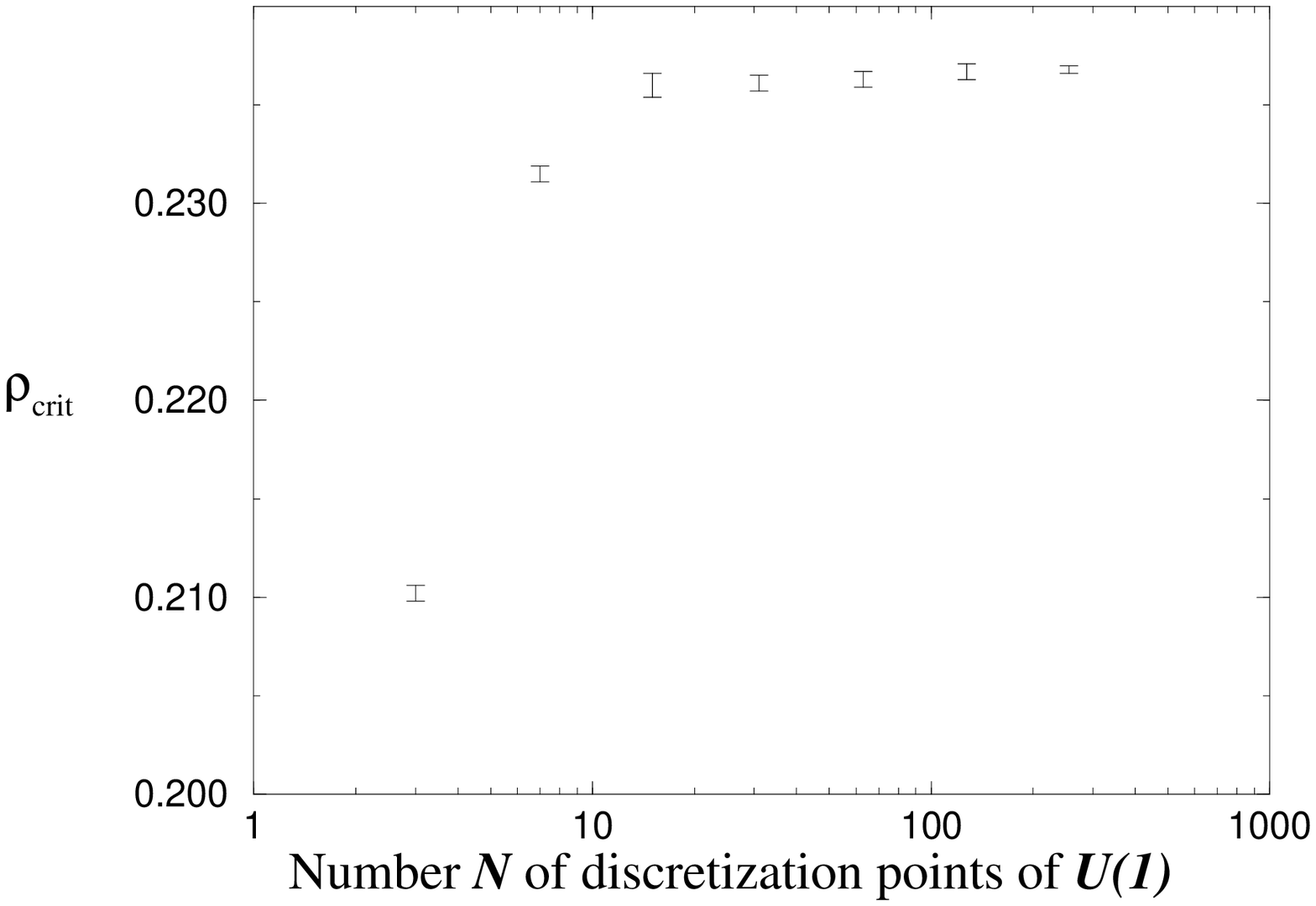}}}\\
\caption{The critical string density (in units of string segments per
tetrahedron) plotted against the number of discretisation points for the
$U(1)$ manifold.}
\label{fig:rho_crit_N}
\end{figure}

In Fig.~\ref{fig:rho_crit_N} we present the values of the critical string
density as a function of the discretisation of $U(1)$. It can be seen
that $N\stackrel{>}{\sim}15$ is already very close to the continuous $U(1)$
limit. In this sense, the ensembles with $N\ge 31$ in
Table~\ref{table:U1} can be interpreted as
the same measurements over several ensembles of the same kind.

\begin{figure}[htb]
\centering
\mbox{~}{\hbox{
\epsfxsize=240pt
\epsffile{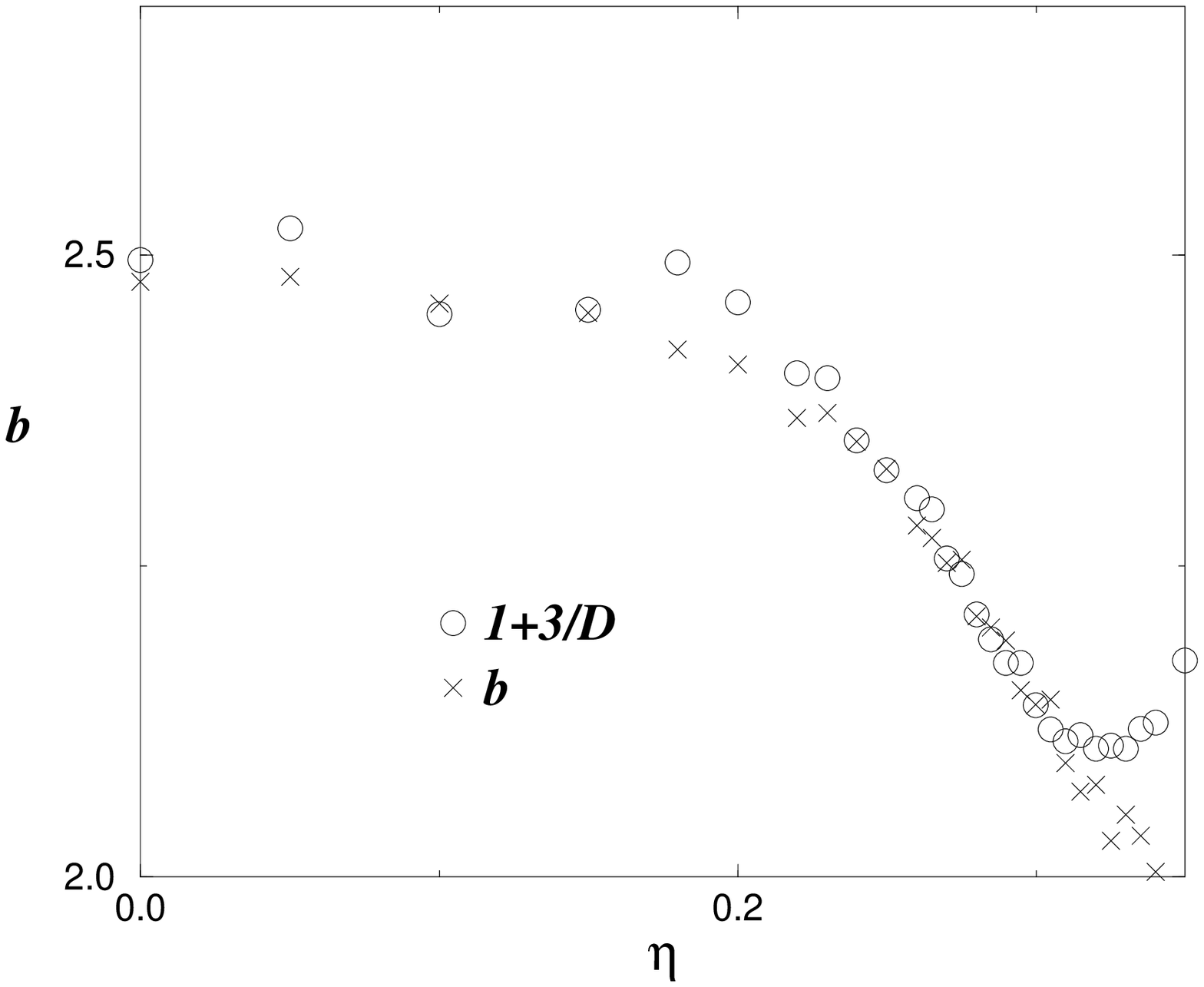}}}\\
\caption{Testing the validity of Eq.~(\ref{eq:scaling}) on a $U(1)$ string
ensemble with $N=63$.}
\label{fig:scaling}
\end{figure}

Lastly, it is worthwhile testing the validity of the scaling relation
for configurational exponents, Eq.~(\ref{eq:scaling}), which -- as we have
discussed before -- implies that on intermediate scales loops show
the same fractal behaviour as infinite strings.
Scaling seems to hold roughly. Considering that the exponent $b$ is usually
taken from quite noisy data (except when close to the percolation
threshold), the agreement seems very good, and is shown
in Fig.~\ref{fig:scaling} for the $N=63$ ensemble. When $c$ is constrained
to be zero, as it was done to obtain Fig.~\ref{fig:scaling},
such plots give a very
good first estimate of the percolation threshold and the Fisher exponent
$\tau$, i.e.~the value of $b$ at criticality, which plays a prominent role
in the next section,
when we will come to use the partition function to describe general
scaling relations.

\subsection{$\R P^2$ strings}

For $\R P^2$ strings our conventions of what the bias means depend
on the discretisation. For the minimally discretised $\R P^2$ strings
(cf.~Fig.~\ref{fig:RP2}), we define a bias $\zeta$ as:
\begin{equation}
p(0)=p(1)=p(2)=\frac{1+\zeta}{6}\,\,\,\,\,,\,\,\,\,\,
p(3)=p(4)=p(5)=\frac{1-\zeta}{6}\,\,\,\,\,,\,\,\,\,\,
\label{eq:zeta}
\end{equation}
such that $\zeta=1$ corresponds to $\rho_{\rm total}=0$, $\zeta=0$ is
a perfect $\R P^2$ symmetry, and $\zeta=-1$ a perfect $U(1)$\footnote{If
all the vacuum values live on the equator, we have the symmetry $U(1)/\Z_2$,
which, is homeomorphic to $U(1)$, and produces exactly
the same results as the minimally discretised $U(1)$ ensembles.}.
For the continuous $\R P^2$ simulations, we define a bias $\mu$ such that
the normalized probability density for the polar angle $\theta\in[0,\pi/2)$
(i.e.~constrained to the upper half--sphere) is
\begin{equation}
d\,p(\theta)=\frac{\mu}{1-e^{-\mu}}e^{-\mu\,\cos\theta}\sin\theta
d\theta\,\,\,\,\,.\,\,\,\,\,
\label{eq:mu}
\end{equation}
The unbiased case corresponds again to $\mu=0$, but
$\mu$ can run from $-\infty$ to $\infty$, where it corresponds to zero string
density or a $U(1)/\Z_2$ symmetry, respectively.

\begin{figure}[htb]
\centering
\mbox{~}{\hbox{
\epsfxsize=240pt
\epsffile{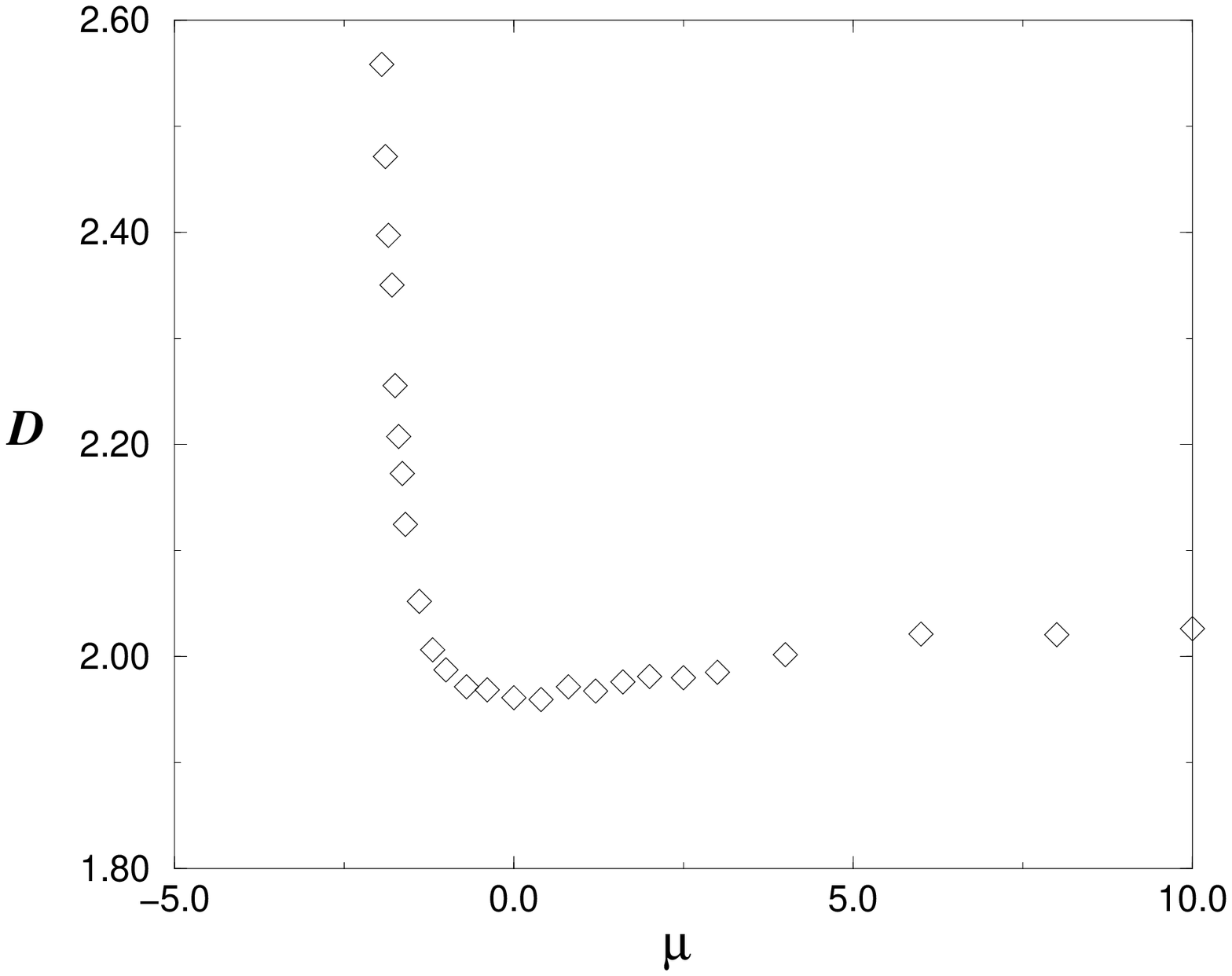}}}\\
\caption{The running of the fractal dimension. At zero bias we measure
$D=1.96$, whereas for large positive bias the $U(1)$ measurements are
recovered asymptotically. As we approach the zero--density limit (large
negative $\mu$), the fractal dimension increases sharply before becoming
ill--defined. The ensemble has 3,000 strings with $\Lambda=10^5$.}
\label{fig:Drunning}
\end{figure}

\begin{figure}[htb]
\centering
\mbox{~}{\hbox{
\epsfxsize=240pt
\epsffile{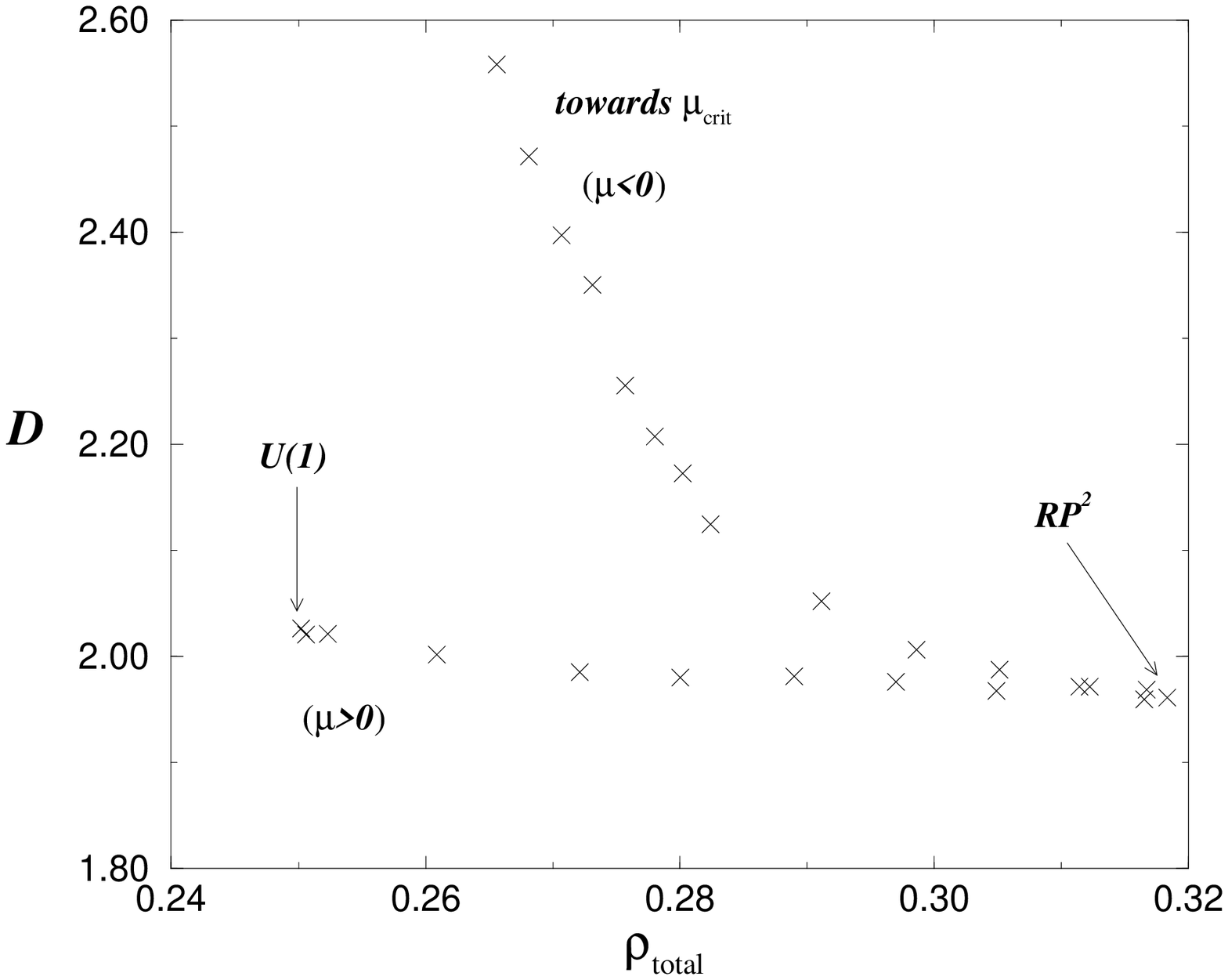}}}\\
\caption{The density of string segments in the continuous $\R P^2$
ensembles as a function of the total string density. The two ends of the
function differ distinctly for positive and negative values of the
bias.}
\label{fig:Dvsdensity}
\end{figure}

Apart from the critical exponents, the interesting feature about biased
$\R P^2$ symmetries is that one can bias them towards the $U(1)$ ensembles
which
we have measured already.
The fractal dimension then runs from what we have measured for $U(1)$ to
values lower than two. This is shown in Fig.~\ref{fig:Drunning}.
If this effect is caused by the running of the
string density, as comparisons of the relative densities seem to suggest
(the average number of penetrated triangles in continuous $U(1)$ is
$\frac{1}{4}$, whereas for continuous $\R P^2$ it is
$\frac{1}{\pi}$~\cite{V91}),
our case for arguing that the string statistics quite generally depend on
the string density in a way opposed to the tendency polymers show, gets
further support from this picture. Vachaspati's model itself would be the
obvious testing ground for such a hypothesis (the penetration probability
per plaquette is $\frac{1}{2}$, the highest known in any model so far),
but unfortunately he quotes neither measurements for $D$ nor for $b$ at
zero bias, and his plots have too large statistical
fluctuations to extrapolate towards zero bias. A more detailed analysis
of this is given later, in Fig.~\ref{fig:Dvsdensity}.

\begin{figure}[htb]
\centering
\mbox{~}{\hbox{
\epsfxsize=240pt
\epsffile{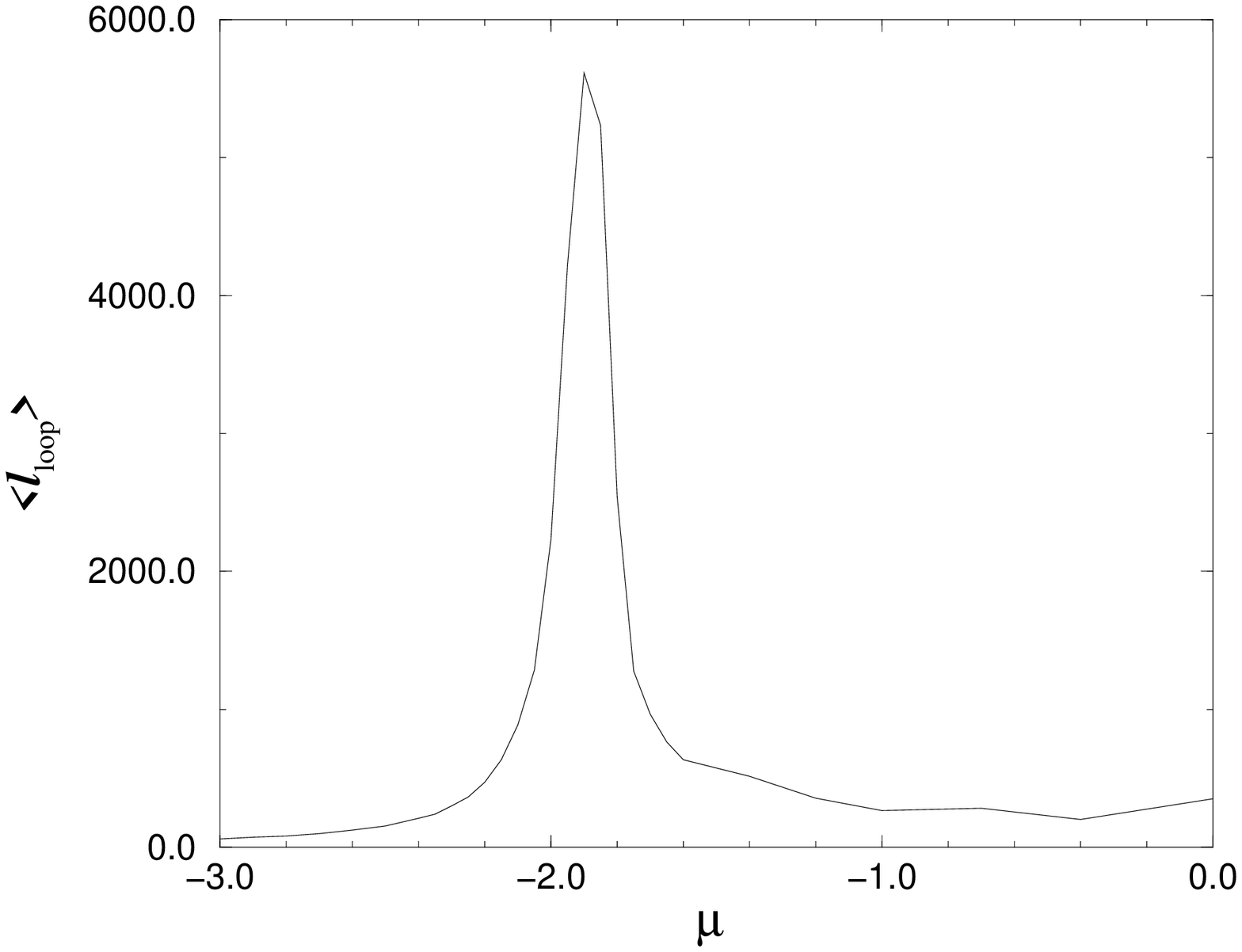}}}\\
\caption{The loop size distribution for negative $\mu$ only. It allows
a very good estimate of the position of $\mu^\star$.}
\label{fig:nlcloop}
\end{figure}

\begin{figure}[htb]
\centering
\mbox{~}{\hbox{
\epsfxsize=240pt
\epsffile{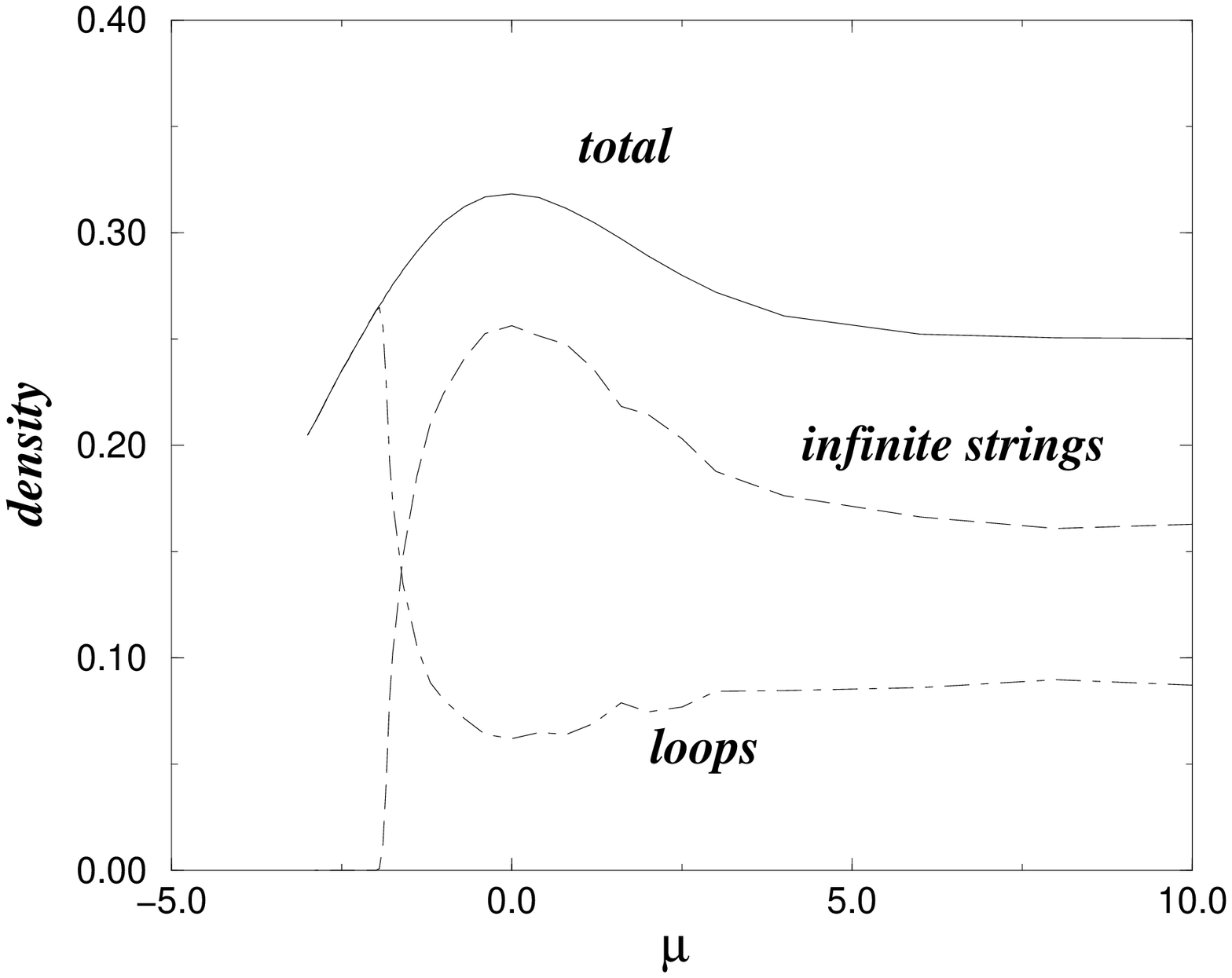}}}\\
\caption{The density of string segments in the continuous $\R P^2$
ensembles as a function of bias. The density in loops is shown as the
dot--dashed line, and the density in infinite strings is the dashed line.}
\label{fig:nlctotaldensity}
\end{figure}

Let us now turn to the measurements of the critical exponents.
The average loop size has a very distinct peak at the critical point,
immediately allowing a rough estimate of the critical bias at $\mu^\star
=-1.875\pm0.025$. This is displayed in Fig.~\ref{fig:nlcloop}.
When we measure $\gamma$, we get the best fits for
$\mu^\star=-1.86\pm0.01$, and
\[
\gamma=1.70\pm0.06\,\,\,\,\,,\,\,\,\,\,
\]
seemingly in very good agreement with the $U(1)$ data. Also
\[
\psi=3.83\pm0.14\,\,\,\,\,,\,\,\,\,\,\sigma=0.435\pm0.013\,\,\,\,\,.\,\,\,\,\,
\]
To obtain $\langle \rho_{\infty}\rangle$, we first calculate
$\rho_{\rm total}$ with a Monte Carlo integration
\[
\rho_{\rm total}=\int\int\int d\Omega_1\,d\Omega_2\,d\Omega_3\,n\,p(\Omega_1)\,
p(\Omega_2)\,p(\Omega_3)\,\,\,\,\,,\,\,\,\,\,
\]
where $\int\,d\Omega$ stands for
$\int_0^{2\pi}\,d\phi\,\int_0^{\pi/2}\,d\theta$,
$p$ is defined by Eq.~(\ref{eq:mu}),
and $n$ is the string flux as a function of the three pairwise dot products,
as defined in the previous section. Then we take again $\rho_{\infty}=
\rho_{\rm total}\,\frac{N_{\infty}}{N_{\rm total}}$, where $N$ with a
subscript stands simply for the total number of appropriate strings in
the ensemble.
The results of this integration are plotted in Fig.~\ref{fig:nlctotaldensity},
together with the mass density in loops and infinite strings.
It can be seen that the loop density is considerably lower than for
$U(1)$, climbing toward the value measured for $U(1)$ strings as
$\mu\rightarrow\infty$, and making up the total string density beyond the
percolation threshold. There are some statistical
errors attached to the lines indicating the loop and infinite string
density, as the ensemble of Fig.~\ref{fig:nlctotaldensity} consists
of only 3000 strings ($\Lambda$=100,000).

{}From the infinite string density we measure
\[
\beta=0.40\pm 0.03\,\,\,\,\,.\,\,\,\,\,
\]
A comprehensive table of measurements can be found
in Table~\ref{table:percolation}.

Let us now come back to a
question raised earlier: {\it is the string density the
fundamental parameter, dictating $D$ and $b$?}
Fig.~\ref{fig:Dvsdensity} shows that this cannot be the case.
There we plot the density vs.~the fractal dimension. Most
of the density regime is reached twice, once for negative $\mu$ and
once for positive $\mu$. The fractal dimension, as we go towards
the critical point, diverges already for string densities {\it higher} than
the densities we can achieve for positive $\mu$, where the string ensemble
just turns into a $U(1)$--set of Monte Carlo data.
The fundamental lesson to learn from this is that, although spoiling the
symmetry may be the only way to change the string density in our lattice
description, it can -- sometimes -- be done in different ways, and lead
to entirely different results.

\begin{figure}[htb]
\centering
\mbox{~}{\hbox{
\epsfxsize=240pt
\epsffile{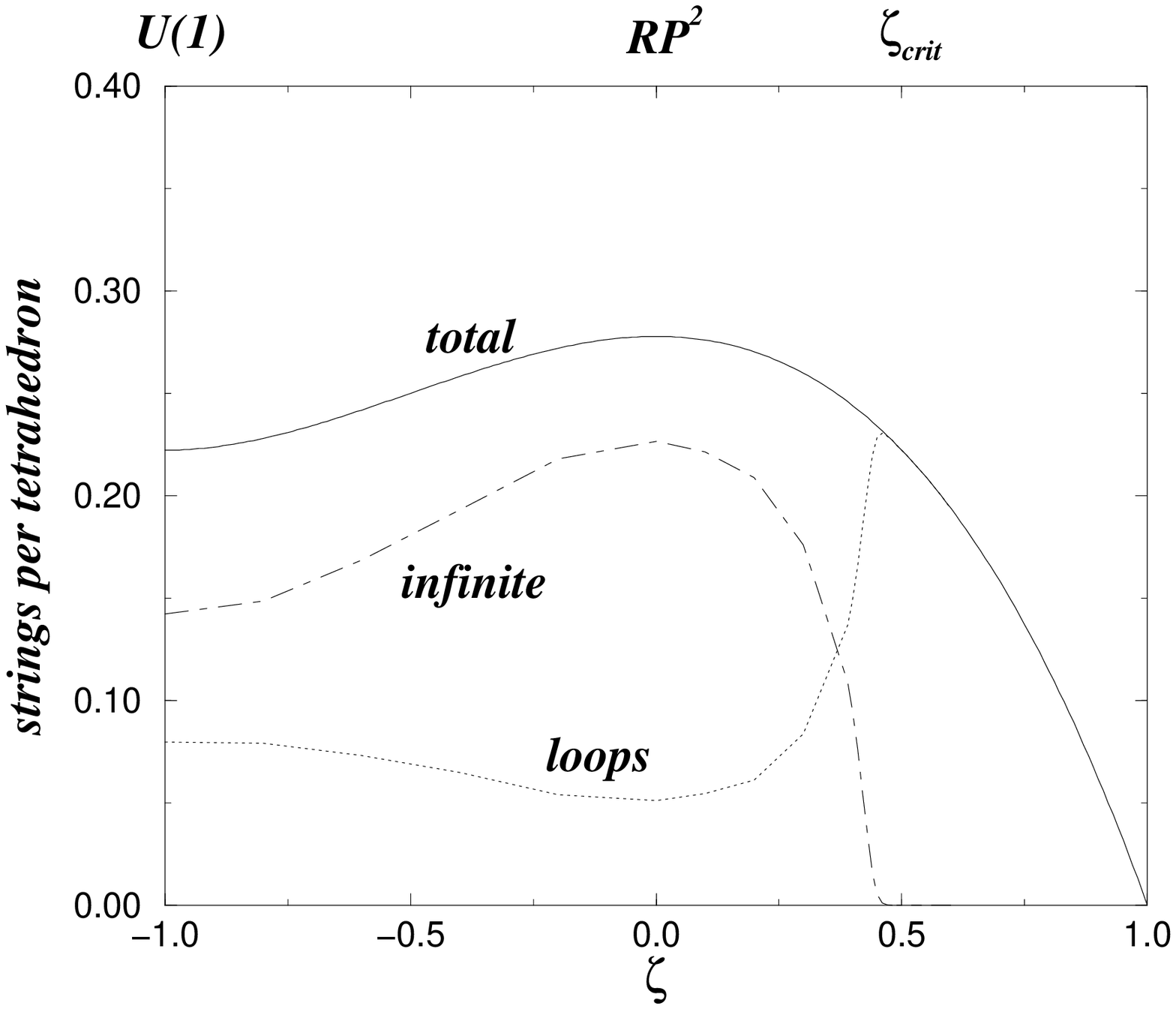}}}\\
\caption{The string densities of the discretised $\R P^2$ ensembles. The
contributions from loops and infinite strings are shown separately.}
\label{fig:nlc3density}
\end{figure}

For the minimal discretisation of $\R P^2$ it is straightforward but lengthy
to derive
\[
\rho_{\rm total} =
\frac{1}{18}\left[5-3\zeta^2-2\zeta^3\right]\,\,\,\,\,.\,\,\,\,\,
\]
It can be seen that it has the right behaviour $\rho_{\rm total}
\rightarrow 0$, if $\zeta\rightarrow +1$ and $\rho_{\rm total}
\rightarrow \frac{2}{9}$, which is the value for the minimal discretisation
of $U(1)$, as $\zeta\rightarrow -1$. It also has an extremum at
perfect symmetry $\zeta=0$. This means it shows all the qualitative features
of the bias dependence we observed in Fig.~\ref{fig:nlctotaldensity} for the
continuous $\R P^2$ strings~\footnote{The success of discretisations
of $\R P^2$ in reproducing qualitatively the behaviour of the continuous
manifold is quite striking: We have tested another icosahedral discretisation,
obtained by viewing a vertex of the icosahedron head--on (such that all the
other five points are equatorial ones). Again, this reproduces the
same qualitative curve for the bias--dependence of the density, and
turns exactly into a representation of the
$(N=5)$--discretisation of $U(1)$ when biased towards
equatorial vacuum states (with the same string density).}.
The density, together with the density in loops and infinite strings,
is shown in Fig.~\ref{fig:nlc3density}. All components have qualitatively
the same behaviour as in the continuous $\R P^2$ ensembles.
{}From this we can conclude that, again, the discretisation
of the vacuum manifold does not spoil the physical picture qualitatively.

On sets of 10,000 strings with $\Lambda=$25,000,
we measure that the percolation threshold is
at $\zeta^\star=0.425\pm 0.004$, $\sigma=0.40\pm 0.03$,
$\beta=0.34\pm 0.04$, $\gamma=1.68\pm 0.13$, and $\psi=3.75\pm 0.25$.

\begin{figure}[htb]
\centering
\mbox{~}{\hbox{
\epsfxsize=240pt
\epsffile{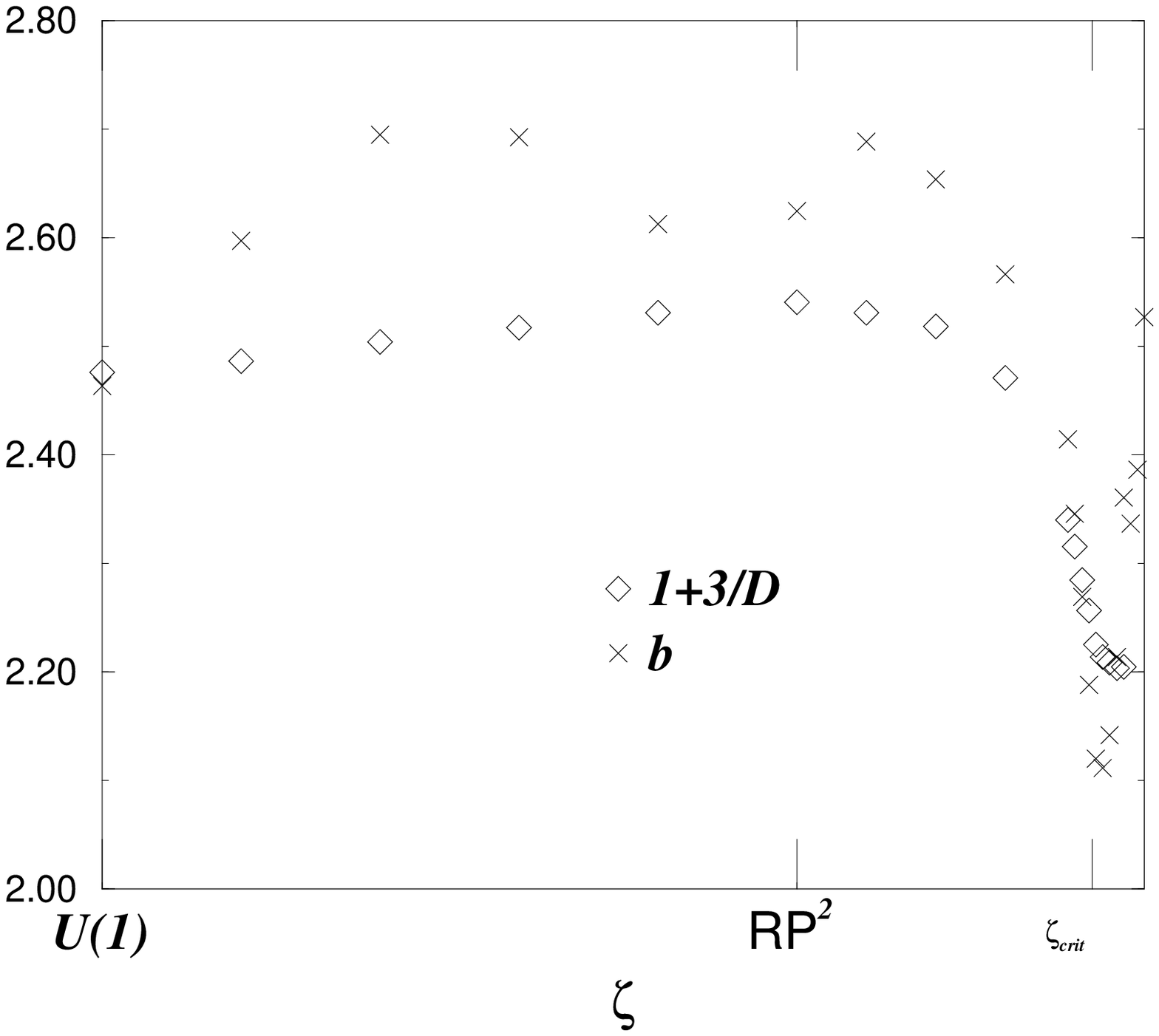}}}\\
\caption{The scaling relation Eq.~(\ref{eq:scaling}) seems to be consistently
violated, even in the low bias regime, except at the critical point and,
perhaps, at a perfect $U(1)$ symmetry. These are the ensembles with the
discretised $\R P^2$ manifold. The loop statistics are not extremely good,
as the number of loops for zero bias (i.e.~where
it is the lowest) is $\approx 1800$.}
\label{fig:RP2scaling}
\end{figure}

\begin{figure}[htb]
\centering
\mbox{~}{\hbox{
\epsfxsize=240pt
\epsffile{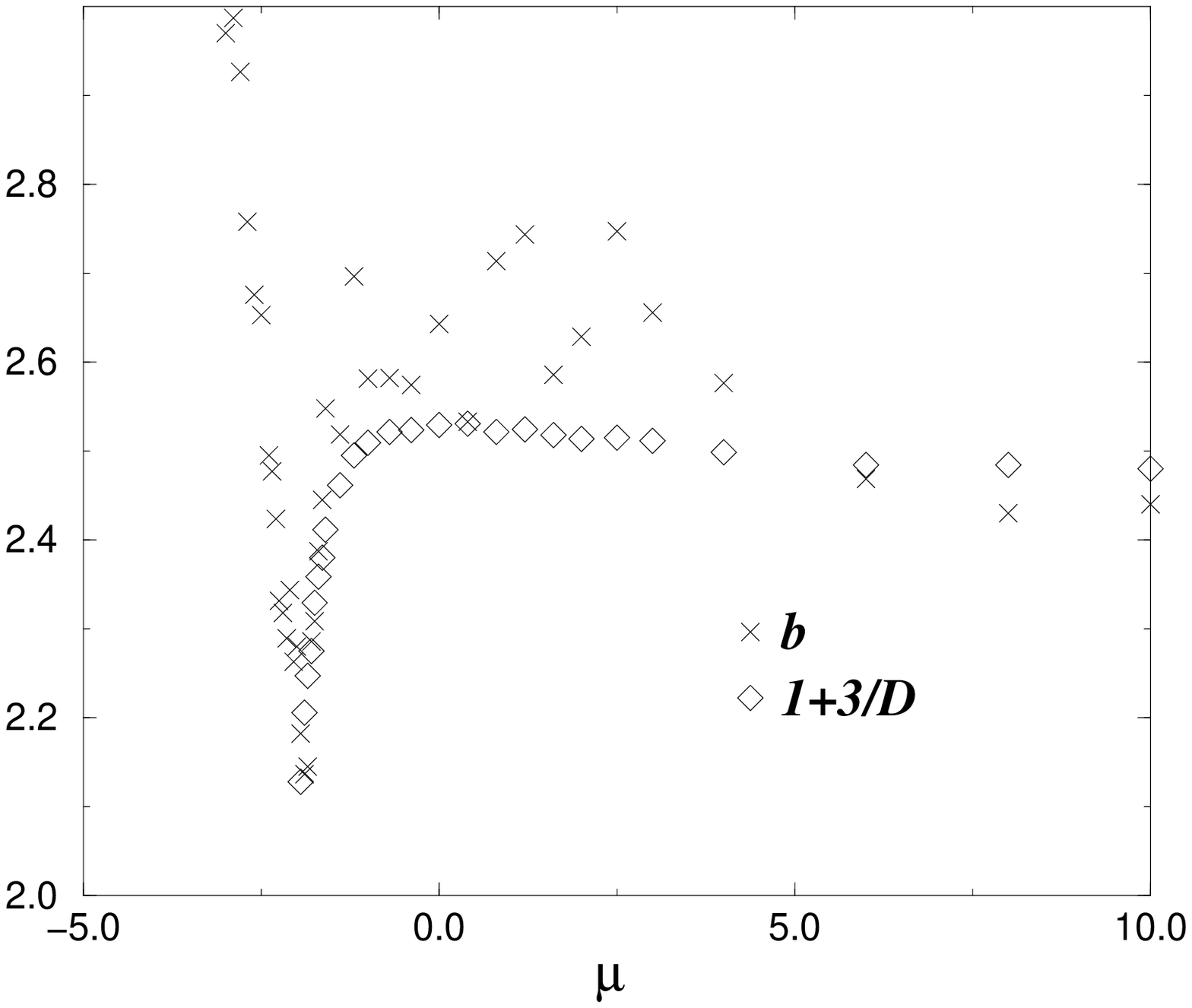}}}\\
\caption{The same plot as in Fig.~\ref{fig:RP2scaling} for the continuous
$\R P^2$ manifold. There are also large statistical errors in $b$,
as the number of loops is extremely small ($\approx 500$) at low bias.}
\label{fig:RP2contscaling}
\end{figure}

What is somewhat unexpected, is that for the $\R P^2$ ensemble, violation
of the `strict' scaling relation Eq.~(\ref{eq:scaling}) seem to be apparent
even in the regime of low bias. Where the
bias turns the symmetry into $U(1)$,
the strict scaling relation Eq.~(\ref{eq:scaling}) seems to hold. For
the other values, although statistical fluctuations are recognisable,
there seems to be a consistent trend towards too large values for $b$,
which disappears again at the critical point (cf.~the same discussion for
biased $U(1)$ strings). Perhaps, the deviations seen for $U(1)$
in Fig.~\ref{fig:scaling} are also not just statistical
errors. An analogous plot for minimally discretised $\R P^2$ strings
is shown in Fig.~\ref{fig:RP2scaling} for the discretised manifold and
in Fig.~\ref{fig:RP2contscaling} for the continuous one.
The obvious source to suspect would be a cutoff--dependence for $b$.
However, this is not the case. We have changed upper and lower cutoffs for
the loop distribution before fitting it to Eq.~(\ref{eq:lengthdistr}), and
the results do not change qualitatively: $b$ stays recognisably
larger than predicted by Eq.~(\ref{eq:scaling}). Clearly, a better analysis
of this violation
of scale-invariance, even in the percolating regime, is needed.
Here, however, we will focus on the non-percolating regime from now on.

\section{Percolation Theory, Universality, and Polymers}
\label{section:polymers}

\subsection{Percolation Theory and Critical Phenomena}

The most notable fact about the critical exponents we have measured in
the previous section is that they do not only agree with each other,
thus indicating universal behaviour for string defects at the critical
density, but also show reasonable agreement
with the exponents obtained by standard site or bond
percolation. This fact has already been noticed for the $N=3$ discretisation
of $U(1)$ strings~\cite{Bradley}.
We give a comparison of the measured exponents, including their
statistical errors, with known results from percolation theory in
Table~\ref{table:percolation}.
\begin{table}[htb]
\centering
\begin{tabular}{|c|c|c|c|c|c|}
\hline
exponent & percolation & minimally & continuous & discrete & continuous \\
& theory & discretised $U(1)$ & $U(1)$ & $\R P^2$ & $\R P^2$ \\
\hline
$\sigma$ & 0.45 & 0.46(2) & 0.42(2) & 0.435(13) & 0.40(3) \\
$\beta$ & 0.41 & 0.54(10) & 0.45(3) & 0.41(3) & 0.34(4) \\
$\gamma$ & 1.80 & 1.59(10) & 1.77(6) & 1.70(6) & 1.68(13) \\
$\psi$ & 4.04 & 3.79(14) & 4.10(15) & 3.83(14) & 3.75(25) \\
\hline
\end{tabular}
\caption{A summary of all the measured critical exponents,
compared with critical exponents in three
dimensional site (or bond) percolation.}
\label{table:percolation}
\end{table}
It is seen that the continuous symmetries show some reasonable correspondence
with percolation data, whereas the minimally discretised $U(1)$
ensemble deviates more strongly.
The $\R P^2$ data has critical exponents which are consistently
slightly lower than the percolation theory exponents.
This allows some fairly wide--ranging conclusions. For instance, we can now
-- without having made the appropriate measurements --
expect Vachaspati's model not only to be ``nearly" a bond percolation problem
as far as the actual percolation threshold is concerned, but it may also
have critical exponents very close to the ones of bond
percolation

This is indeed somewhat surprising: from the point of view of
percolation theory,
we have constructed a very awkward definition of ``clusters". Not
only are they topologically defined, but also definable through
continuous, as well as discrete, manifolds of possible lattice states.

There is another comment to be made: From Table \ref{table:percolation},
one could hypothesize that there may be a trend for higher--dimensional
manifolds to deviate more strongly from percolation data. However,
Vachaspati's previously discussed model describes strings produced from
a $S^{\infty}/\Z_2$ ground state manifold. We have shown above that this
model corresponds closely to site percolation, such that one
expects especially the Vachaspati model to be close to a percolation problem.
One would therefore not expect the differences between the $\R P^2$ data and
standard percolation exponents to be a generic consequence of the higher
dimensionality of the manifold.

However, our measurements certainly do not seem to compellingly suggest exact
agreement of our critical exponents with bond (or site) percolation in
three dimensions. Nevertheless, the relative smallness of the
deviations should make no difference when ones tries to understand the
possible physical relevance of the string statistics near the critical
point. Let us therefore try and learn some more lessons from percolation
theory.

\subsection{Fisher Exponent and Partition Function}

The first step towards some indications that there are some universal
relations between the configurational exponents (and that, therefore,
we may extract more exponents than we have actually measured), has been
done already by defining a partition function
in Eq.~(\ref{eq:Z}), which we can view as a
generating function for the loop size distribution, such that
\begin{equation}
\langle l_{\rm
loop}^n\rangle=Z^{-1}\left(-\frac{d}{d\,c}\right)^nZ\,\,\,\,\,.\,\,\,\,\,
\label{eq:lloop}
\end{equation}
This partition function
has the same functional form as the one derived in \cite{Cop+}
in the dilute free--string approximation. If we want to get back
to a physical interpretation of the partition function, it is perhaps
natural to expect some of the ingredients of the results of
ref.~\cite{Cop+} to enter the description.
In this way, one is led to interpret $c$ as an effective
string tension $\sigma_{\rm eff}$, divided by a temperature, and the
Hagedorn transition occurs when the effective string tension (or the
effective energy per unit length of string) is zero. Another paper
by the same authors~\cite{Cop+2} then provides the understanding for
the mechanism by which the Hagedorn transition occurs in the case of
biased symmetries: we have seen from the microscopic details of the
lattice description that, for biased symmetries, the strings tend to
become more crumpled as they wind around the sparsely spread lattice points
with disfavoured vacuum values. In~\cite{Cop+2} it has been shown that
for the reverse situation, when one tries to incorporate stiffness,
the effective string tension for fixed temperature behaves as
\[
\sigma_{\rm eff}=\sigma-\frac{{\rm ln}\left[(1+z)\,e^{-\beta\,u}\right]}
{l_o\,\beta}\,\,\,\,\,,\,\,\,\,\,
\]
where $u$ is the energy cost of turning a string through a right angle,
and $z$ is the number of directions available to the string at every
lattice point.
{}From the details of the lattice description (and also from an intuitive
point of view in the continuum case), this energy cost becomes negative
for strongly
biased symmetries\footnote{One only needs to consider the limit in which
the disfavoured vacuum values occupy only very small regions of space to
understand the trend of negative energy cost associated with crumpling
the string.}, so that for some value of the bias $\sigma_{\rm eff}=0$
will be achieved, and a Hagedorn transition occurs. However, this
correspondence is qualitative at best, since we do not deal with string
dynamics\footnote{For this reason, the exponent $b$ in the partition
function for example also differs from the one obtained in~\cite{Cop+}.},
and the actual string tension right at the phase transition is zero
by definition (at least in the case of a second--order phase transition).
A more stringent proof that such a thermal interpretation of the partition
function yields nonsensical results is developed in the Appendix.
Since ``our" partition function describes the initial conditions for
dynamical calculations,
one should not view it as being more than
the generating function of the moments of the loop length distribution.

If we know how $b$ and $c$ scale at the critical point,
we can derive the critical exponents for all the moments of the loop
length distribution. From the previous section it is clear that $b$
depends only weakly on the bias, whereas $c$ scales as $c\propto
|\eta-\eta^\star|^{-\gamma}$. In agreement with the percolation theory
literature, we define $\tau=b(\eta^\star)$ and will call it the Fisher
exponent.
The critical exponents for the moments of the loop distribution are then
derived by
\[
\langle l_{\rm loop}^n\rangle = Z^{-1}\left(-\frac{d}{d\,c}\right)^n\,
\left[\sum_l l^{-\tau+1} e^{-cl}\right]\propto\sum_l l^{-\tau+1+n} e^{-cl}
\]
Approximating the sum by an integral, we have
\begin{equation}
\langle l_{\rm loop}^n\rangle \propto \int l^{-\tau+1+n} e^{-cl}dl\propto
c^{\tau-2-n}\int z^{2-\tau}e^{-z}dz
\propto |\eta-\eta^\star|^{\frac{\tau-2-n}{\sigma}}\,\,\,\,\,,\,\,\,\,\,
\label{eq:loopmoments}
\end{equation}
where we have approximated $b$ by the constant $\tau$ in the first step,
and used $c\propto|\eta-\eta^\star|^{1/\sigma}$ in the last step.
The integral over $z$ is just a numeric constant.
Thus we find that the following scaling relations should hold for our
string ensembles
\begin{equation}
\gamma=\frac{3-\tau}{\sigma}\,\,\,\,\,,
\,\,\,\,\,\psi=\frac{4-\tau}{\sigma}\,\,\,\,\,,\,\,\,\,\,
\label{eq:gamma_scaling}
\end{equation}
and so on for higher moments of the loop distribution. In this way,
one can use two of the critical exponents as the fundamental ones,
and derive the others from them\footnote{Note that, after the measurements
of the previous section, the partition function
is no longer just a {\it model} of how the moments of the loop distribution
should scale: it is a strict consequence of the fact that
Eq.~(\ref{eq:lengthdistr2}) is satisfied so well by the data.}.
The fact only {\it two} critical exponents can be independent is also
a consequence of the existence of just {\it one} crossover length scale.
This is discussed in ref.~\cite{scaling}.
The trouble with our definition of a generating function is that it yields
the loop length distribution only for $\eta>\eta^\star$.
What one is interested in for low bias is the density in infinite strings.
The critical exponents thereof cannot be extracted as straightforwardly
as the other exponents of the loop distribution, seeing
that $\langle l_{\rm loop}^0 \rangle$ would formally correspond to 1.
A prescription which is generally workable in percolation theory~\footnote{This
can also be made more precise by introducing a general scaling function which
is meant to describe both the long and short range limits exactly. We chose
not to introduce it here, since we do not have enough data to extract such
a scaling function.}, is to argue that, since
\[
\langle l_{\rm loop}\rangle\propto
\frac{d}{d\,c}\langle l_{\rm loop}^0 \rangle\,\,\,\,\,,\,\,\,\,\,
\]
we can write
\[
\rho_{\infty}={\rm const_1}+{\rm const_2}\,\langle l_{\rm loop}\rangle\,c
\,\,\,\,\,,\,\,\,\,\,
\]
and therefore
\[
\rho_{\infty}\propto|\eta-\eta^\star|^{\frac{\tau-3}{\sigma}+
\frac{1}{\sigma}}\,\,\,\,\,,\,\,\,\,\,
\]
such that we have
\begin{equation}
\beta=\frac{\tau-2}{\sigma}\,\,\,\,\,.\,\,\,\,\,
\label{eq:beta_scaling}
\end{equation}
One may wonder why this gives the behaviour for infinite strings rather
than the total mass density or the loop mass density. To answer this, one
can just remind oneself that we have extended arguments which hold
only for loops at
$\eta>\eta^\star$. In this regime, gradually longer and longer
loops dominate the loop size distribution, such that, if we extend the
arguments beyond the percolation threshold, it is only the infinite strings
which contribute to the moments of of this distribution~\footnote{This
can easily be seen by the fact that all moments $\langle l\rangle$ and
higher, if defined through the partition function, diverge in the whole
percolating regime.}.

The scaling relations (\ref{eq:gamma_scaling}) and (\ref{eq:beta_scaling})
seem to be very well obeyed by the critical exponents we measure, and
seem to give a value of $\tau$ consistent with what we observe in
Fig.~\ref{fig:scaling}, and consistent with each other. The different
predictions of $\tau$, to make the scaling relations consistent with each
other, are listed in Table~\ref{table:taupredictions}.
\begin{table}[htb]
\centering
\begin{tabular}{|c|c|c|c|c|}
\hline
& minimally & continuous & discrete & continuous \\
& discretised $U(1)$ & $U(1)$ & $\R P^2$ & $\R P^2$ \\
\hline
$\tau=\beta\sigma+2$ & 2.25(6) & 2.19(2) & 2.18(2) & 2.14(3) \\
$\tau=3-\gamma\sigma$ & 2.27(8) & 2.26(6) & 2.26(5) & 2.33(11) \\
$\tau=4-\psi\sigma$ & 2.26(15) & 2.28(15) & 2.33(11) & 2.50(22) \\
\hline
\end{tabular}
\caption{Testing the scaling relations Eqs.~(\ref{eq:gamma_scaling}) and
(\ref{eq:beta_scaling}) by comparing
the values of the Fisher exponent predicted
by them.}
\label{table:taupredictions}
\end{table}

\subsection{The Average Loop Size}

Encouraged by this, we can go on to another cosmologically perhaps more
relevant parameter, which is not directly measurable from our ensembles, since
we did not record string--string correlation functions. It is another
correlation length, call it $\chi$, which measures the average separation
between two points on the same string. This gives a clearer picture of the
actual physical size of a collapsed string lump beyond the percolation
transition. Define
\[
\chi^2=\frac{\sum_r r^2 g(r)}{\sum_r g(r)}\,\,\,\,\,,\,\,\,\,\,
\]
where $g(r)$ is the probability of finding, at a distance $r$ from the
origin, a segment of string belonging to the same string as the segment at
the origin.
If we call the average squared distance between two string segments
$R_l^2$, we have to weigh it with the number of string segments in loops
of given length to obtain the same definition for $\chi$ in terms of
$R_l^2$
\[
\chi^2=\frac{2\sum R_l^2\,n_l\,l^{2-\tau}e^{-cl}}{\sum n_l\,
l^{2-\tau}e^{-cl}}\,\,\,\,\,.\,\,\,\,\,
\]
Thus, apart from numerical prefactors, $\chi$ is the radius of those loops
which give the main contribution to $\langle l\rangle$. Since we have
already established that, at least close to the percolation threshold,
Eq.~(\ref{eq:scaling}) is satisfied and loops have part in a proper definition
of the fractal dimension, we can substitute $R_l\propto s^{1/D}$, and
the numerator scales, according to Eq.~(\ref{eq:loopmoments}), with a
critical exponent $(3-\tau+2/D)/\sigma$. The denominator is just
$\langle l \rangle$, such that, for the longest loops (i.e.~the ones
which exhibit fractal behaviour on intermediate scales)
\[
\chi\propto|\eta-\eta^\star|^{-\nu}\,\,\,\,\,,\,\,\,\,\,
\]
with
\[
\nu=\frac{1}{D\sigma}\,\,\,\,\,,\,\,\,\,\,
\]
where $D$ is now the fractal dimension at criticality $D\approx2.5$
We have thus a reasonably good understanding of how the statistics of
stringy lumps changes as we approach the Hagedorn transition.

One interesting aspect of this may relate for instance to axion cosmology:
It is known~\cite{Georgi+,KamMar} that the $U(1)_{\rm PQ}$ Peccei-Quinn
symmetry
arising in axion models may never have been a perfect symmetry. If this is
the case, this may solve the domain wall problem arising in thermal
axion scenarios (i.e.~scenarios with no inflation below the Peccei-Quinn
scale where the Peccei-Quinn symmetry $U(1)_{PQ}$ arises)
with colour anomaly $N>1$.
The network of axionic domain walls, arising at the QCD scale, bounded by
axionic strings, which formed at the PQ scale, may never have been an
infinite domain wall network. This would solve the domain wall problem
in thermal axion scenarios.

\section{Open Questions}
\markboth{}{{\it
OPEN QUESTIONS}
\hrulefill}

\subsection{Features Generic to Properties of the Vacuum Manifold}
\label{section:6.1}

We have seen that, at perfect symmetry, $\R P^2$ strings and $SO(3)$ strings
tend to have lower fractal dimension and, therefore, higher values of $b$
than the $U(1)$ strings. Within the high statistical errors, and considering
the extremely low value of $\Lambda$ used in ~\cite{KibZ2}, it is impossible
to conclude with certainty that
there are significant differences in these parameters
between $\R P^2$ strings and $SO(3)$ strings, although the mean value of
$D$ measured by Kibble for $SO(3)$ strings is $1.95$, i.e.~it is lower
than for $\R P^2$ strings. It would be worth investigating
whether the lower fractal dimension is generic to $\Z_2$ strings, to
non--abelian strings, or indeed -- as our hypothesis goes -- to the higher
string densities one achieves with these symmetry groups.
Having data for so few vacuum manifolds only, this question
remains unanswered at present.

Furthermore, the fraction of mass in string loops is lower in $SO(3)$ strings
($\approx$~5\%) \cite{KibZ2}, to be compared with caution, since a cubic
lattice
was used) than in $\R P^2$ strings ($\approx$~19\%) or $U(1)$ strings
($\approx$~35\%). The same questions as for the symmetry dependence of
$D$ and $b$ arise. Nevertheless, the tendency to have less loops with
increasing string density is apparent. Vachaspati's graphs~\cite{V91} for
the loop density, when extrapolated to zero bias, are extremely close to
zero.

One way of addressing this would be an approach similar to the one presented
in Fig.~\ref{fig:Drunning}, starting with a symmetry manifold of high
dimensions and transforming it smoothly into several different
lower--dimensional symmetries. This is planned in the future.
Ideally, one would want some analytic arguments for why such a
symmetry--group dependence should occur, if it can be observed consistently
for several different vacuum manifolds.

A partial answer can, however, be given right now: since Vachaspati's model is
equivalent to an $\R P^{\infty}$$\equiv S^{\infty}/\Z_2$ symmetry, the
sequence $\{U(1)\simeq S^1/\Z_2$,
$\R P^2$$\simeq S^2/\Z_2$, $SO(3)\simeq S^3/\Z_2$,
$\ldots$,$S^{\infty}/\Z_2\}$
has been probed to some extent, and it seems that, as we move along
this series, the string density increases {\it with} the dimensionality
of the vacuum manifold. For continuous representations of the symmetry
group, the exact string densities have been calculated by Vachaspati \cite{V91}.

We now know that, at least among the $\R P^N$ symmetries, the lowest string
density and lowest proportion of density in infinite strings is probably
achieved for a $U(1)$ symmetry. Now let us remind ourselves that
the smallest allowed string loops (in lattice units of the dual lattice)
on the tetrakaidekahedral lattice are
probably shorter than on any other sensible regular lattice, and that the
fraction of string mass in loops for a fixed vacuum manifold is therefore
likely to be higher on a tetrakeidekahedral lattice than on any other one.
We therefore suspect that the measured
proportion of infinite string density for $U(1)$ strings is
the lowest one can find on any regular lattice, and for any
$\R P^N$ vacuum manifold.
It is certainly the lowest found to date in VV simulations at perfect
symmetry on a regular lattice.
All vacuum manifolds - other than $U(1)$ -- with nontrivial $\pi_1$
have more dimensions that $U(1)$. If the dimensionality of the vacuum
manifold turns out to be the relevant paramter, the measured $63$\% for
$U(1)$ strings would actually be the lowest mass fraction of infinite
strings for {\it any} vacuum manifold on {\it any} regular lattice, if
measured with the VV algorithm.

However, when comparing these symmetry groups, even at perfect symmetry, one
observes that they just continue the trend one observes for spoiled
symmetries when moving away from the percolation threshold: as the string
density increases, not only the loop density, but also the
fractal dimension decreases.
This seems to suggest that the total string density may play
a fundamental role in deciding the fractal dimension of the strings,
even in an unbiased symmetry. We are lacking an analytic
argument for this hypothesis, and Fig.~\ref{fig:Dvsdensity} introduces
the additional complication that biased symmetries do not scale their
string statistics the same way as unbiased ones do, although the trend
of an increase in the
fractal dimension with decreasing density is maintained in
either case. Group theory alone does not seem to tell
much about where the strings go. Perhaps a renormalization group description,
generalisable to different symmetries, would be useful in shedding some
light on this question. This brings us to the next unsolved problem:
we do not have {\it any} renormalization group description which allows
to put in (and conserve) the string density as the fundamental parameter.

\subsection{Renormalization Group Arguments}
\label{subsection:RGA}

Having observed that the scaling relation Eq.~(\ref{eq:scaling}) holds
reasonably well through a whole range of values of the bias,
we would like to understand how this comes about, since percolation
theory does not really help us there (except when close to the threshold).
Because we are dealing
with a critical phenomenon, one would hope that renormalization
group arguments can shed light on this question. In this section
we will mention the difficulties with such attempts, and why we have
been unsuccessful (so far) in finding non--trivial fixed points.

It is straightforward, but tedious, to derive renormalization group arguments,
derived from RG ideas in percolation theory,
with vacuum field averaging (at least in the lattice description). The
technical parts of this appear in Appendix~\ref{chapter:RG}. Here we just
take the renormalization group polynomial as given.
Let us group the sites of a unit cell of the bcc lattice (i.e.~a cell with
spanning vectors $(1,0,0)$, $(0,1,0)$, and $(\frac{1}{2},
\frac{1}{2},\frac{1}{2})$) into a single super--site, such that the
coarse--grained lattice is again a bcc lattice with twice the lattice
spacing of the original lattice $a'=2a$. If we take a $U(1)$ symmetry
discretised by $\Z_3$, at arbitrary bias $\eta$, such that the probabilities
for the three vacuum values are $p(0)=p$ and $p(1)=p(2)=\frac{1}{2}(1-p)$,
with a sensible renormalization procedure (see Appendix~\ref{chapter:RG}),
the renormalized bias becomes
\[
p'=\frac{105}{8}p^8-55p^7+\frac{315}{4}p^6-\frac{63}{2}p^5-\frac{175}{8}
p^4+\frac{35}{2}p^3\,\,\,\,\,.\,\,\,\,\,
\]
The fixed points are the solutions to $p'(p)=p$, and are, as expected,
at 0, $\frac{1}{3}$, and 1. Because the renormalization procedure in
Appendix~\ref{chapter:RG} was chosen such that the field values over
the group of sites are averaged in some sensible way, this means that the
interpretation of the vacuum values, living at the sites of the tetrahedral
lattice, as horizon--volume average over the field value is sensible,
and that, even if the description is chosen to be more or less
coarse--grained than one correlation length per lattice spacing, we would
get the same scaling laws for any such lattice description at zero bias
($p=\frac{1}{3}$) or at maximum bias ($p=0,1$, which are trivial fixed
points, as no strings appear at all, and the field is in its true ground
state).

This is comforting in itself, but clearly this renormalization does not
preserve the features we have measured for small bias: $p=\frac{1}{3}$
is the {\it only} non--trivial fixed point, but is unstable
($dp'/dp>0$ at $p=\frac{1}{3}$), which means that {\it any}
series of coarse--grainings will end up at a trivial fixed point if
we start with any small bias. If this was a sensible discretisation procedure,
we would have to conclude that large loop sizes are exponentially suppressed
for {\it any} non--zero bias.
This renormalization procedure is clearly inappropriate when
we try to understand string percolation near the percolation threshold,
because $p_c\ne\frac{1}{3}$. For $p_c<p<\frac{1}{3}$ the simple averaging will
always produce a homogeneous background on large scales and therefore predict
no infinite strings at all.

For biases near the string percolation threshold, a meaningful
renormalization procedure would have to keep information on the percolation
properties of the least likely vacuum value. In that case, however, the
obvious approach is to take a standard percolation renormalization.
This would mean we gain nothing that we do not know yet, as the best we could
do to develop some intuitive understanding of the Hagedorn transition on
the lattice was to take percolation theory results, before we even started
contemplating about the renormalization group.

However, it turns out that the renormalization group in three--dimensional
percolation is never exact~\cite{Stauffer} and needs systematic improvement,
so that the coarse--graining does not reconnect disconnected clusters or
disconnect connected ones\footnote{On some level, this problem will always
reappear. Therefore the systematic improvement does not yield an exact
renormalization group either.}. From Fig.~\ref{fig:Vach} it is obvious
(at least for Vachaspati's $\R P^{\infty}$ model) that such systematic
improvement would
differ, depending on whether we want to improve the renormalization group
for a string description or for a bond percolation problem. This explains
on a more formal basis why the Vachaspati model is so close to a percolation
problem but not quite identical to it.

Clearly, finding a systematically improved renormalization description
starting from a percolation picture would enhance our analytic understanding of
the Hagedorn transition and associated critical exponents. Depending on
how the details of this procedure will turn out to work, it may then also
be possible to extend it to groups other
than $S^{\infty}/\Z_2$ and the minimally discretised $U(1)$, for which the
percolation theory picture was so effective.

It is not at all obvious what variables should be conserved in such a
renormalization group description. We have seen in this section that
preserving vacuum field averages alone is not enough.
Towards the end of Appendix~\ref{chapter:RG} we show
that preserving the string flux alone is not enough, either.

\section{Conclusion}
\markboth{}{{\it
CONCLUSION}
\hrulefill}

We have investigated how generally known concepts of statistics of topological
line defects are affected if a symmetry--breaking phase transition occurs in
such a fashion that the remaining symmetry in the manifold
of possible ground states is
only approximate. Such a concept has been familiar as a solution to
the domain wall problem. There it leads naturally
to a percolation theory understanding
of defect statistics, because the sets of possible ground states are
disconnected. So far this concept has not been worked out properly
for continuous symmetries, mainly perhaps because the
naive correspondence
with standard percolation theory breaks down for continuous symmetries.

Out of the variety of possible defects exhibited by non--simply connected
continuous vacuum manifolds,
we concerned ourselves in this work with string defects in both perfect, and
approximate symmetries.
With improvements in our algorithm, explained in detail in ref.~\cite{thesis}
and based on the methods of ref.~\cite{MHKS},
we have been able to go much further with the measurements presented here,
and are able to provide answers to a number of important questions.

{\em Is the fractal dimension of infinite strings precisely 2?}

For infinite strings in the U(1) models the answer, summarised in
Table \ref{table:DU1}, is yes (within the statistical errors of about
0.8\%\footnote{To reach this accuracy we have used a simple extrapolation to
cope with finite size effects}).  It should be noted, however, that this is the
infinite-lenght limit of a running effective fractal dimension (compare
ref.~\cite{Bradley}). For strings in the $\R P^2$ models
(see Table \ref{table:rpcontperf}) the answer appears to be no:
we have good evidence from
the low cut-off high-statisitics simulations of the continuous $\R P^2$
manifold that the fractal dimension is slightly lower than 2.  More
evidence is displayed in Figure \ref{fig:Drunning}, which
shows that the fractal
dimension of discretised $\R P^2$ strings is less than that of U(1) strings.

{\em Are strings scale invariant in the percolating phase?}

Scale invariance in our sense means that the loop size distribution follows
Eq.~\ref{eq:scaling_function}, and that Eq. \ref{eq:scaling} is satisfied.
Figure \ref{fig:scaling} shows that the assumption of
scale invariance
is consistent with our measurements for U(1) strings.  However, the
evidence for $\R P^2$ strings in Figures \ref{fig:RP2scaling} and
\ref{fig:RP2contscaling} is against
scale invariance, except at the percolation transition.
In both cases, the fractal dimension increases
as the string density decreases: the string gets more crumpled.
This is implicit in Vachaspati's results \cite{V91}.

{\em Are the critical exponents of the string percolation transition
universal?}

We have investigated several different representations of two
vacuum manifolds, U(1) and $\R P^2$, and for these models at least
we have found good evidence, summarise in Table \ref{table:percolation},
that the
answer is yes.  Furthermore, the critical exponents are those of
three-dimensional site or bond percolation.

Our claims of universality are aided by the construction of a
two-parameter generating function for the loop size distribution,
which has some of the qualitites of a partition function (but
no thermal intepretation).  The existence of such a
distribution implies that certain scaling relations charactersitic
of universality should be satisfied -- as indeed they
are (Table \ref{eq:beta_scaling}).

Our results
extend the work of Bradley et al \cite{Bradley} on trichord percolation, who
first noted the correspondance of the critical exponents with that
of site percolation.  With our variety of models of the manifolds, we
provide evidence that the correspondence is not confined to their
three-colour model (equivalent to our three point triangulation
of U(1)).

The simulations presented here
model the initial conditions of condensed matter
systems with a non--conserved order parameter after a rapid quench.
Dynamical simulations have been performed on systems where the non--conserved
order parameter is a complex scalar field $\phi$, both with
$\langle\phi\rangle=0$
(corresponding to zero bias), and $\langle\phi\rangle\ne 0$ \cite{MonGol92}.
It is
found that the introduction of a bias in the initial expectation value
of the order parameter results in the eventual departure from dynamical
scaling, with the density of the string network going as
\begin{equation}
\rho(t) \sim t^{-1} \exp(-gt^{3/2}),
\end{equation}
where $g$ depends approximately quadratically on the initial bias
$\langle\phi\rangle$. This is due to the network breaking up into isolated
loops, with an exponentially suppressed size distribution. When we published
ref.~\cite{MHKS}, it was not
clear whether this is due to the initial conditions possessing no
infinite string, or whether the infinite string somehow manages to chop
itself up into an infinite number of loops. If it had been the former, the
percolation transition would have to happen at very small bias, perhaps
even at $\langle\phi\rangle=0$,
for the departure from power-law scaling in
$\rho(t)$ was observed from rather small biases, down to
$\langle\phi\rangle=0.001$.
Having extended the work of ref.~\cite{MHKS} to (a) better statistics,
(b) much larger upper cutoff lengths,
(c) one more symmetry group, and (d) continuous representations of the
ground--state manifold, we have now ascertained that this is not the case.
The existence of infinite strings holds
for a range of biased initial conditions.

By the example of string defects,
this work has provided some intuitive, as well as some quantitative
understanding of how defects in
systems with continuous symmetries are affected by a lifting of the
degeneracy of the possible ground states. Since these effects also change the
late--time dynamics of systems with non--conserved order parameter, the hope
is that understanding the initial conditions carries us one step further in
understanding these dynamics. It should also provide
some incentive to study other defects under similar conditions, in theory,
and in the laboratory.

\section*{Acknowledgments}
KS is supported by PPARC Fellowship GR/K94836, and
MH  by PPARC Advanced
Fellowship  B/93/AF/1642.  Further support is provided by
PPARC grant GR/K55967, by
the European Commission under the Human Capital and Mobility
programme, contract no.~CHRX-CT94-0423, and by a Royal Society Research
Grant.
The authors would like to thank Julian Borril, Mark Bradley,
Ray Rivers, James Robinson, and Andy Yates for their comments and
discussions.

\newpage
\section*{\normalfont\LARGE\bfseries Appendix\\}
\appendix
\section{Failure of the Dodecahedral Discretisation of $\R P^2$}
\markboth{}{{\it
APPENDIX: THE DODECAHEDRAL DISCRETISATION OF $\R P^2$}
\hrulefill}
\label{chapter:failure}

\begin{figure}[htb]
\centering
\mbox{~}{\hbox{
\epsfxsize=160pt
\epsffile{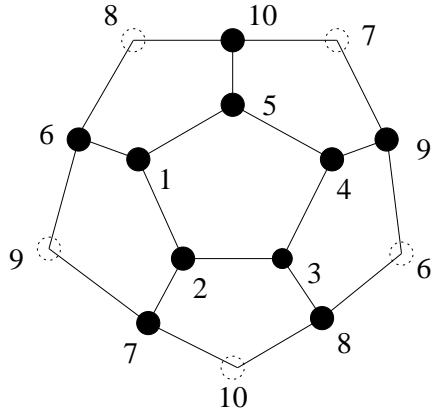}}}
\caption{The convention for
the numbering of the vertices of the dodecahedron in
Appendix~\ref{chapter:failure}. Points lying on the lower half--sphere are
drawn as dotted circles. The projections of the points 1-5 to the lower
half--sphere are not drawn, but all their links can be seen here.}
\label{fig:pentagon}
\end{figure}

Here we will prove that a discretisation of $\R P^2$ achieved by using only
those points which are the
vertices of a dodecahedron (DH) embedded into the sphere does
not force self--avoidance of the resulting string defects.
Let us take the upper half of the dodecahedron,
with its vertices numbered according to Fig.~\ref{fig:pentagon}.
Not all the point pairs are directly linked to each other, but all are
at most two links apart. There are no ambiguities in applying the geodesic
rule, because no pairs can be connected in more than one way, if we only
allow connection lengths of up to two links. For any assignment of such
a restricted selection of elements of $\R P^2$ to the vertices of a triangle,
we can therefore identify uniquely
whether the triangle is penetrated by
a string or not. This is equivalent to the statement that we could use the
same flux definition as before, because none of the projections to the
upper half--sphere of vectors pointing to
the vertices of the DH are at right angles to each other. This correspondence
will make the proof easier, because we can just check whether a given shortest
path between two points
crosses the equator and assign it a value of $-1$ if it does, or $1$ otherwise.
The product of all links will then indicate whether there is a string
present in a triangle (if the product is negative) or not.
Now let us take e.g.~the assignments 2, 5, 8, and 9 for the respective
vacuum fields on the vertices of a tetrahedron. Then the links
have the following ``equator--crossing indicators" assigned to them:
\[
\begin{array}{c}
(25)\rightarrow +1\\
(28)\rightarrow +1\\
(29)\rightarrow -1\\
(58)\rightarrow -1\\
(59)\rightarrow +1\\
(89)\rightarrow +1\\
\end{array}
\]
Since the two links which have $-1$ assigned to them are disconnected,
all triangles border exactly one of the two links, and all the link--variable
products are negative. There are therefore four string segments entering
the tetrahedron.\mbox{~~}{\Large $\Box$}

\section{Renormalization Groups}
\label{chapter:RG}

\subsection{Vacuum Field Averaging}

\begin{figure}[htb]
\centering
\mbox{~}{\hbox{
\epsfxsize=240pt
\epsffile{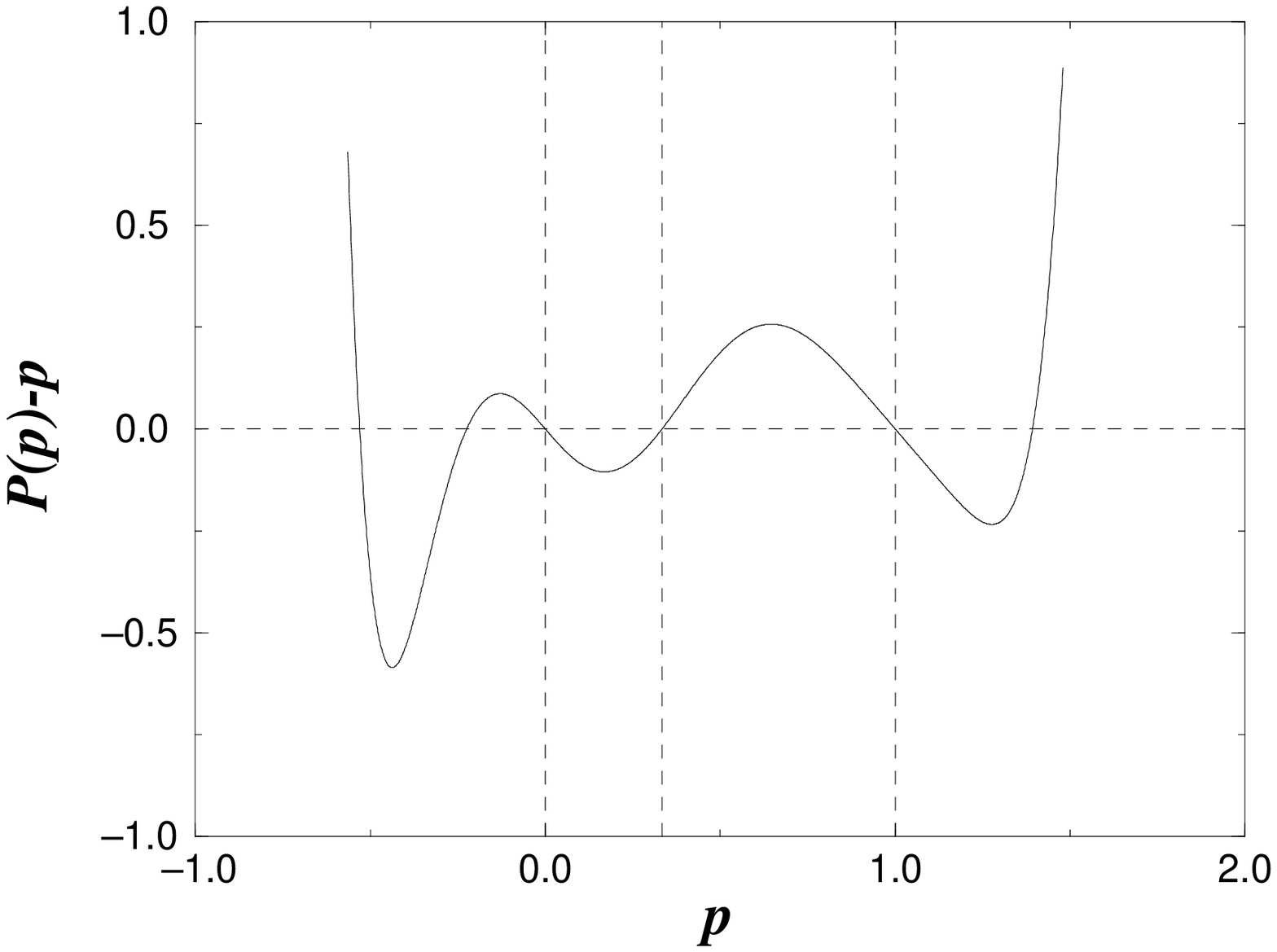}}}\\
\caption{The renormalization polynomial $P(p)-p$ of Eq.~(\ref{eq:renormU1}).}
\label{fig:3dimrenorm.point}
\end{figure}

Let us group the eight points of the bcc lattice cells spanned by
the vectors $(1,0,0)$, $(0,1,0)$, and $(\frac{1}{2},
\frac{1}{2},\frac{1}{2})$ such that the most frequently occurring vacuum
value will be the value assigned to the super--site. In case of ambiguities,
i.e.~if two values occur three times each or four times each, we take the
smaller value\footnote{For consistency, this means $0<1$, $1<2$, and $2<0$,
to preserve the periodicity of the circle.}.
Let us define the respective probabilities for the vacuum field values as
\begin{equation}
p(0)=p\,\,\,\,\,,\,\,\,\,\, p(1)=p(2)=\frac{1-p}{2}\,\,\,\,\,.\,\,\,\,\,
\label{eq:B.1}
\end{equation}
The possible configurations yielding, for the super-site, a field value of
zero, then have respective probabilities as listed
in Table \ref{table:RG}.
\begin{table}
\centering
\begin{tabular}{|c|c|c|}
\hline
field values & number of & probability \\
(+permutations) & permutations &\\
\hline
(0,0,0,0,0,0,0,0) & 1 & $p^8$ \\
(0,0,0,0,0,0,0,a) & $8\cdot 2$ & $p^7\left(\frac{1-p}{2}\right)$ \\
(0,0,0,0,0,0,a,a) & $27\cdot 2$ & $p^6\left(\frac{1-p}{2}\right)^2$ \\
(0,0,0,0,0,0,1,2) & 56 & $p^6\left(\frac{1-p}{2}\right)^2$ \\
(0,0,0,0,0,a,a,a) & $56\cdot 2$ & $p^5\left(\frac{1-p}{2}\right)^3$ \\
(0,0,0,0,0,a,a,b) & $168\cdot 2$ & $p^5\left(\frac{1-p}{2}\right)^3$ \\
(0,0,0,0,1,1,1,1) & 70 & $p^4\left(\frac{1-p}{2}\right)^4$ \\
(0,0,0,0,a,a,a,b) & $280\cdot 2$ & $p^4\left(\frac{1-p}{2}\right)^4$ \\
(0,0,0,0,1,1,2,2) & 420 & $p^4\left(\frac{1-p}{2}\right)^4$ \\
(0,0,0,1,1,1,2,2) & 560 & $p^3\left(\frac{1-p}{2}\right)^5$ \\
\hline
\end{tabular}
\label{table:RG}
\caption{The configurations of the eight sites grouped together to
a supersite, which yield a vacuum value of 0 for the super--site, together
with their respective probabilities, according to Eq.~(\ref{eq:B.1}).
It is assumed that $a\ne b$ and neither is equal to 0.}
\end{table}
This gives, for the probability of the renormalized super--site to have
a zero vacuum field
\begin{equation}
\begin{array}{rcl}
p'& = & P(p)\\[3mm]
& = & p^8+8p^7(1-p)+28p^6(1-p)^2+56p^5(1-p)^3+\frac{525}{8}
p^4(1-p)^4+\frac{35}{2}p^3(1-p)^5 \\[3mm]
& = & \frac{105}{8}p^8-55p^7+\frac{315}{4}p^6-\frac{63}{2}p^5-\frac{175}{8}
p^4+\frac{35}{2}p^3\,\,\,\,\,.\,\,\,\,\,
\end{array}
\label{eq:renormU1}
\end{equation}
The polynomial $P(p)-p$ is plotted in Fig.~\ref{fig:3dimrenorm.point}.
The zeros of this plot correspond to fixed points. Fixed points outside
the range $[0,1]$ have no interpretation in terms of probabilities.
We see that the only physical fixed points are at 0, $\frac{1}{3}$, and 1.
They are all expected, as discussed in section~\ref{subsection:RGA}.
What is not discussed there is that the stability of the fixed points
is determined by the sign of $dp'/dp=dP/dp$ at the fixed points.
As expected again, the fixed points 0 and 1 are the only stable ones, so
that this renormalization procedure would get rid of all infinite strings
for any $p\ne\frac{1}{3}$, which is clearly different from our measurements.
A renormalization procedure that merely performs field averages is therefore
not sufficient to describe the Hagedorn transition in the initial
conditions (but it may perhaps describe the disappearance of infinite
strings in the subsequent dynamical evolution, as observed in
ref.~\cite{MonGol92}).

We stress that we haven't been able to find any non-trivial fixed points
for various generalisations of Eq.~\ref{eq:renormU1}. It is
straightforward to write down the renormalisation polynomial for general
groupings of $n$ sites to a supersite, by generalising the criteria leading
to valid entries in Table~\ref{table:RG}
\begin{eqnarray}
p' & = & P^{(n)}(p)\\
& = & \sum_{\begin{array}{c}{\scriptscriptstyle l,k=0}\\[-2mm]
{\scriptscriptstyle n-l\ge 2k}\label{eq:RG2}\\[-2mm]
{\scriptscriptstyle n-k\ge 2l}\end{array}}^{l,k\le n/2}
\frac{n!}{l!\,k!\,(n-l-k)!}\,p^{n-l-k}\left(\frac{1-p}{2}\right)^{k+l}
\left(1-\frac{1}{2}\delta_{n-k-l,l}-\frac{1}{2}\delta_{n-k-l,k}
+\frac{1}{3}\delta_{n,3l}\delta_{l,k}\right),
\nonumber
\end{eqnarray}
which does not seem to have any physical fixed points other than
the trivial ones found with the polynomial in Eq.~\ref{eq:renormU1}.
The Kronecker deltas just divide out factors of two or three, depending
on whether the most frequent vacuum phases are more than one, as e.g.~in the
fourth to last and last row of Table~\ref{table:RG}.
We have tried the same approach for a minimally discretised $\R P^2$
symmetry where it is in fact much easier to write down the renormalisation
polynomial for the bias, but again, all the fixed points are the trivial
ones, and the percolation threshold cannot be located in this way.
For completeness, the easily derivable renormalisation polynomial,
if $p(0)=p(1)=p(2)=p/3$ and $p(3)=p(4)=p(5)=(1-p)/3$, and if we group
a number $n$ of sites to become a supersite of the renormalised lattice,
becomes
\[
p'=\sum_{k=0}^{[n/2]}\left[\left(\begin{array}{c}
n\\k\end{array}\right)p^{n-k}(1-p)^{k}\right]-\frac{1}{2}\delta_{2[n/2],n}
\left(\begin{array}{c}n\\n/2\end{array}
\right)p^{n/2}(1-p)^{n/2}\,\,\,\,\,.\,\,\,\,\,
\]
The bracket $[n/2]$ denotes the largest integer not larger than $n/2$, such
that the term containing the Kronecker-delta corrects for the double counting
of symmetric terms if $n$ is even. Again, the only physical fixed points for
all of these polynomials (with $n>2$) are the trivial ones at
$p=0,\frac{1}{2},1$, and no percolation threshold can be identified in this
way.

\subsection{Renormalizing the Vachaspati Model with Conserved Flux}

\begin{figure}[htb]
\centering
\mbox{~}{\hbox{
\epsfxsize=240pt
\epsffile{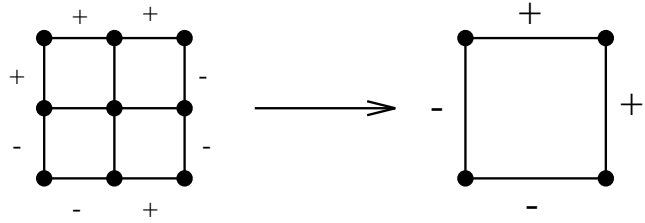}}}\\
\caption{The renormalization procedure for a grouping such that the
coarse--grained lattice spacing is twice the spacing on the original lattice.
The assignments to the links inside the finer lattice is irrelevant to the
total string flux. Changing any of the outer links of the fine lattice
changes the string flux, and changes one of the links of the coarse--grained
lattice, changing the string flux also in the renormalized description.}
\label{fig:RP2renorm}
\end{figure}

As the next obvious step, one could try a renormalization procedure which
conserves string flux. Clearly, this is most easily done for $\Z_2$ strings,
where the string flux can only be 0 or 1 (or better: ``even" or
``odd")\footnote{If the first homotopy group of the vacuum manifold is
e.g.~$\Z$,
then the string--flux through multiply renormalized super--sites can
become arbitrarily large (if it is to be conserved), such that in every
renormalization step the lattice
definition of the string flux would have to change together with the
probabilities of the relevant parameters.}.
As seen in all the models for
$\Z_2$ strings (i.e.~the Vachaspati model, our calculations for $\R P^2$, and
the calculations of Kibble for $SO(3)$ strings~\cite{KibZ2}), this is
essentially done by identifying lattice links as having one of the two
values $\pm 1$ assigned to them, and taking the product over a closed
contour.
The obvious approach is therefore to associate
two consecutive links with a super--link, and assigning to it the product
of the two link values. On any lattice (cubic or tetrahedral), this
preserves the string flux in the sense that the string flux through a
super--plaquette is zero (even) if the sum of the string lines penetrating the
constituent plaquettes is even, and odd (one) otherwise.
The procedure is outlined in Fig.~\ref{fig:RP2renorm}
The third dimension does not have to be drawn, since the renormalization
procedures for differently oriented links are independent from each other
(the second dimension is only necessary to show how changes in the string flux
survive the renormalization procedure).
Let us define the probability of finding a link with value $+1$ by $p_{+}=p$,
such that $p_{-}=1-p$.
Naively, one expects that, as the size of the super--plaquettes increases
with every renormalization step, the string flux for very large plaquettes
is 0 or 1 with equal probability, irrespective of the initial bias,
as long as the average number of expected (non--renormalized)
$(-1)$--links on the edge of the super--plaquette is $\gg 1$, i.e.~for any
non--trivial bias. The probabilities to have $-1$ or $+1$ assigned to
a link converge towards $p_{-}=p_{+}=\frac{1}{2}$, i.e.~$p_{+}=\frac{1}{2}$
should be a stable fixed point.
This is exactly what happens: $p'$ consists of two contributions: both links
of the non--renormalized lattice have to have the same value assigned to them
in order to make the super--link carry the value $+1$. Therefore
\[
p'=p^2+(1-p)^2\,\,\,\,\,.\,\,\,\,\,
\]
It is easy to see that the only fixed points are $\frac{1}{2}$ and
$1$\footnote{If the entire non--renormalized lattice consists of links
with $-1$ (i.e.~$p=0$), the renormalized lattice will consist entirely
of links with $+1$, such that $p=0$ is not a fixed point.}.
Furthermore, $dp'/dp=4p-2$, such that $p=1$ is an unstable fixed point.
$p=\frac{1}{2}$ is a stable fixed point and is approached exponentially
fast for any sequence of renormalization steps\footnote{This fast approach
is also understandable: as soon as the probability of having a value of $-1$
on any of the original links spanned by a super--link becomes appreciable,
the probabilities of having either $+1$ or $-1$ assigned to the super--link
are very nearly equal. The probability of encountering a $(-1)$--link
approaches 1 exponentially with the size $l$ of the super--link as $1-p^l$.}.

\begin{figure}[htb]
\centering
\mbox{~}{\hbox{
\epsfxsize=140pt
\epsffile{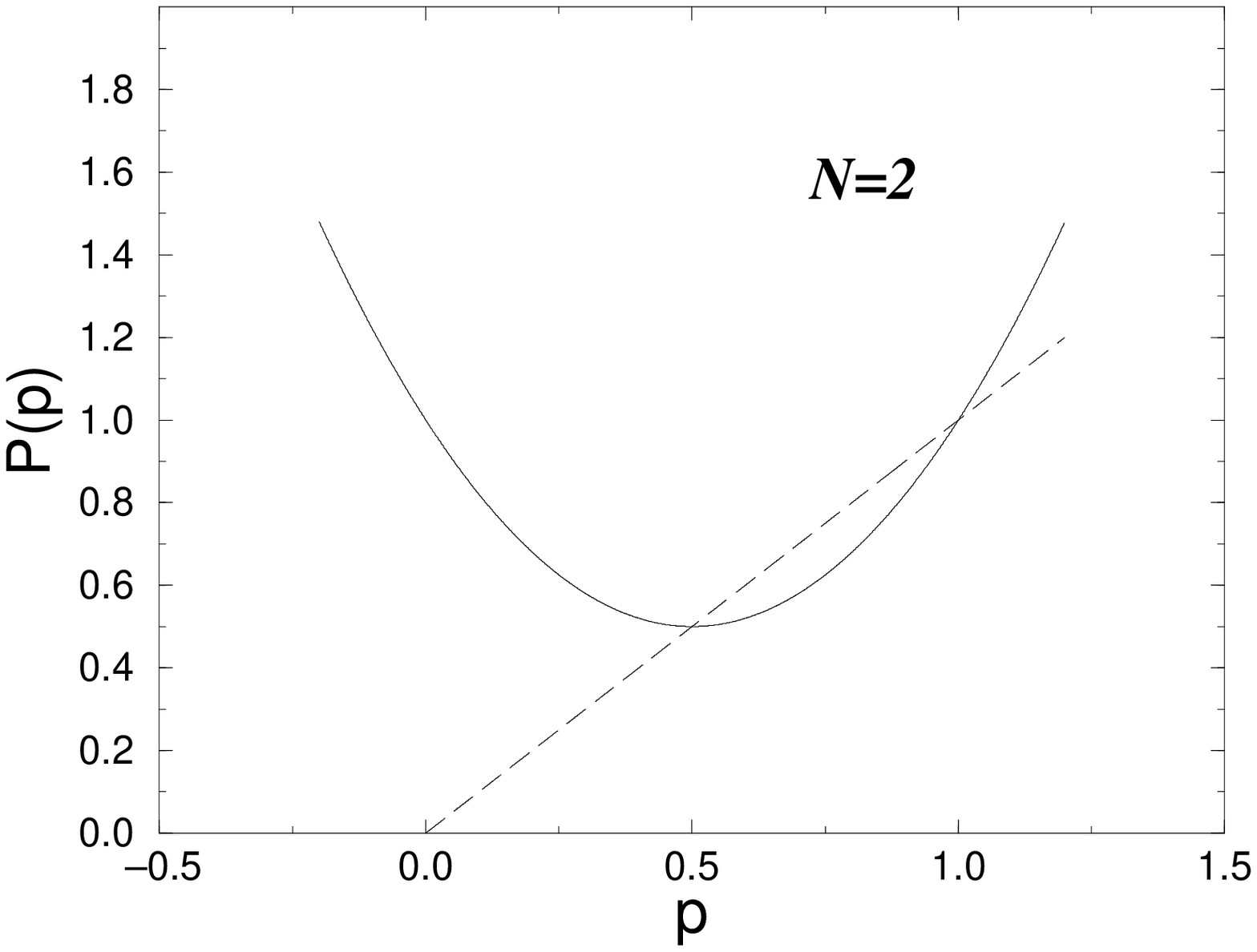}
\epsfxsize=140pt
\epsffile{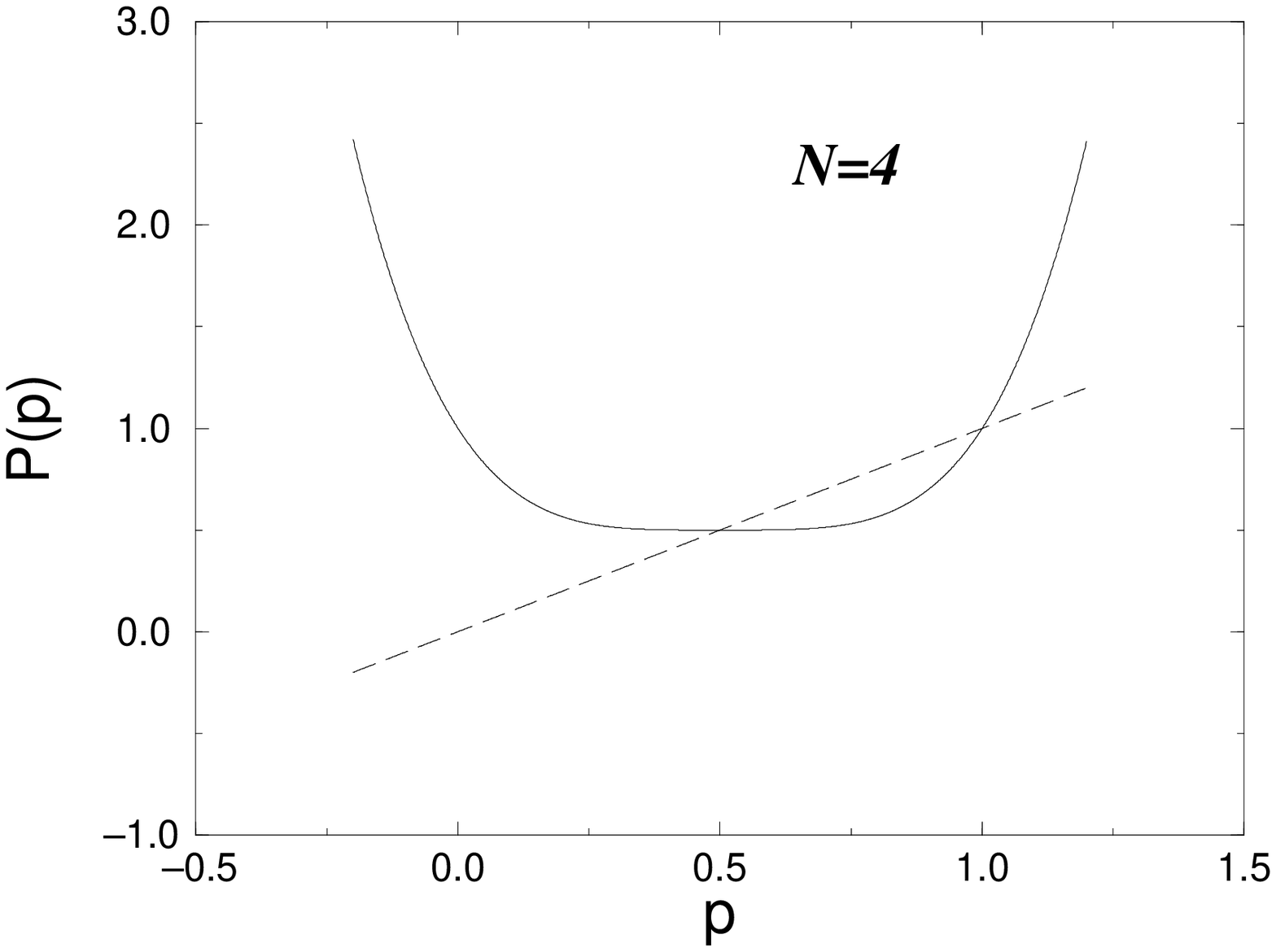}
\epsfxsize=140pt
\epsffile{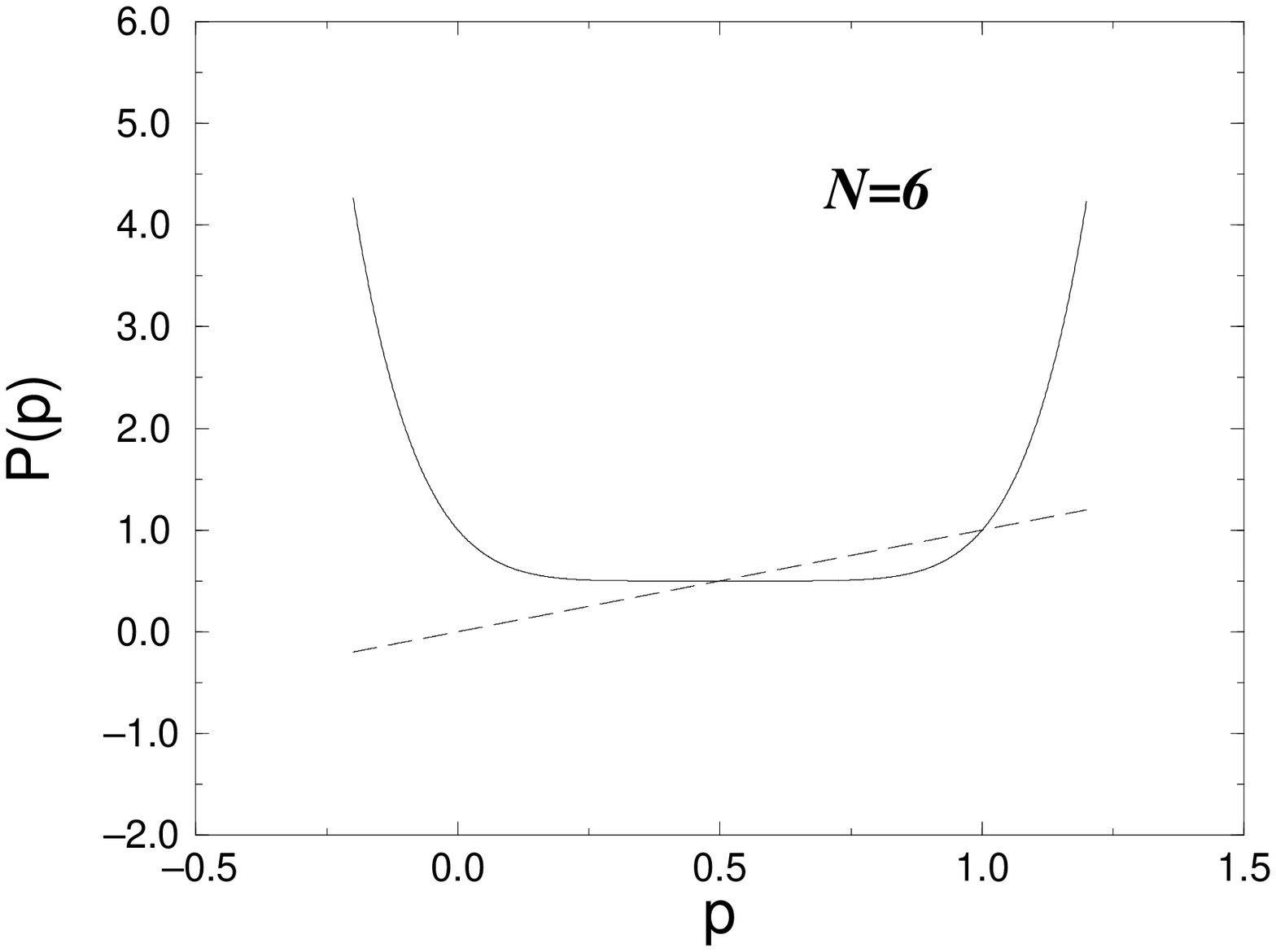}}}\\
\mbox{~}{\hbox{
\epsfxsize=140pt
\epsffile{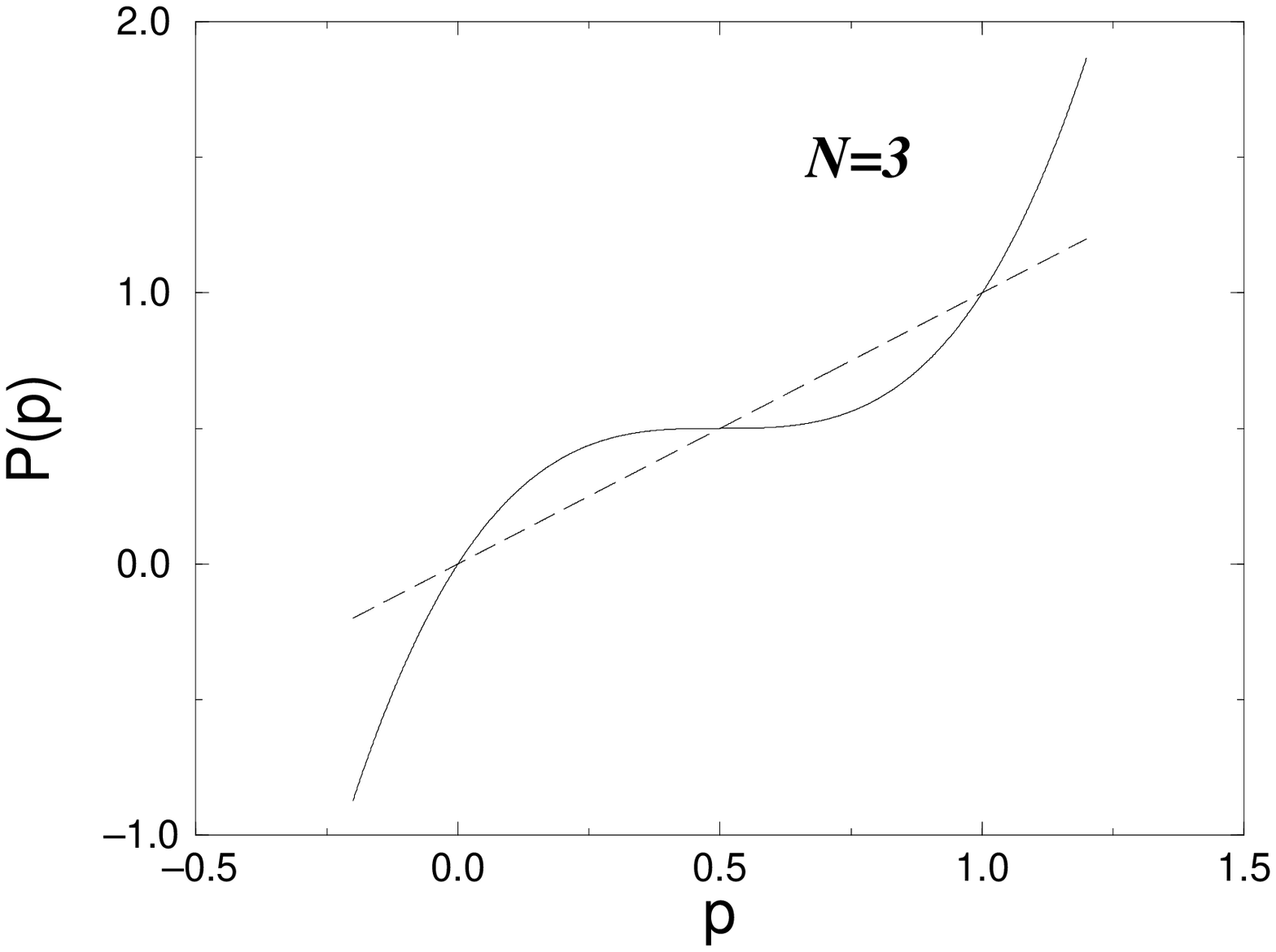}
\epsfxsize=140pt
\epsffile{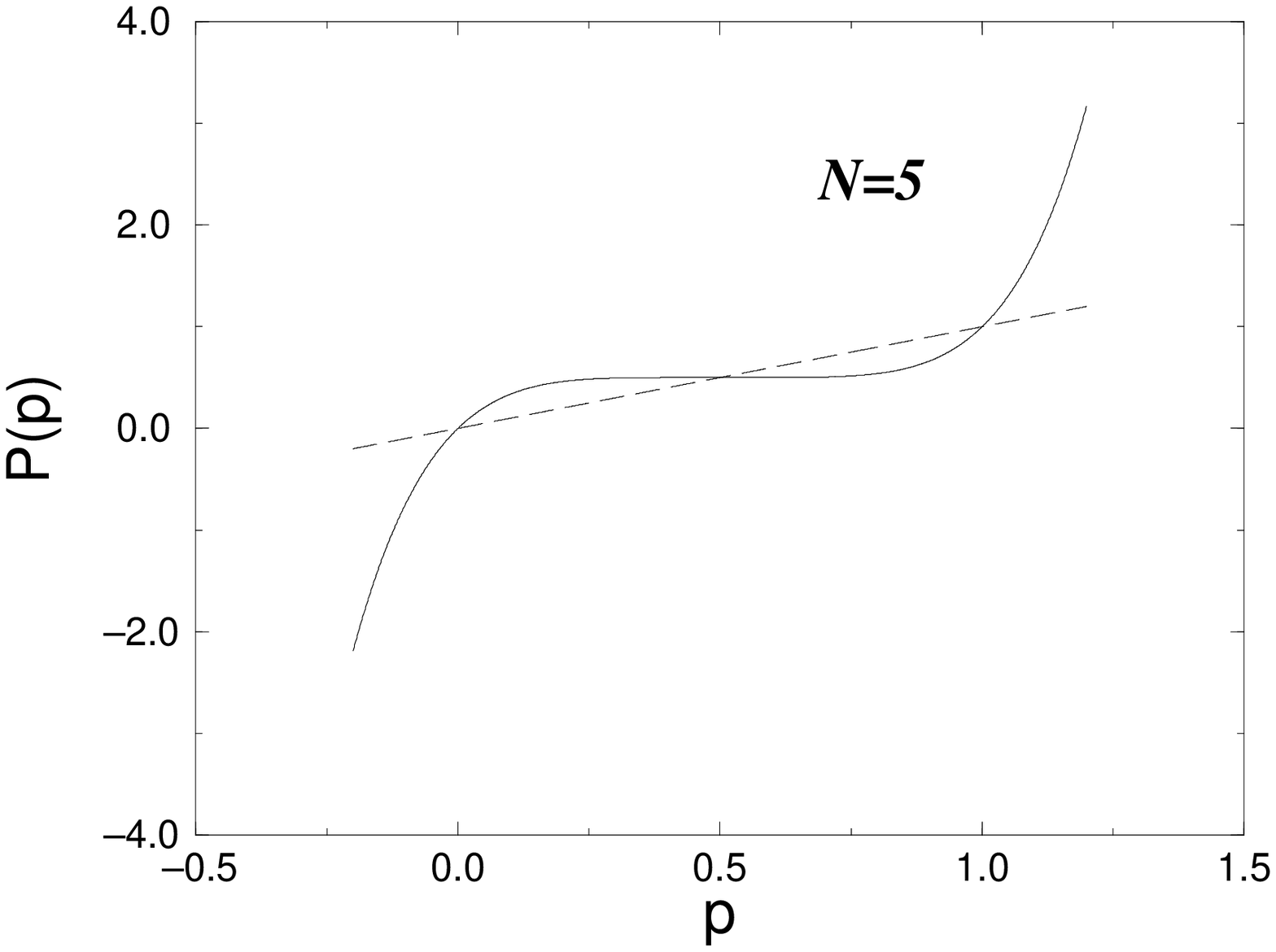}
\epsfxsize=140pt
\epsffile{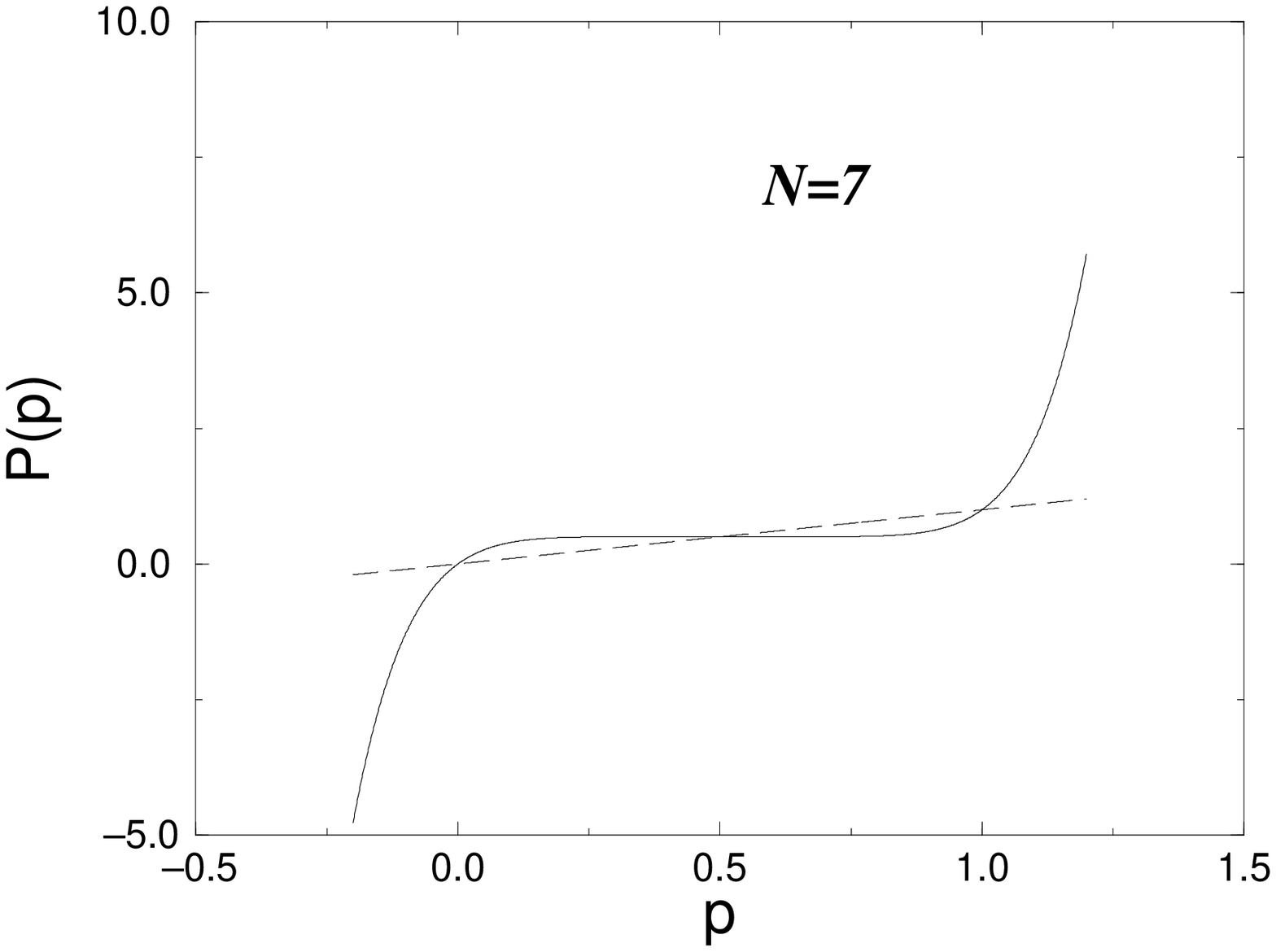}}}\\
\caption{The renormalization polynomial for orders 2 to 7. There is a clear
tendency for the potential to flatten out near $p=\frac{1}{2}$,
making the speed of
convergence towards the stable fixed point $p=\frac{1}{2}$ faster for larger
$N$. There is no tendency for the emergence of any fixed points other
than 0 (for odd polynomials only), $\frac{1}{2}$, and 1. The dotted line
is the line $p=p$. The intersections define the fixed points.}
\label{fig:P_N}
\end{figure}

This procedure cannot be refined by combining more than two links to a
super--link in the next step, since the arguments leading to the very
stable fixed point $p=\frac{1}{2}$ stay valid for any such renormalization.
This has been tested up to coarse--grainings by factors of seven.
The renormalization polynomial in any of these coarse--graining procedures
is (if the lattice is coarse--grained by a factor of $N$)
\[
p'=P_N(p)=\sum_{k=0}^{k\le N/2}
\left(
\begin{array}{c}
N\\
(N{\rm mod}\, 2) + 2k\\
\end{array}
\right)
\,p^{N-2k}\,(1-p)^{2k}
\,\,\,\,\,.\,\,\,\,\,
\]
For which it is trivially true that $p=1$ and $p=\frac{1}{2}$ is a fixed point.
Using
\[
1
+\left(\begin{array}{c}n\\ 2\\ \end{array}\right)
+\left(\begin{array}{c}n\\ 4\\ \end{array}\right)
+\ldots=2^{n-1}\,\,\,\,\,,\,\,\,\,\,
\left(\begin{array}{c}n\\ 1\\ \end{array}\right)
+\left(\begin{array}{c}n\\ 3\\ \end{array}\right)
+\left(\begin{array}{c}n\\ 5\\ \end{array}\right)
+\ldots=2^{n-1}\,\,\,\,\,,\,\,\,\,\,
\]
it is easy to see that $p=\frac{1}{2}$ is also a fixed point for all of
these renormalization procedures. Also, for odd numbers of $N$, $p=0$ is
a fixed point, as one expects (an odd number of $(-1)$--links gets
renormalized to a $(-1)$--super--link).
we have tested $P_N(p)$ for values up to $N=7$, and the mentioned fixed
points are really the only solutions of $P_N(p)=p$ for all of these.
Furthermore,
the speed of divergence of $P_N(p)$ outside the interval $[0,1]$ tends
to increase, so that it seems that the physical fixed points are in fact
the only real ones, and all other solutions to $P_N(p)=p$ are complex
numbers. Surely it should be possible to find a rigorous proof for this
statement, from the general form of $P_N(p)$, but instead we refer to
Fig.~\ref{fig:P_N}, which seems
to suggest very strongly a tendency for the $P_N(p)$ not to develop
any other physical fixed points as we increase $N$.

What this means is that the here proposed renormalization procedure is
also inappropriate for telling us about the scaling of the string network,
as all it tells us is that through very large surfaces (on super--horizon
scales), the probability of finding a $\Z_2$--string flux 1 is $\frac{1}{2}$.
In particular, it does not enable us to locate a string percolation
threshold.

Summarising both sections of this appendix, the question of how to
formulate a renormalization group
transformation which identifies the Hagedorn transition point as a fixed
point remains unanswered, as does the question of which variables one should
preserve in such a transformation.

\section{Comparison to Thermal Quantities}
\label{chapter:thermal}

In ref.~\cite{MHKS} we attempted to extract a statistical
temperature definition from
the probability distribution of strings. This can of course only be done
by analogy to statistical mechanics arguments.
It will turn out that the thus obtained temperature is meaningless in
physical terms, and that we arrive at a circular argument, giving us
that $c$ should be interpreted as an inverse temperature, which is not
the case, as we argued before. Thus, comparisons with thermal ensembles
or thermal partition functions are not appropriate.

If we define a probability density that a given string segment belongs to
a loop in the length interval $[l,l+dl]$ by $P(l)\propto(l\,{dn}/{dl})$,
we can define an information entropy, contained
in the probability distribution function by
\[
S=-\sum_l P(l) \ln P(l)\,\,\,\,\,.\,\,\,\,\,
\]
We can for instance try to relate the average length of string loops to an
average energy of a thermal ensemble. It would then make sense to define
a temperature
$\theta$ as
\[
\frac{1}{\theta}=\frac{\partial S}{\partial \langle l
\rangle}\,\,\,\,\,,\,\,\,\,\,
\]
and see how the system behaves under changes of the bias. In ref.~\cite{MHKS}
we did not have good enough statistics to extract that parameter, but with
the partition function we do not need to get involved with the numerically
highly unstable division of two differentials, because, without
normalization,
\[
P(l)=l^{-\tau+1}e^{-cl}\,\,\,\,\,.\,\,\,\,\,
\]
The entropy can then be split into two parts
\[
S=-\sum_l l^{-\tau+1}e^{-cl}\left[(-\tau+1){\rm ln}\,l -cl\right]=S_1+S_2
\,\,\,\,\,.\,\,\,\,\,
\]
Then
\[
S_2\propto c\langle l_{\rm loop}\rangle\propto c^{\tau-2}\,\,\,\,\,.\,\,\,\,\,
\]
$S_1$, by partial integration, behaves as
\[
\begin{array}{rcl}
S_1 & \propto & \int {\rm ln}\,(l)\,l^{-\tau+1}e^{-cl}\,dl \\
& \propto & c^{\tau-2}\left[{\rm const}_1-{\rm const}_2\,{\rm ln}c\right]
\,\,\,\,\,,\,\,\,\,\,
\end{array}
\]
Thus, with
\[
dS\propto(\Delta\eta)^{\frac{\tau-2-\sigma}{\sigma}}[1+{\rm const}
\,{\rm ln}\eta]d(\Delta\eta)
\,\,\,\,\,,\,\,\,\,\,
\]
and
\[
d\langle l_{\rm loop}\rangle\propto(\Delta\eta)^{\frac{\tau-3-\sigma}{\sigma}}
d(\Delta\eta)\,\,\,\,\,,\,\,\,\,\,
\]
we obtain
\[
\theta\propto(\Delta\eta)^{-\frac{1}{\sigma}}[1+{\rm const}
\,{\rm ln}\eta]\,\,\,\,\,,\,\,\,\,\,
\]
and the thus obtained temperature diverges, up to logarithmic corrections,
as $1/c$, and the Hagedorn transition happens at infinite $\theta$. Thus,
the correspondence of our partition function with a thermal one does not hold.
\mbox{~~}{\Large $\Box$}

\def\APJ#1{Ap.~J.~{\bf #1}}
\def\CMP#1{Comm.~Math.~Phys.~{\bf #1}}
\def\JPA#1{J.~Phys.~A {\bf #1}}
\def\MNRAS#1{Mon.~Not.~R.~Ast.~Soc.~{\bf #1}}
\def\MP#1{Mod.~Phys.~{\bf #1}}
\def\MPL#1{Mod.~Phys.~Lett.~{\bf #1}}
\def\NC#1{Nuovo Cim.~{\bf #1}}
\def\NPB#1{Nucl.~Phys.~{\bf B#1}}
\def\PA#1{Physica A~{\bf #1}}
\def\PL#1{Phys.~Lett.~{\bf #1}}
\def\PLB#1{Phys.~Lett.~{\bf B#1}}
\def\PRep#1{Phys.~Rep.~{\bf #1}}
\def\PRD#1{Phys.~Rev.~{\bf D#1}}
\def\PRL#1{Phys.~Rev.\ Lett.~{\bf #1}}
\def\RMP#1{Rev.~Mod.\ Phys.~{\bf #1}}
\def\YF#1{Yad.~Fiz.~{\bf #1}}
\def\ZETF#1{Zh.~Eksp.~Teor.~Fiz.~{\bf #1}}

\def\rf#1#2#3#4#5{#1, {#3} {#4} (#2) #5}

\end{document}